\documentclass[prb,preprint,titlepage,amsmath,amssymb,longbibliography,aps]{revtex4-2}

\newcommand{\arm}{\xi}

\usepackage[inner=1.5cm,outer=1.5cm,bottom=2cm,tmargin=1cm]{geometry}
\usepackage{graphicx}

\usepackage[colorlinks]{hyperref}
\hypersetup{
    colorlinks=true,
    linkcolor=blue,
    filecolor=blue,      
    urlcolor=blue,
    citecolor=blue
}

\usepackage{mathtools}
\usepackage{sansmath}
\newcommand{\iu}{\text{i}}
\newcommand{\T}{\text{T}}
\renewcommand{\vec}[1]{\boldsymbol{#1}}
\DeclareMathOperator{\cov}{cov}
\DeclareMathOperator{\sinc}{sinc}
\DeclareMathOperator{\laguerre}{L}
\newcommand{\qFp}[2]{{}_#1 F_#2}
\newcommand{\e}{e}
\DeclareMathOperator{\diag}{diag}
\newcommand{\one}{\mathbb{I}}
\DeclareMathOperator{\rect}{rect}
\newcommand{\hc}{\text{H.c.}}
\DeclareMathOperator{\sign}{sign}
\newcommand{\PI}[1]{#1^+}

\usepackage{physics}
\usepackage{siunitx}
\begin{document}
\sansmath

\title{Time-domain field correlation measurements enable tomography of highly multimode quantum states of light}

\author{Emanuel Hubenschmid}
\email[]{emanuel.hubenschmid@uni-konstanz.de}
\affiliation{Department of Physics, University of Konstanz, D-78457 Konstanz, Germany}

\author{Guido Burkard}
\email[]{guido.burkard@uni-konstanz.de}
\affiliation{Department of Physics, University of Konstanz, D-78457 Konstanz, Germany}

\begin{abstract}
Recent progress in ultrafast optics facilitates the investigation of the dynamics of highly multimode quantum states of light, as demonstrated by the application of electro-optic sampling to quantum states of the electromagnetic field.
Yet, the complete tomographic reconstruction of optical quantum states with prior unknown statistics and dynamics is still challenging, since state-of-the-art tomographic methods require the measurement of many orthogonal and distinguishable modes.
Here, we propose a tomography scheme based on time-domain quadrature correlation measurements and theoretically demonstrate its ability to reconstruct highly multimode Gaussian states.
In contrast to (eight-port) homodyne detection, the two local oscillator pulses are shorter in time and are (independently) time-delayed against the pulsed quantum state.
The distinguishable mode structure is obtained in post-processing from the correlation measurement data by orthogonalization.
We show that the number of reconstructable modes increases with the number of time delays used and decreases with the temporal extent of the local oscillator.
We extend and optimise our proposed correlation measurement to electro-optic sampling by adding a nonlinear crystal prior to the homodyne detection, potentially achieving subcycle resolution in the mid-infrared to THz regime.
By analysing the (quantum) correlations present in the measurement data, we show how thermalisation of the quantum state during detection leads to the requirement of correlation measurements.
The thermalisation is especially pronounced in the strong squeezing limit, for which we developed a non-perturbative theory.
Furthermore, we open an avenue to extending our tomography scheme to non-Gaussian states by theoretically establishing the complete joint statistics of the correlation measurement and showing how to obtain spectral information about pulsed Fock states from the joint statistics.
\end{abstract}

\maketitle

Pulsed quantum states of light are vastly complex and rich objects with numerous degrees of freedom - a property which makes these states desirable or even unavoidable for many applications spanning from fundamental investigation of the dynamics of quantum systems \cite{Gulla2021,Yanagimoto2024} to the transmission of quantum information over large distances \cite{Weedbrook2012,Braunstein2005,Weedbrook2004,Lance2005,Madsen2012,Usenko2015,Diamanti2015,Hosseinidehaj2019,Silberhorn2002,Hillery2000}.
Full quantum state tomography of the radiation field, i.e., the complete characterization encompassing the statistics of pulsed (free space) quantum states poses a challenge even for state-of-the-art experiments \cite{Slusher1987,Hirano1990,Smithey1992,Smithey1993a,Smithey1993b,Zavatta2002,Zavatta2005,Haderka2009,Okubo2008,Ansari2017,Tiedau2018,Ansari2018,GilLopez2021,Kalash2023,Serino2023}, since the information describing the state can be distributed among a high number of optical modes.

In this work, we propose a new method, which we term \textit{correlation tomography}, that only relies on time-domain correlation measurements of two electromagnetic field quadratures. These measurements allow for individual time delays between a pulsed quantum state and the detection time window.
The key insight is that the covariances between distinguishable modes of quantum states can be recovered from temporally overlapping correlation measurements by orthogonalizing the measurements in post-processing.
We find that the number of reconstructable modes scales with the bandwidth of the detection and the number of time delays used in the measurement.
We compare two implementations of correlation tomography: One based on homodyne detection and one on electro-optic sampling. The latter comprises a nonlinear crystal before the homodyne detection.
By optimising the pulse driving the nonlinear interaction in the crystal, one can shift the reconstructed modes toward lower frequencies, potentially resolving dynamics below a single optical oscillation and entering the subcycle regime.
We analyse signs of (quantum) correlations present in the measurement data and argue how thermalisation due to entanglement breakage leads to the requirement of correlation measurements even for pure states such as the squeezed vacuum.
Furthermore, we open an avenue to the extension to non-Gaussian states by providing the full joint statistics of time-domain quadrature measurements and show how to obtain information about the dynamics of a Fock state occupying a single temporal mode from correlation measurements.

The complete dynamics of a pulsed quantum state (for some fixed polarization) can be described in phase space, spanned by the electric-field-related quadrature $\hat{p}_{\omega} = \frac{\iu}{\sqrt{2}}(\hat{a}_{\omega}^\dagger - \hat{a}_{\omega})$ and its conjugate $\hat{x}_{\omega} = \frac{1}{\sqrt{2}}(\hat{a}_{\omega}^\dagger + \hat{a}_{\omega})$ with (angular) frequency $\omega$ and corresponding Bosonic annihilation operator $\hat{a}_{\omega}$ \cite{Mrowczynski1994,Roux2019}. Defining a quantum state in the phase space of continuous frequency $\omega$ can lead to divergences \cite{Virally2019,Roux2020}, whereas discretizing the (free space) phase space using a basis of orthogonal functions $f_i(\omega)$ allows one to describe the quantum state by a multimode Wigner function \cite{Adesso2014,Weedbrook2012}.
Since the Fourier transforms $f_i(t)$ of the mode functions allow a description in the time domain, the corresponding modes are called temporal modes \cite{Raymer2020,Brecht2015}.
To describe low-frequency and broadband modes, we introduced the subcycle mode basis.
The fundamental mode of the subcycle basis is described by the frequency scaling and cycle parameters $\sigma_0$ and $k_0$, determining the number of optical cycles completed during the temporal extent of the pulse (the pulse is subcycle for $0<k_0\le 1$, see Methods for more details).
The parameters $\sigma_0$ and $k_0$ can be used to calculate the centre frequency $\bar{\omega}_0$ and bandwidth $\Delta \omega_0$.
The orthogonal functions define two sets of orthogonal quadrature operators $\hat{x}_i = \int_0^\infty f_i(\omega) \hat{x}_\omega \dd \omega$ and $\hat{p}_i = \int_0^\infty f_i(\omega) \hat{p}_\omega \dd \omega$, each pair $(\hat{x}_i,\hat{p}_i)$ acting on the same temporal mode.
If we collect the eigenvalues of the operators $\hat{\vec{\zeta}} = (\hat{x}_1, \hat{x}_2, \ldots, \hat{p}_1, \hat{p}_2, \ldots)^\T$ in a vector $\vec{\zeta}$, the pulsed quantum state can be represented by the multimode Wigner function $W(\vec{\zeta})$.
For multi-mode Gaussian states, the Wigner function,
\begin{equation}\label{eq:gaussian_wigner}
	W(\vec{\zeta}) = \frac{1}{\sqrt{\det(\cov)}}\exp[-\frac{1}{2}(\vec{\zeta} - \vec{\mu})^\T \cov^{-1} (\vec{\zeta} - \vec{\mu})]
,\end{equation}
is completely characterized by the expectation value $\vec{\mu}$ and the (symmetric) covariance matrix $\cov$, with the determinant of the covariance matrix $\det(\cov)$ \cite{Adesso2014,Weedbrook2012}.
Tomography of optical quantum states is usually accomplished by reconstructing the Wigner function (or a smoothed out, positive phase space density like the Husimi function), from individual quadrature measurements at different directions in phase space \cite{Walker1986,Freyberger1993,Leonhardt1993,Zucchetti1996,Rehacek2015,Hubenschmid2022} (or simultaneous measurement of two noncommuting quadratures \cite{Vogel1989,Smithey1993a,Smithey1993b,Leonhardt1994,Wallentowitz1996,Breitenbach1997,Zavatta2002,Zavatta2005,Luis2015,Tiedau2018,Bohmann2018,Knyazev2018,Olivares2019,Kalash2023}).

Quadrature measurements can be implemented using homodyne detection where the quantum pulse to be measured is interfered at a beam splitter with a strong, coherent reference pulse, called local oscillator (LO).
The difference between the photon number detected at each output port constitutes a quadrature measurement in the temporal mode defined by the local oscillator pulse.
Yet, the reconstruction of the multimode Wigner function $W(\vec{\zeta})$ would require the measurement of the complete high-dimensional phase space statistics and even in the Gaussian case of equation~\eqref{eq:gaussian_wigner} would require the measurement of all covariances between temporal modes.
One approach is to measure in a quadrature basis for which the covariance matrix $\cov$ is diagonal \cite{Ansari2017,Tiedau2018,Ansari2018,GilLopez2021,Serino2023}.
Measuring in the eigenbasis of $\cov$ is optimal in the number of measurements required for a complete reconstruction, i.e., for the determination of the variances corresponding to the most significant modes.
While being very efficient, this approach requires knowledge of the most significant modes constituting the pulsed quantum state.

In contrast to matching the (temporal) mode of the local oscillator to the quantum state, as described above, recent works have proposed a quantum state tomography in the time domain, accessing the (time local) phase-space dynamics of free-space quantum states with a high temporal resolution \cite{Hubenschmid2024,Yang2023,Onoe2023,Lordi2024,BeneaChelmus2025}. 
Inspired by the electro-optic sampling of (squeezed) vacuum fluctuations \cite{Riek2015,Riek2017,BeneaChelmus2019,Moskalenko2015,Kizmann2019,Guedes2019,Kizmann2022,Onoe2022,Guedes2023}, time-domain quantum state tomography uses an ultrashort local oscillator pulse to sample the phase space statistics of the quantum pulse.
By repeating the reconstruction for different time delays between the local oscillator and quantum pulse, the (time-local) dynamics of the quantum state can be scanned through.
Electro-optic sampling can be understood as homodyne detection with the local oscillator pulse in a higher frequency range as the sampled state \cite{Namba1961,Gallot1999,Leitenstorfer1999,Sulzer2020,Moskalenko2015,Riek2015,Riek2017,Kempf2024,Guedes2019,Kizmann2019,BeneaChelmus2019,Lindel2020,Lindel2021,Virally2021,Beckh2021,Onoe2022,Kizmann2022,Guedes2023,Settembrini2022,Settembrini2023,Lindel2023,Lindel2024,Schubert2014,Langer2016,BeneaChelmus2025}.
We refer to the coherent pulse involved in the nonlinear interaction as probe and in the subsequent homodyne detection as local oscillator.
In our description probe and local oscillator pulse can differ.
To enable the interference of the high-frequency probe and the lower-frequency state, the state is upconverted to the frequency range of the probe with the aid of a nonlinear interaction.
In general, the frequency conversion can involve sum-frequency (SF) and difference-frequency (DF) processes.
In many cases, the DF processes are suppressed by a wave-vector mismatch between the involved photons; however, they are matched in the subcycle regime and have to be accounted for in electro-optic sampling.
The simultaneous presence of both SF and DF processes affects the quadrature correlations generated in the nonlinear interaction \cite{Hubenschmid2024}, similar to the effect of losses during the interaction \cite{Kopylov2024}.
In the Supplementary Section~V, we present a new approach that still allows the identification of the most significant modes contributing to the nonlinear interaction.
Since electro-optic sampling is an indirect measurement, it exhibits shot noise depending on the probe photon number and thermalisation due to entanglement breaking in the detection \cite{Hubenschmid2024}.
However, by correlating many degrees of freedom, the nonlinear interaction has been shown to increase the resilience against losses in quadrature measurements \cite{Kalash2023, Hubenschmid2024, Tziperman2024}.
Yet, if the probe pulse is sufficiently strong to suppress shot noise and the local oscillator is frequency-filtered appropriately after the nonlinear interaction to reduce thermalisation, electro-optic sampling is mostly equivalent to homodyne detection.
The advantage of electro-optic sampling compared to homodyne detection is the ability to sample low frequencies, usually in the mid-infrared to THz range, with a broadband probe at optical frequencies, thus approaching a subcycle temporal resolution.
In this frequency range, the dynamics is too fast for electronics, but the photon energy is comparable to thermal energies at room temperature, making efficient photo-detection challenging.
Therefore, electro-optic sampling and homodyne detection can be used complementarily, covering a frequency range from microwave to optical, with electro-optic sampling accessing the mid-infrared to THz range \cite{Namba1961,Gallot1999,Leitenstorfer1999,Sulzer2020,BeneaChelmus2025}, filling the gap between the microwave \cite{Yurke1984,Virally2019} and optical \cite{Walker1986,Breitenbach1997} frequency ranges that homodyne detection is available for.
Electro-optic sampling has recently been extended to temporal quadrature correlation measurements and applied to the ground state of the electromagnetic field \cite{BeneaChelmus2019,Lindel2020,Lindel2021,Settembrini2022,Settembrini2023,Lindel2023,Lindel2024}.
A similar method based on the time-domain correlation measurement method has been applied to investigate magnetic properties exploiting the magneto-optic instead of the electro-optic effect \cite{Weiss2023}.
However, the application of temporal quadrature correlation measurements to more intricate, highly multimode (Gaussian) quantum states of the electromagnetic field is still missing.
We fill this gap with our proposal to reconstruct the orthogonal mode structure of pulsed Gaussian quantum states from overlapping time-domain correlation measurements.

\section{Time-domain quadrature correlation measurements}\label{sec:td_correlation}
\begin{figure}
	\includegraphics[width=\textwidth]{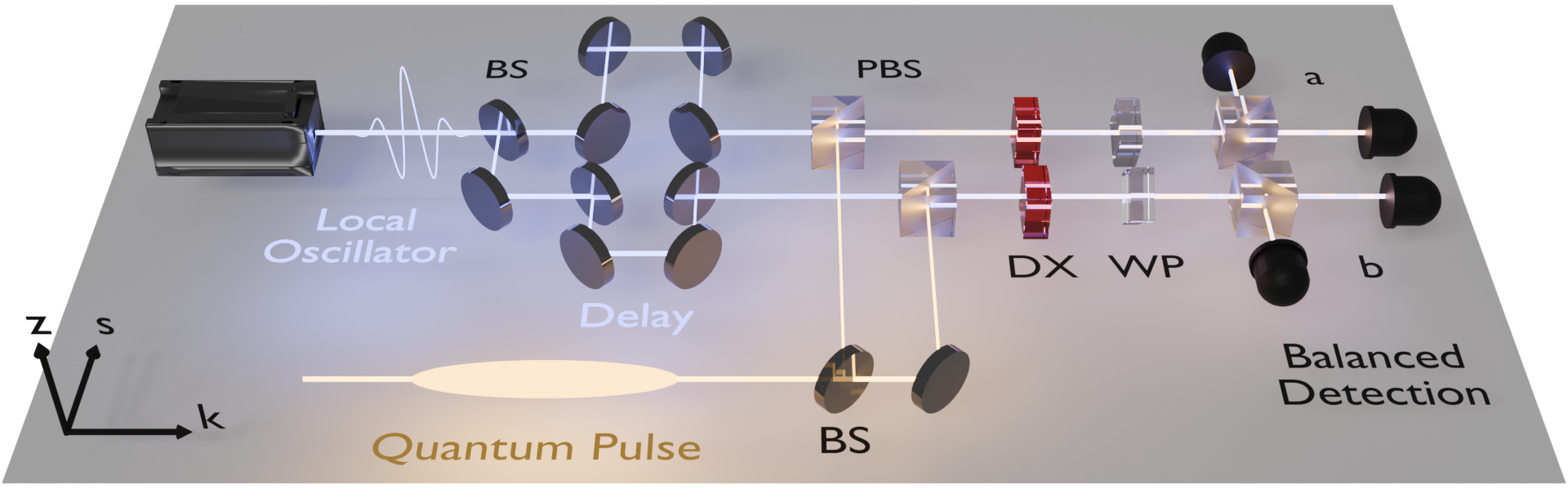}
	\caption{%
		\label{fig:setup}
		\textbf{Schematic of the proposed quadrature correlation measurement.}
		The $ z$-polarised, short local oscillator (LO) pulse (blue) and $ s$-polarised quantum pulse (orange) are subdivided by beam splitters (BS) and directed towards the two detection arms referred to as $a$ and $b$.
		Before combining in a polarising beam splitter (PBS) with the quantum pulse, each local oscillator pulse traverses a delay stage, adding the time delays $\Delta t_a$ or $\Delta t_b$ between the quantum state and LO.
		If the detection stage is implemented through electro-optic sampling, the quantum pulse is at lower frequencies (usually THz to mid-infrared) than the LO (usually optical frequencies), which is called the probe pulse in this context.
		The quantum pulse is upconverted to the probe's frequency range by interacting in a nonlinear (detection) crystal (DX).
		Subsequently, the quadrature of the (potentially upconverted) quantum pulse in the temporal mode of the LO is measured by homodyne detection.
		Different from usual homodyne detection, the quantum and classical pulses are mixed in a $\Phi$-wave plate (WP) with its optical axis rotated by $\theta$.
		Subsequently, on each detection stage, the $s$- and $ z$-polarised contributions are separated by a PBS.
		A balanced photon detection results in a measurement signal that amounts to the difference between the photon numbers in the $s$- and $ z$-polarisation.
		If the detection stage is implemented as homodyne detection, the nonlinear crystal is removed.
		}
\end{figure}

The time-domain quadrature correlation measurement we propose is schematically depicted in Fig.~\ref{fig:setup}.
To enable correlation measurements, the local oscillator and quantum pulse are subdivided using 50:50 beam splitters and are directed towards two detection arms, which we will refer to as arm $a$ and arm $b$.
The two local-oscillator pulses traverse a delay stage, which adds the time delays $\Delta t_a$ and $\Delta t_b$ relative to the pulsed quantum state to the respective local oscillators.
Subsequently, the local oscillator and quantum pulse combine in a polarizing beam splitter (PBS) and enter the detection stage.

Each detection stage constitutes a homodyne measurement and can thus be related to a quadrature measurement of the temporal mode defined by the local oscillator.
One subtle difference to usual homodyne detection is the orthogonal polarization of the local oscillator and the quantum pulse in the $z$- and $s$-direction.
This requires a $\Phi$-wave plate with its optical axis rotated by $\theta$ to produce interference between the two pulses, as well as a polarizing beam splitter to separate the $s$- and $z$-polarisation direction for balanced photodetection.
The combined effect of the wave plate and polarizing beam splitter replaces the beam splitter and unrotated wave plate in usual homodyne detection \cite{Hubenschmid2022,Hubenschmid2024}.
The wave plate leads to a phase shift $\varphi = \varphi(\Phi, \theta)$ of the local oscillator (see Methods).
The use of a half-wave plate rotated by $\SI{22.5}{\degree}$ results in an $\hat{x}$-quadrature measurement with $\varphi = 0$, while choosing a quarter-wave plate rotated by $\SI{45}{\degree}$ results in a $\hat{p}$-quadrature measurement with $\varphi = \pi / 2$.
Setting up homodyne detection to be between different polarizations enables the extension to electro-optic sampling by including a nonlinear crystal in front of the wave plate, referred to as the detection crystal (DX).
After the nonlinear interaction, the probe pulse can be replaced by a new local oscillator pulse to optimise the detection.

The initial state of the local oscillator pulse at detection arm $\arm = a,b$ directly after the first beam splitter, can be represented by the vector $\vec{\zeta}^\T_{\text{LO}}$ in the high-dimensional phase space with respect to the mode basis $f_i(\omega)$ (see Methods for definition).
Since all components of the setup in Fig.~\ref{fig:setup} correspond to a quadratic Hamiltonian, we can describe the time evolution as a symplectic transformations, $M(t_2,t_1)$, of the quadratures, e.g., $\vec{\zeta}(t_2) = M(t_2,t_1) \vec{\zeta}(t_1)$ \cite{Weedbrook2012,Adesso2014}.
Thus, at arm $\arm = a,b$ during the balanced detection, the local oscillator is described by the vector $\vec{\zeta}^\T_{\text{LO}}(\Delta t_\arm, \alpha_{\text{DX}}, \varphi_\arm) = M_{\text{DLY}}^\T(\Delta t_\arm) M^\T_{\text{NL}}(\alpha_{\text{DX}}) M_{\text{WP}}^\T(\varphi_\arm) \vec{\zeta}_{\text{LO}}$, depending on the time delay $\Delta t_\arm$ added by the delay stage, described by $M_{\text{DLY}}(\Delta t)$, the probe amplitude $\alpha_{\text{DX}}$ driving the nonlinear interaction, $M^\T_{\text{NL}}(\alpha_{\text{DX}})$, and phase $\varphi_\arm$ added to the local oscillator by the wave plate, $M_{\text{WP}}^\T(\varphi)$.
For more details, see Methods.

As a final step to relate the quadrature of the quantum pulse at the input to the quadrature measured by one of the balanced detections, the effect of the beam splitter has to be considered.
While one input port of the beam splitter is occupied by the pulsed quantum state (with quadrature operators $\hat{\vec{\zeta}}$), the other input port couples to the vacuum (with quadrature operators $\hat{\vec{\zeta}}_{\text{vac}}$).
The vacuum at the empty port leads to additional noise to the detected photon-count difference.
This additional noise is fundamentally unavoidable in the simultaneous detection since it ensures the positivity of the probability distribution over the photon-difference counts \cite{Hubenschmid2022, Hubenschmid2024}. 
Thus, the detection arm $\arm = a,b$ measures the quadrature,
\begin{equation}\label{eq:quadrature_measured}
	\hat{q}_\arm(\Delta t_\arm, \alpha_{\text{DX}}, \varphi_\arm) = \vec{\zeta}_{\text{LO}}^\T(\Delta t_\arm, \alpha_{\text{DX}}, \varphi_\arm) \hat{\vec{\zeta}} + \vec{\zeta}_{\text{LO}}^\T(\Delta t_\arm, \alpha_{\text{DX}}, \varphi_\arm + \Delta \varphi_\arm) \hat{\vec{\zeta}}_{\text{vac}}
.\end{equation}
The above equation can be understood as projecting the high-dimensional, $s$-polarized quadrature operators $\hat{\vec{\zeta}}$ and $\hat{\vec{\zeta}}_{\text{vac}}$ at the inputs of the beam splitter onto the transformed local oscillator state.
The additional phase $\Delta \varphi_a = \pi / 2$ for detection arm $a$ and $\Delta \varphi_b = -\pi / 2$ for arm $b$ is due to the phase shift acquired in the reflection at the beam splitter.
The phase is on the vacuum port since $\varphi_\arm$ is already adjusted to the reflection phase shift (see Methods).
Equation~\eqref{eq:quadrature_measured} is only valid for a large local oscillator amplitude.
We assume the amplitude of the two probes both equal $\alpha_{\text{DX}} \neq 0$ in the case of electro-optic sampling and $\alpha_{\text{DX}} = 0$ for homodyne sampling.

\section{Reconstruction of highly multimode Gaussian quantum states}\label{s:reconstruction}
To gain statistical information about the sampled quantum state and its dynamics, different combinations of time delays in the two detection arms of the setup in Fig.~\ref{fig:setup} have to be measured.
Yet, the statistical information contained in the covariance matrix, $\cov$, of the pulsed quantum state is encoded in orthogonal, distinguishable modes, while the local oscillator pulses at different time delays can overlap and are thus partially indistinguishable.
We present an algorithm capable of extracting information about the orthogonal mode structure of the pulsed quantum state from the correlation measurements. The algorithm contains the following steps:
\begin{enumerate}
	\item Measure all possible combinations from the series of $N$ time delays $(\Delta t_i)_{i \leq N} = \{\Delta t_1, \Delta t_2, \ldots, \Delta t_N\}$ and two possible wave plate settings $\varphi = 0, \pi / 2$, i.e., a half- and quarter-wave plate,
	\begin{equation}\label{eq:parameter_collection}
		\Gamma = \Big((\Delta t_1, \alpha, 0), (\Delta t_2, \alpha, 0), \ldots, (\Delta t_1, \alpha, \pi / 2), (\Delta t_2, \alpha, \pi / 2), \ldots\Big)
	.\end{equation}
	Collect all the measurement results for the different time delays and wave plate settings in the correlation matrix
    \begin{equation}\label{eq:correlations}
	\text{corr}_{i,j} = \frac{1}{2} \ev{\{\hat{q}_a(\Gamma_i), \hat{q}_b(\Gamma_j)\}}
    ,\end{equation}
    where the quadrature operators $\hat{q}_\arm$ are defined in equation~\eqref{eq:quadrature_measured} and $\{\hat{A},\hat{B}\}=\hat{A}\hat{B}+\hat{B}\hat{A}$ denotes the anticommutator.	

    \item The states of the local oscillators for the corresponding measurements in the first step are assembled in the matrix
    \begin{equation}\label{eq:lo_combinations}
		Z_{\text{LO}} = (\vec{\zeta}_{\text{LO}}(\Gamma_i) \;\; | \;\; i \leq 2 N)
	.\end{equation}

	\item Use the definitions in equation~\eqref{eq:quadrature_measured}-\eqref{eq:lo_combinations} to relate the covariance matrix of the pulsed quantum state, $\cov_{\hat{\rho}}$, as well as the covariance matrix of the vacuum at the empty input port of the beam splitter, $\cov_{\text{vac}} = \frac{1}{2}\mathbb{I}$ ($\mathbb{I}$ being the identity matrix), to the covariance matrix of the measurement results, 
    \begin{equation}\label{eq:matrix_eq}
	\text{corr} = Z^\T_{\text{LO}} \cov_{\hat{\rho}} Z_{\text{LO}} - Z^\T_{\text{LO}} \cov_{\text{vac}} Z_{\text{LO}}
    .\end{equation}
	Orthogonalise the local oscillator states and transform the correlation matrix with the aid of the singular value decomposition, $Z_{\text{LO}} = U \Sigma V^\T$, and the Moore-Penrose pseudoinverse of $\Sigma$, denoted by $\Sigma^+$ \cite{Moore1920,Penrose1955}. 
	We define the diagonal projector $P = \Sigma \Sigma^+$ and the transformed covariance matrix $\cov_{\hat{\rho}, U} = U^\T \cov_{\hat{\rho}} U$. After subtracting the vacuum contribution and transforming equation~\eqref{eq:matrix_eq} with $V \Sigma^+$, we get the (partially) reconstructed covariance matrix
    \begin{equation}\label{eq:projected_cov}
	P \cov_{\hat{\rho}, U} P = \left[V \Sigma^+\right]^\T \text{corr} V \Sigma^+ + \frac{1}{2} UPU^\T
    .\end{equation}
	To avoid numerical instability, we set all singular values of the matrix $Z_{\text{LO}}$ to zero if they are smaller than $10^{-3}$ times the maximal singular value.
\end{enumerate}
\begin{figure}
	\includegraphics[width=\textwidth]{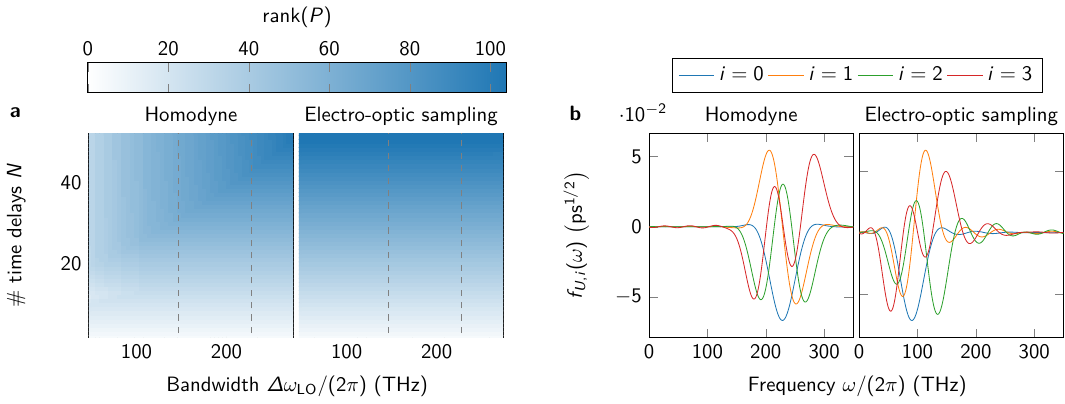}
	\caption{%
		\label{fig:temporal_resolution}
		\textbf{Temporal resolution and reconstructed mode basis.}
		\textbf{a}, Number of (temporal) modes, given by $\rank(P)$, reconstructed with the algorithm presented in Sec.~\ref{s:reconstruction} in dependence of the local oscillators bandwidth $\Delta \omega_{\text{LO}}$ and the number of time delays $N$ used in the correlation measurements for homodyne detection and electro-optic sampling respectively.
        The centre of the local oscillator pulse is fixed at $\bar{\omega}_{\text{LO}} / (2\pi)= \SI{230}{\tera\hertz}$ and the vertical dashed lines indicate the transition from multicycle to single cycle ($k_{\text{LO}} = 3$) and to subcycle pulses ($k_{\text{LO}} = 1$).
		\textbf{b}, First four (orthogonalized) mode functions in the frequency domain reconstructed by the correlation measurement.
		In the case of homodyne detection, the bandwidth of the local oscillator is $\Delta \omega_{\text{LO}} / (2 \pi) = \SI{59}{\tera\hertz}$ and thus multicycle with $k_{\text{LO}} = 20.8$.
		For electro-optic sampling, the probe involved in the nonlinear interaction is centred at $\bar{\omega}_{\text{DX}} / (2\pi) = \SI{200}{\tera\hertz}$ with the bandwidth $\Delta \omega_{\text{DX}} = \SI{118}{\tera\hertz}$ ($k_{\text{LO}} = 4$), but is filtered after the nonlinear interaction to $\Delta \omega_{\text{LO}} / (2 \pi) = \SI{59}{\tera\hertz}$.
		}
\end{figure}
%
In general, the reconstructed covariance matrix in equation~\eqref{eq:projected_cov} corresponds to the marginal distribution of a Gaussian state with covariance matrix $\cov_{\hat{\rho}, U}$.
Thus, the rank $r = \rank(P)$ of the projector determines the dimension of the phase space reconstructable from the measurement data. 
The number of reconstructable phase space dimensions increases with the number of time delays $N$ and the bandwidth $\Delta \omega_{\text{LO}}$ of the local oscillators, as shown in Fig.~\ref{fig:temporal_resolution}~\textbf{a} for the case of homodyne detection ($\alpha_{\text{DX}} = 0$) and electro-optic sampling ($\alpha_{\text{DX}} = -1.94 \cdot 10^6$, optimised according to Supplementary Section~V).
We assume a equal temporal distribution of the time delays $(\Delta t_i)_{i \leq N}$ over the time interval $I_{\Delta t} = [\SI{-12}{\femto\second}, \SI{12}{\femto\second}]$ in the case of homodyne detection and $I_{\Delta t} = [\SI{-17}{\femto\second}, \SI{17}{\femto\second}]$ for electro-optic sampling.
The corresponding orthogonalized (local oscillator) mode functions, $f_{U,i}(\omega)$, are obtained from the rows of $U$, introduced in step 3 of the algorithm, and are shown in Fig.~\ref{fig:temporal_resolution}~\textbf{b}.
In the case of homodyne detection, the number of reconstructable modes increases linearly with the number of time delays until reaching a plateau.
A behaviour explained by Fig.~\ref{fig:orthogonalization}~\textbf{a} and \textbf{c}.
\begin{figure}
	\centering
	\includegraphics[width=\textwidth]{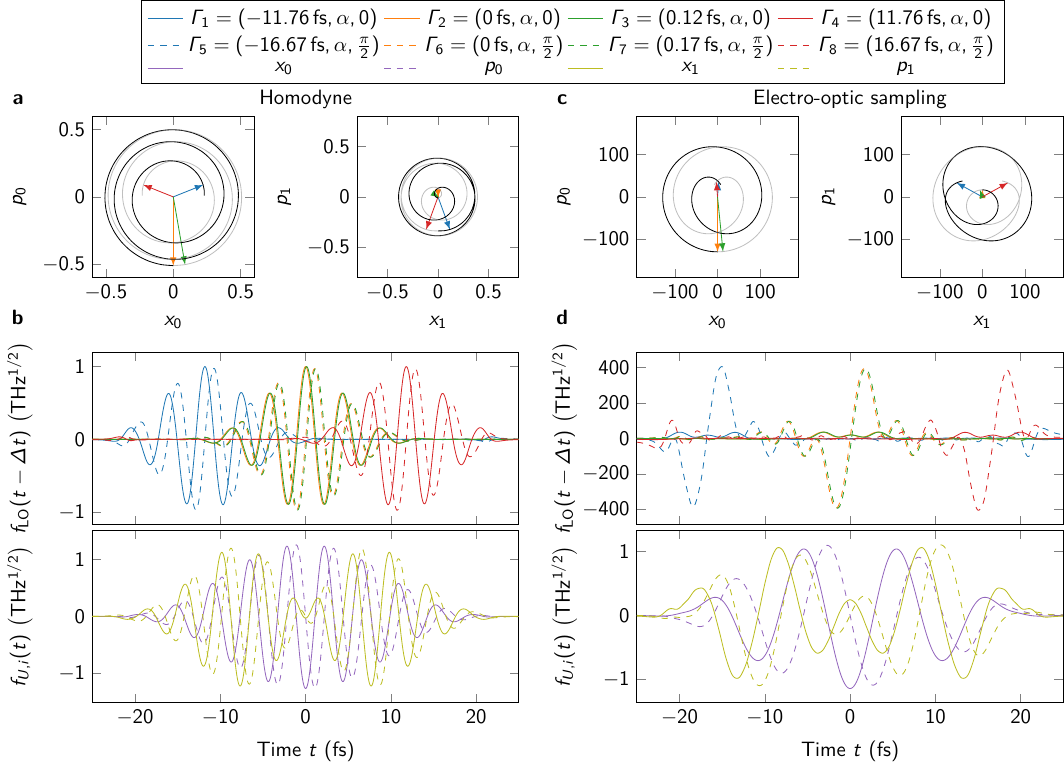}
	\caption{%
		\label{fig:orthogonalization}
		\textbf{Dynamics of the local oscillator in phase space and in the time domain.}
		\textbf{a}, State of the local oscillator for different time delays in the phase space spanned by the first two orthogonalized modes obtained from a homodyne measurement with a quarter-wave plate.
		The black (gray) line indicates the evolution for negative (positive) time delays.
		\textbf{b}, Waveform of the time-delayed local oscillator, $f_{\text{LO}}(t)$, and first two orthogonalized modes. $f_{U,i}(t)$, obtained from homodyne detection, in the time domain.
		The symmetric function corresponds to the $\hat{x}$ quadrature, while the antisymmetric function relates to the $\hat{p}$ quadrature.
		The dynamics in \textbf{a} can be obtained from the overlap of the time-delayed local oscillator and the orthogonal quadrature basis.
		\textbf{c}, Phase space representation of the local oscillator in electro-optic sampling after the nonlinear interaction (see Supplementary Section~V.).
		\textbf{d}, Time-domain picture of the time-delayed local oscillator after the nonlinear interaction and the first two corresponding orthogonalized modes.
		While the $\hat{x}$ quadrature of the probe is suppressed exponentially by the nonlinear interaction, it is still possible to sample the $\hat{x}_i$ quadrature, since for some time delays the overlap of $p$ quadrature of the local oscillator with the $x_i$ quadrature of the orthogonalized mode basis dominates, agreeing with the dynamics shown in \textbf{c}.
        The parameters for the probe and local oscillator pulse are the same as in Fig.~\ref{fig:temporal_resolution}.
	}
\end{figure}
As the time delay changes, the vector describing the state of the local oscillator moves along spirals through the high-dimensional phase space, spanned by the rows of $U$.
The vectors describing the local oscillators at two different time delays are linearly independent and, in principle, orthogonalizable.
Yet, if the difference in time delay is small, then the vectors are almost linearly dependent, as is the case in Fig.~\ref{fig:orthogonalization} \textbf{a} for $\Delta t = \SI{0}{\femto\second}$ (orange) and $\Delta t = \SI{0.12}{\femto\second}$ (green).
The cutoff introduced at the end of the third step of the algorithm will parallelise these vectors. 
The cutoff explains the plateau in Fig.~\ref{fig:temporal_resolution} \textbf{a} and can only be increased by increasing the bandwidth of the local oscillator.
A local oscillator shorter in time can resolve more features of the orthogonal mode basis, as evident from Fig.~\ref{fig:orthogonalization} \textbf{b} and \textbf{d} by considering the overlap from the local oscillator and the orthogonal modes in the time domain.
In electro-optic sampling, the modes are centred at much lower frequencies reaching the mid-infrared (MIR) range (c.f. Fig.~\ref{fig:temporal_resolution} \textbf{b}).
This frequency range is particularly interesting.
While homodyne detection is available for higher or lower frequencies, the lack of efficient photo detectors makes homodyne detection unsuitable for this frequency range.
Another striking feature of Fig.~\ref{fig:temporal_resolution} \textbf{a} is the dependence of the reconstructed phase space dimension on the bandwidth of the local oscillator in the case of electro-optic sampling.
The nonlinear interaction correlates a broad band of frequencies to the detected ones. This band only weakly depends on the bandwidth of the local oscillator.
Thus, the bandwidth of the detection can be much smaller than the sampled frequency range.
In the Supplementary Section~V we show how to choose optimised parameters to correlate the lower frequency modes to the detected ones.
Furthermore, the half-wave-plate measurement is exponentially suppressed by squeezing in the nonlinear crystal \cite{Sulzer2020}, as can be seen in Fig.~\ref{fig:orthogonalization} \textbf{d}.
Yet, this is not detrimental for correlation measurements, since depending on the time delay, the local oscillator overlaps more with the $x$- or $p$-quadrature of the orthogonal modes. (c.f. Fig.~\ref{fig:orthogonalization}~\textbf{c} and \textbf{d}).
The time domain picture in Fig.~\ref{fig:orthogonalization} also shows the different time scales of the dynamics captured by homodyne detection and electro-optic sampling.

As a final remark, we would like to mention that this approach is not limited to time-delayed pulses.
In principle, any set of linearly independent pulses, in the time or frequency domain, can be used.

\section{Analysing correlations in the time domain}
\begin{figure}
	\centering
	\includegraphics[width=\textwidth]{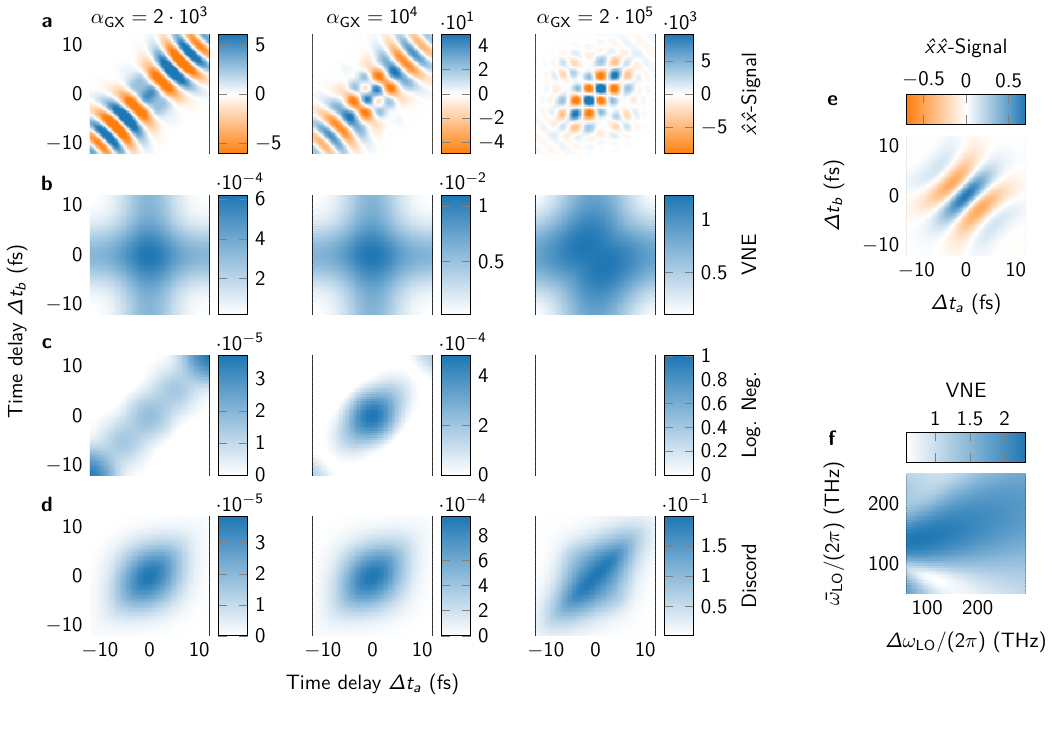}
	\caption{%
		\label{fig:correlations}
		\textbf{Correlations in the sampled state.}
		\textbf{a}, Correlated signal, $g(\Delta t_a, \Delta t_b, 0, 0) = \ev{\hat{q}_a(\Delta t_a, 0, 0)\hat{q}_b(\Delta t_b, 0, 0)}$, for a $\hat{x}\hat{x}$-quadrature measurement as a function of the time delays $\Delta t_a$ and $\Delta t_b$ on the two detection arms $a$ and $b$, for a weakly squeezed state, $\alpha_{\text{GX}} = 2 \cdot 10^3$, moderately squeezed, $\alpha_{\text{GX}} =10^4$, and strongly squeezed state, $\alpha_{\text{GX}} = 2 \cdot 10^5$.
		The squeezing is implemented by the same free-space nonlinear interaction used in Fig.~\ref{fig:setup} (see Supplementary Information~V).
		\textbf{b}, Von Neumann entropy (VNE) of the detected state defined by all 16 (co-)variances of the two possible wave-plate settings at the two arms, i.e., quarter- and half-wave plate, and given time delays $\Delta t_a$ and $\Delta t_b$.
		The input states are the same squeezed states as in \textbf{a}.
		\textbf{c}, Entanglement between the two detection arms quantified by the logarithmic negativity of the detected state, described above, with the bipartition between the two detection arms $a$ and $b$.
		\textbf{d}, Quantum correlations, beyond the entanglement in \textbf{c}, quantified by the quantum discord with the bipartition as above.
		\textbf{e}, Correlated signal for $\hat{x}\hat{x}$-quadrature measurement for a thermal state with temperature $T = \SI{1000}{\kelvin}$ (see supplementary information~IX).
		\textbf{f}, Von Neumann entropy of the strongly squeezed state at fixed time delays $\Delta t_a = \Delta t_b = 0$ as a function of the two parameters of the local oscillator, i.e., the bandwidth $\Delta \omega_{\text{LO}}$ and centre frequency $\bar\omega_{\text{LO}}$.
	}
\end{figure}

Time-local phase-sensitive measurements are insufficient to reconstruct thermal states since they do not carry any phase information.
The same argument cannot be made for the squeezed vacuum, since the variance depends on the phase and, in turn, on time.
Yet, in the following, we show the thermalisation of squeezed states during a time-domain measurement due to entanglement breaking between measured and unmeasured modes, leading to the requirement of correlation measurements for the reconstruction of multimode squeezed states.
We assume that the pulsed squeezed state is generated by squeezing the vacuum using a nonlinear interaction equivalent to the one used for electro-optic sampling (see Methods).
The strength of the squeezing interaction is determined, among other parameters, by the photon-number content of the pump $N_{\text{GX}} = \abs{\alpha_{\text{GX}}}^2$, with $\alpha_{\text{GX}}$ being the coherent amplitude.
The average $\hat{x}\hat{x}$-signal for a measurement with two half-wave plates, i.e., $g(\Delta t_a, \Delta t_b, 0, 0) = \ev{\hat{q}_a(\Delta t_a, 0, 0)\hat{q}_b(\Delta t_b, 0, 0)}$, for a weakly squeezed state with $\alpha_{sq} = 10^3$ is shown in Fig.~\ref{fig:correlations} \textbf{a}.
The signal is dominated by alternating squeezing and antisqueezing along the time-local axis, i.e., $\Delta t_a = \Delta t_b$.
The maxima of the signal are shifted along this axis, which originates from the frequency oscillations of the phase matching function \cite{Hubenschmid2024}.
As the pump strength is increased from $\alpha_{\text{GX}} = 10^3$ via $\alpha_{\text{GX}} = 10^4$ to $\alpha_{\text{GX}} = 2 \cdot 10^5$, oscillations along the orthogonal axis, $\Delta t_a = - \Delta t_b$, appear.
These oscillations along the time-nonlocal axis are a sign of thermalization, as a comparison with the signal of a thermal state in Fig.~\ref{fig:correlations} \textbf{e} illustrates (see Supplementary Section~IX for details).

The thermalisation of the detected state can be validated with the von Neumann entropy (VNE) shown in Fig.~\ref{fig:correlations}~\textbf{b}.
We define the detected state via the covariance matrix of the observable 
\begin{equation}
    \vec{q} = (\hat{q}_a[\Delta t_a, 0, 0], \hat{q}_a[\Delta t_a, 0, \pi / 2], \hat{q}_b[\Delta t_b, 0, 0], \hat{q}_b[\Delta t_b, 0, \pi / 2])^\T
,\end{equation}
the Wigner function of which corresponds to a valid quantum state.
Since the initial state is assumed to be pure and the evolution unitary, a finite von Neumann entropy can only originate from entanglement between the detected and undetected parts of the quantum state.
The presence of entanglement in the detected state can be determined using the logarithmic negativity (see Methods), shown in Fig.~\ref{fig:correlations}~\textbf{c}.
We chose the bipartition of the detected state to be between the two detection arms $\arm=a,b$.
The logarithmic negativity results from an interplay between entanglement within and purity of the quantum state.
For weak squeezing, the entanglement and thermalisation (quantified by the von Neumann entropy in Fig.~\ref{fig:correlations}~\textbf{b}) are low, resulting in a small logarithmic negativity.
Increasing the squeezing, the entanglement will increase as well, which in turn leads to more thermalisation, increasing the logarithmic negativity until thermalisation overshadows the entanglement and the logarithmic negativity drops to zero.
However, the quantum correlations of the thermalised state can still be captured by the quantum discord (see Methods), even in the presence of thermalisation, and shows a clear increase with the squeezing strength, as can be seen in Fig.~\ref{fig:correlations}~\textbf{d}.
Since the quantum correlations are low in the weak squeezing limit, the signal can be understood as a time-varying noise pattern along the time local axis, $\Delta t_a = \Delta t_b$ (see Supplementary Section~VIII). For stronger squeezing the quantum nature of the multimode squeezed state becomes relevant and correlation measurements are necessary.
Fig.~\ref{fig:correlations}~\textbf{f} shows the von Neumann entropy of the detected state at $\Delta t_a = \Delta t_b = 0$, at which point the local oscillator is matched best to the quantum pulse, as a function of the parameters describing the mode of the local oscillator, i.e., the centre frequency $\bar\omega_{\text{LO}}$ and the bandwidth $\Delta \omega_{\text{LO}}$.
The minimum of the von Neumann entropy in Fig.~\ref{fig:correlations}~\textbf{f}, corresponds to the local oscillator mode matched closest to the first principal mode of the squeezed state (see Supplementary Section~VI).
Yet, in time-domain sampling, the modes of the local oscillator and quantum state are never matched since the detection is faster than the dynamics of the state, leading to the breakage of entanglement and in turn to thermalisation of the state.
Therefore, the reconstruction of the (thermalised) multimode quantum state requires correlation measurements, at least in the strong squeezing regime in which thermalisation becomes relevant.

\section{Correlations measurements of non-Gaussian states}
\begin{figure}[h]
	\centering
	\includegraphics[width=\textwidth]{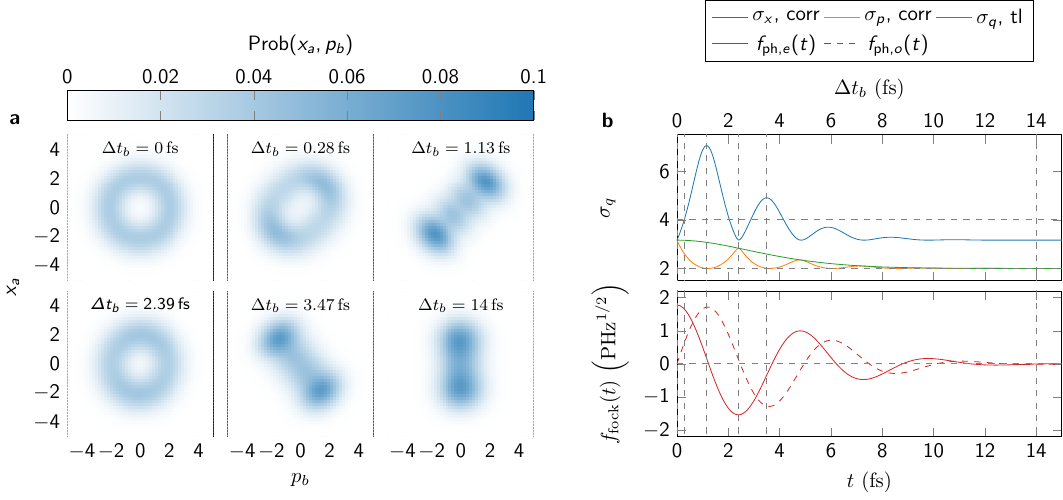}
	\caption{%
		\label{fig:prob_dist_fock_state_submode}
		\textbf{Joint probability distribution for correlation measurements of a Fock state.} 
		\textbf{a}, Joint probability distribution for a $\hat{x}$-quadrature measurement at detection arm $a$ and a $\hat{p}$-quadrature measurement at detection arm $b$ with fixed time delay $\Delta t_a = 0$ and different $\Delta t_b$ for each distribution.
		The joint probability distribution is taken over the possible measurement outcomes, which are the eigenvalues $x_a$ and $p_b$ of the operators $\hat{q}_a(\Delta t_a, 0, 0)$ and $\hat{q}_b(\Delta t_b, 0, \pi / 2)$.
		The input state used in this example is a three-photon Fock state in a temporal mode centred at $\bar{\omega}_{\text{ph}} = \SI{230}{\tera\hertz}$ with a bandwidth $\Delta \omega_{\text{ph}} = \SI{59}{\tera\hertz}$.
        The local oscillator is characterized by $\bar{\omega}_{\text{LO}} = \SI{200}{\tera\hertz}$ and $\Delta \omega_{\text{LO}} = \SI{118}{\tera\hertz}$.
		\textbf{b}, The singular values $\sigma_q$ with $q = x,p$ quantify the compatibility of the two measurements compared to the waveform $f_{\text{ph},e}(t)$ and $f_{\text{ph},o}(t)$ related to the $\hat{x}$ and $\hat{p}$ quadratures of the pulsed Fock state.
        For a time local (tl) measurement, $\Delta t_a = \Delta t_b$, or for a difference between the time delays equal to integer multiple of half an optical cycle of the quantum states waveform, the singular values are equal $\sigma_x \approx \sigma_p$ and the measurement is incompatible leading to a measurement of the Husimi function, e.g., for $\Delta t_b = \SI{0}{\femto\second}$ or $\Delta t_b = \SI{2.47}{\femto\second}$.
        For a correlation (corr) measurement with a difference between the time delays equal to odd multiples of a quarter optical cycle, one of the singular values approaches 2. 
        In this case, the same quadrature is measured by both detection arms and the measurements are compatible. 
        The compatibility of the measurement allows the extraction of phase information about the, rotationally symmetric, Fock state.
		}
\end{figure}

Since Gaussian states are completely described by their expectation values and covariance matrix, two-point correlation measurements can fully characterize even highly multimode Gaussian states.
Non-Gaussian states such as Fock states, on the other hand, can exhibit higher-order correlations and require going beyond the second moment.
In the following, we will show how correlation measurements can be used to obtain phase information about non-Gaussian states, even if they are rotationally symmetric in phase space, such as the Fock states. The joint probability distribution,
\begin{equation}\label{eq:full_statistcs}
	Prob(x_a, p_b) = \int \ldots \int_{-\infty}^\infty K(x_a, p_b | \vec{\zeta}) W(\vec{\zeta}) \dd^{2 i_{\text{max}}} \zeta
,\end{equation}
of the correlated quadrature measurement with outcomes $x_a$ for $\hat{q}_a(\Delta t_a, 0, 0)$ and $p_b$ for $\hat{q}_b(\Delta t_b, 0, \pi / 2)$ is given by convolving the $2 i_{\text{max}}$-dimensional, multimode Wigner function $W(\vec{\zeta})$, describing the pulsed quantum state with a multivariate Gaussian kernel $K(x_a, p_b | \vec{\zeta})$ defined in equation~\eqref{eq:kernel}, accounting for the effect of the measurement (for details see the Supplementary Section~X).
Generally, inverting equation~\eqref{eq:full_statistcs} is a nontrivial task.
However, in the case of a Gaussian Wigner function, we can recover equation~\eqref{eq:matrix_eq} from the convolution, the inversion of which is demonstrated in equation~\eqref{eq:projected_cov}.
Another example for which equation~\eqref{eq:full_statistcs} proves useful is the reconstruction of $n$-photon Fock states.
Figure~\ref{fig:prob_dist_fock_state_submode} \textbf{a} shows the result of the convolution in equation~\eqref{eq:full_statistcs} for a single-mode, three-photon Fock state.

We can distinguish between two special cases of the joint probability distribution by scrutinising the two singular values $\sigma_q$ with $q=x,p$, quantifying the measurement and preparation uncertainty along two orthogonal directions in the phase space of the Fock state (see Supplementary Material X for more details).
The singular values $\sigma_q$ are shown in Fig.~\ref{fig:prob_dist_fock_state_submode}~\textbf{b} as a function of the time delay $\Delta t_b$.

In the first case, the two singular values, and thus the measurement uncertainties, are equal, $\sigma_x \approx \sigma_p$. In this case, the two quadrature measurements are maximally incompatible and the measurement statistics are determined by a phase space distribution.
If the mode functions of the local oscillator and the pulsed quantum state are matched, the singular values approach 4, $\sigma_q \to 4$, and the phase space distribution converges to the Husimi function of the quantum pulse.
Equal measurement uncertainties can be observed along the time local (tl) axis with $\Delta t_a = \Delta t_b$.
The Fock state appears to the detection as an incoherent mixture of Fock states with occupations $0$ to $n$ distributed binomially, with probability $p = \sigma_q / 4$, reproducing the result from \cite{Hubenschmid2024}.

In the second case, one singular value tends to 2, while the other stays above 2. In this case, the measurement uncertainty of one measurement is decreased at the expense of the other measurement, resulting in a smoothed-out quadrature distribution.
An example of this special case can be observed at large differences in time delay, one detection arm measures the vacuum and the other (partially) the Fock state. 
In this case, the probability distribution disintegrates into the quadrature distribution of the vacuum and the Fock state, given by a Hermite-Gauss polynomial, convolved with a Gaussian describing the vacuum contributions from the empty beam splitter port.

Neither case reveals any phase information about the quantum state.
However, the singular values $\sigma_q$ for correlation (corr) measurements oscillate with twice the carrier frequency of the quantum pulse, as shown in Fig.~\ref{fig:prob_dist_fock_state_submode} \textbf{b}.
At a difference between the time delays corresponding to even multiples of half an optical cycle of the quantum pulse, orthogonal quadratures of the quantum pulse are measured and the singular values are equal.
At a difference between the time delays corresponding to odd multiples of half an optical cycle, the same quadrature is measured and one singular value tends to 2.
As one of the time delays is varied while the other one is fixed, the measurement statistics interpolates between the two cases and oscillates with twice the carrier frequency of the quantum pulse, as can be seen in Fig.~\ref{fig:prob_dist_fock_state_submode} \textbf{b}.
The oscillation between compatible and incompatible measurements allows the extraction of spectral information about the pulsed Fock states.

\section{Conclusion}
While time-domain quadrature correlation measurements have already been used to scrutinise the statistics of the ground state of the electromagnetic field, the reconstruction of (unknown) pulsed quantum states has not been addressed to this point.
Here, we present a full quantum state tomography scheme able to reconstruct highly multimode Gaussian quantum states.
Different from existing tomography schemes, our proposed reconstruction only relies on time-delayed pulses, which are orthogonalised in post-processing.
We compare two implementations of the correlation measurement, one based on homodyne detection suited for multicycle modes in microwave or optical frequencies and another, based on electro-optic sampling suited for subcycle modes in the THz to mid-infrared frequency range.
Accompanying the latter, we develop a non-perturbative description of electro-optic sampling which allows to optimise the probe strength in the strong interaction regime.
Furthermore, we show how thermalisation during the detection makes correlation measurements necessary, by analysing the sign of quantum correlations in the measurement data.
Whereas the above achievements are limited to Gaussian states, we also present the joint measurement statistics for the correlation measurement of non-Gaussian states and show how correlation measurements can improve the reconstruction of Fock states.
Thus, we are opening a new avenue to investigate quantum phenomena of light on an ultrafast time scale.

\vspace{0.5cm}
\textit{Note added.} While finishing writing this article we became aware of a related work, exploring the reconstruction of Gaussian states using electro-optic sampling in the perturbative regime \cite{Yang2025}.

\section*{Methods}\label{methods}

\subsection{Discretization and the calculation (subcycle) mode basis}\label{m:discretization}
To discretise the continuous Hilbert space of the free-space electromagnetic field, we introduce the subcycle mode basis,
\begin{equation}\label{eq:mode_functions}
	f_i(\omega) = f_0(\omega) \tilde{L}_i(\omega)
,\end{equation}
with two parameters, $\sigma_0, k_0 > 0$ based on the fundamental mode
\begin{equation}\label{eq:fundamental_mode_function}
	f_0(\omega) = \left[\frac{2 / \sigma_0}{\Gamma(k_0 + 1 / 2)}\right]^{\frac{1}{2}} \left(\frac{\omega}{\sigma_0}\right)^{k_0} \exp(-\frac{\omega^2}{2\sigma_0^2})
\end{equation}
centred at $\bar{\omega}_0 \approx \sigma_0 \sqrt{k_0 + 1/\pi}$ with bandwidth $\Delta \omega_0 \approx \sqrt{2\ln(2)} \sigma_0$ and the (scaled) generalized Laguerre polynomials
\begin{equation}\label{eq:polynomial}
	\tilde{L}_i(\omega) = \left[\frac{\Gamma(i + 1) \Gamma(k_0 + 1 / 2)}{\Gamma(i + k_0 + 1 / 2)}\right]^{\frac{1}{2}} L_i^{(k_0 - 1 / 2)}\left(\frac{\omega^2}{\sigma_0^2}\right)
.\end{equation}
We used $\sigma_0 = \SI{100}{\tera\hertz}$ and $k_0 = 0.5$.
See Supplementary Section~I for further details about the subcycle mode basis.
Using the relation $\hat{a}(\omega) \approx \sum_{i=0}^{i_\text{max}} f_i^\ast(\omega) \hat{a}_i$, we can discretize the quadrature operator,
\begin{equation}\label{eq:msnt_q_1}
	\hat{q}(0, 0, \varphi^\prime) \approx \int_0^\infty \left[e^{\iu\varphi^\prime}\alpha_{\text{LO}}(\omega)\{\hat{a}_s^\prime(\omega)\}^\dagger + e^{-\iu\varphi^\prime}\alpha_{\text{LO}}^\ast(\omega)\hat{a}_s^\prime(\omega)\right] \dd \omega
,\end{equation}
measured in a homodyne detection with a local oscillator described by a frequency-dependent coherent amplitude $\alpha_{\text{LO}}(\omega)$ given by the function defined in equation~\eqref{eq:fundamental_mode_function} with parameters $\sigma_{\text{LO}}$ and $k_{\text{LO}}$ and a phase shift $\varphi^\prime$ due to a $\Phi$-wave plate rotated by $\theta$.
By choosing
\begin{equation}
	\varphi^\prime = -\arctan(\sqrt{\frac{2\cos^2(\Phi / 2)}{1 - 2\cos^2(\Phi / 2)}})
,\end{equation}
the effective phase can be calculated using
\begin{equation}
    \theta = \frac{1}{2}\arccos(\sqrt{\frac{1 - 2\cos^2(\Phi / 2)}{2\sin^2(\Phi / 2)}})
.\end{equation}
A detailed derivation is given in Supplementary Section~II.
We collect the coefficients resulting from the discretisation in the vector $\vec{\zeta}_{\text{LO}}$, with
\begin{equation}
	[\vec{\zeta}_{\text{LO}}]_i = \begin{cases}
	    \Re\left[\int_0^\infty \alpha_{\text{LO}}(\omega)f_i(\omega) \dd \omega\right] & 0 \leq i < i_\text{max} \\
        \Im\left[\int_0^\infty \alpha_{\text{LO}}(\omega) f_{i - i_{\text{max}}}(\omega) \dd \omega\right] & i_{\text{max}} \leq i < 2 i_{\text{max}}
	\end{cases}
.\end{equation}
With
\begin{equation}
	R(\varphi^\prime) = \begin{pmatrix} 
		\cos(\varphi^\prime) \mathbb{I} & -\sin(\varphi^\prime) \mathbb{I} \\
		\sin(\varphi^\prime) \mathbb{I} & \cos(\varphi^\prime) \mathbb{I} \\
	\end{pmatrix}
\end{equation}
describing the effect of the wave plate, the measured quadrature can be expressed as 
\begin{equation}
	\hat{q}  = \left(R^\T(\varphi^\prime)\vec{\zeta}_{\text{LO}}\right)^\T \hat{\vec{\zeta}}
.\end{equation}
The vector of mode operators $\hat{\vec{\zeta}}$ is defined in the introduction.
With the matrix elements $\tilde{\omega}_{ij} = \int_0^\infty \omega f_i(\omega) f_j(\omega) \dd \omega$ of the free-field Hamiltonian and the definition
\begin{equation}
	G_{\Delta t} = \Delta t\begin{pmatrix} 
		0 & \tilde{\omega} \\
		-\tilde{\omega} & 0
	\end{pmatrix}
\end{equation}
we can describe the time delay using the symplectic matrix $M(\Delta t) = \exp(G_{\Delta t})$.
The effect of the nonlinear crystal can be described with a squeezing and beam-splitting interaction of photons at (angular) frequency $\Omega$ and $\omega$, described by the kernels derived in the Supplementary Section~III,

\begin{align}
	S(\Omega, \omega) &= \frac{1}{\hbar} (2\pi)^{3/2}\left(\frac{\hbar}{4\pi\varepsilon_0 c A}\right)^{3/2} \sqrt{\frac{\omega + \Omega}{n(\omega + \Omega)}} f_{\alpha}(\omega + \Omega) \sqrt{\frac{\omega\Omega}{n(\omega)n(\Omega}} \widehat{\lambda}[\Delta k(\Omega, \omega)], \label{eq:squeezing_nl} \\
	B(\Omega, \omega) &= \frac{1}{\hbar} (2\pi)^{3/2}\left(\frac{\hbar}{4\pi\varepsilon_0 c A}\right)^{3/2} \sqrt{\frac{\omega + \Omega}{n(\omega + \Omega)}} \left[f_{\alpha}^\ast(\omega - \Omega) + f_{\alpha}(\Omega - \omega)\right] \sqrt{\frac{\omega\Omega}{n(\omega)n(\Omega}} \widehat{\lambda}[\Delta k(\Omega, \omega)] \label{eq:beam_splitting_nl}
,\end{align}
with $c$, $\hbar$, $\varepsilon_0$ being the speed of light in vacuum, reduced Planck constant, and the vacuum permittivity.
Furthermore, $A = \pi (\SI{3}{\micro\meter})^2$ is the beam waist area and $\widehat{\lambda}[\Delta k(\Omega, \omega)]$ the Fourier transform of the transversal profile of the $\chi^{(2)}$ interaction at the wave-vector mismatch $\Delta k(\Omega, \omega)$.
In case of a zinc-telluride crystal of length $L = \SI{20}{\micro\meter}$ in free space, we have $\widehat{\lambda}(k) = \lambda\frac{L}{\sqrt{2\pi}}\sinc(k L / 2)$ with $\lambda = A \varepsilon_0 d / 2$, the interaction parameter $d = - n^4(\ev{\omega}) r_{41}$ and the electro-optic coefficient of zinc telluride, $r_{41} = \SI{4}{\pico\meter\per\volt}$ \cite{Boyd2019}.
The refractive index $n(\omega)$ is taken from \cite{Marple1964}.
The function $f_{\alpha}(\omega)$ described the spectrum of the coherent pump/probe pulse with amplitude $\alpha$ driving the nonlinear interaction.
We assume the spectrum to be described by equation~\eqref{eq:fundamental_mode_function} with parameters $\sigma_{p} = \SI{100}{\tera\hertz}$ and $k_{p} = 4$.
Discretizing the kernel again leads to the matrix elements
\begin{align}
	S_{ij} &= 2 \iint_0^\infty S(\Omega, \omega) f_i^\ast(\Omega) f_j^\ast(\omega) \dd \Omega \dd \omega, \\
	B_{ij} &= 2 \iint_0^\infty B(\Omega, \omega) f_i(\Omega) f_j^\ast(\omega) \dd \Omega \dd \omega
,\end{align}
and by defining the matrix
\begin{equation}\label{eq:nl_hamiltonian_matrix}
	G_{\text{NL}}  = \begin{pmatrix} 
		-\Re(S - B) & \Im(S - B) \\
		\Im(S + B) & \Re(S + B)
		\end{pmatrix}
,\end{equation}
we can describe the effect of the nonlinear interaction using the symplectic matrix $M_{\text{NL}}(\alpha) = \exp[\abs{\alpha} G_{\text{NL}}]$.
By applying the transformation of the beam splitter to the measured quadrature, we obtain equation~\eqref{eq:quadrature_measured}.

\subsection{Correlation analysis}\label{m:correlations}

Since the detection takes place at two outputs of a beam splitter, the detected modes commute and the operators $\vec{q} = (\hat{q}_a[\Delta t_a, 0, 0], \hat{q}_a[\Delta t_a, 0, \pi / 2], \hat{q}_b[\Delta t_b, 0, 0], \hat{q}_b[\Delta t_b, 0, \pi / 2])^\T$ define a covariance matrix,
\begin{equation}
    [\cov_{\hat{\rho}, \text{d}}]_{ij} = \frac{1}{2}\ev{\{\vec{q}_i, \vec{q}_j\}}_{\hat{\rho}\otimes\ket{\text{vac}}\bra{\text{vac}}}
,\end{equation}
corresponding to a valid quantum state, which we call the detected state.
The von Neumann entropy of the sampled state can be calculated from the symplectic spectrum $\{\nu_i\}_{i=1,2}$ using $S(\cov_{\hat{\rho},\text{d}}) = \sum_{i=1}^2 s(\nu_i)$
with 
\begin{equation}
    s(x) = \left(\frac{x + 1}{2}\right)\log_2\left(\frac{x + 1}{2}\right) - \left(\frac{x - 1}{2}\right)\log_2\left(\frac{x - 1}{2}\right)
\end{equation}
according to \cite{Weedbrook2012}.
Correlation measures can be calculated using the bipartition between the mode measured at arm $a$ and $b$.
Thus, the subsystem $A$ is spanned by the mode operators in $\vec{q}_A = (\hat{q}_a[\Delta t_a, 0, 0], \hat{q}_a[\Delta t_a, 0, \pi / 2])^\T$, and the subsystem $B$ by $\vec{q}_B = (\hat{q}_b[\Delta t_b, 0, 0], \hat{q}_b[\Delta t_b, 0, \pi / 2])^\T$.
Moreover, by defining the time-reversal operator for the subsystem $B$, $\Lambda = \text{diag}(1, 1, 1, -1)$, the logarithmic negativity, 
\begin{equation}
    L(\cov_{\hat{\rho},\text{d}}) = \max\{0, -\log_2(\min\{\tilde{\nu}_i\}_{i=1,2})\}
,\end{equation}
can be calculated from the symplectic spectrum $\{\tilde{\nu}_i\}_{i=1,2}$ of the covariance matrix $\Lambda \cov_{\hat{\rho}, \text{d}} \Lambda$ \cite{Adesso2004, Weedbrook2012}.
The same bipartitian is used to calculate the quantum discord according to \cite{Adesso2010}.

While not being independent of the local oscillator mode, analysing correlations in the time domain could offer new insights to the entanglement of Gaussian states beyond correlations in the orthogonal mode basis and could be used to define a mode independent notion of entanglement precent in Gaussian quantum states \cite{Sperling2019}.

\subsection{The joint probability distribution}\label{m:full_statistics}
A detailed derivation of the full joint measurement statistics of correlation homodyne detection (i.e., $\alpha = 0$) is given in the Supplementary Section~X.
With the definitions
\begin{equation}
	P_{\text{LO}} = \vec{\zeta}_{\text{LO}}(\Delta t_a, \pi / 2)[\vec{\zeta}_{\text{LO}}(\Delta t_a, \pi / 2)]^\T + \vec{\zeta}_{\text{LO}}(\Delta t_b, 0)[\vec{\zeta}_{\text{LO}}(\Delta t_b, 0)]^\T
,\end{equation}
and
\begin{align}
	\vec{\zeta}_{d}(x_a, p_b) = 4\frac{P_{\text{LO}} [x_a\vec{\zeta}_{\text{LO}}(\Delta t_a, \pi / 2) - p_b \vec{\zeta}_{\text{LO}}(\Delta t_b, 0)]}{\norm{\vec{\zeta}_{\text{LO}}(\Delta t_a, \pi / 2) - \vec{\zeta}_{\text{LO}}(\Delta t_b, 0)}^2}
,\end{align}
as well as the covariance matrix ($\Omega$ defining the symplectic matrices)
\begin{align}
	\cov_{\text{d}}^{-1} &= 4\frac{\Omega^\T P_{\text{LO}}^2 \Omega}{\norm{\vec{\zeta}_{\text{LO}}(\Delta t_a, \pi / 2) - \vec{\zeta}_{\text{LO}}(\Delta t_b, 0)}^2}
,\end{align}
we can express the integration kernel in equation~\eqref{eq:full_statistcs} as
\begin{align}\label{eq:kernel}
	K(x_a, p_b | \vec{\zeta}) &= \frac{2\sqrt{2}\exp[-2\frac{\norm{x_a\vec{\zeta}_{\text{LO}}(\Delta t_a, \pi / 2) - p_b \vec{\zeta}_{\text{LO}}(\Delta t_b, 0)}^2}{\norm{\vec{\zeta}_{\text{LO}}(\Delta t_a, \pi / 2) - \vec{\zeta}_{\text{LO}}(\Delta t_b, 0)}^2}]}{\norm{\vec{\zeta}_{\text{LO}}(\Delta t_a, \pi / 2) - \vec{\zeta}_{\text{LO}}(\Delta t_b, 0)}}  \exp[\vec{\zeta}_{\text{d}}^\T(x_a, p_b) \vec{\zeta} - \frac{1}{2}\vec{\zeta}^\T \cov_{\text{d}}^{-1} \vec{\zeta}]
.\end{align}
We assume the quantum pulse is an $n_{\text{ph}}$-photon Fock state in a single temporal mode with mode function according to equation~\eqref{eq:fundamental_mode_function} parametrized by $\Delta \omega_{\text{ph}} = \SI{59}{\tera\hertz}$ and $\bar{\omega}_{\text{ph}}/(2\pi) = \SI{202}{\tera\hertz}$ ($k_{\text{ph}} = 16$).
By defining the projector $P_{\text{ph}}$ of the Fock states phase space, we can calculate the Schur complement
\begin{equation}
	\cov_{\text{schur}}^{-1} = P_{\text{ph}}(\cov_{\text{vac}}^{-1} + \cov_{\text{d}}^{-1}) P_{\text{ph}} - P_{\text{ph}} \cov_{\text{d}}^{-1} P_{\text{r}} [P_{\text{r}}(\cov_{\text{vac}}^{-1} + \cov_{\text{d}}^{-1})P_{\text{r}}]^{-1} P_{\text{r}} \cov_{\text{d}}^{-1} P_{\text{ph}}
,\end{equation}
with singular values $\sigma_x$, $\sigma_p$ and define
\begin{equation}
	\vec{\zeta}_{\text{d,ph}}(x_a, p_b) = P_{\text{ph}}[\mathbb{I} - \cov_{\text{d}}^{-1} P_{\text{r}} \{ P_{\text{r}} (\cov_{\text{vac}}^{-1} + \cov_{\text{d}}^{-1}) P_{\text{r}} \}^{-1} P_{\text{r}}] \vec{\zeta}_{\text{d}}(x_a, p_b)
.\end{equation}
as well as the normalization envelope
\begin{align}
	&N(x_a, p_a) = \frac{2\sqrt{2}\exp[-2\frac{\norm{x_a\vec{\zeta}_{\text{LO}}(\Delta t_a, \pi / 2) - p_b \vec{\zeta}_{\text{LO}}(\Delta t_b, 0)}^2}{\norm{\vec{\zeta}_{\text{LO}}(\Delta t_a, \pi / 2) - \vec{\zeta}_{\text{LO}}(\Delta t_b, 0)}^2}]}{\norm{\vec{\zeta}_{\text{LO}}(\Delta t_a, \pi / 2) - \vec{\zeta}_{\text{LO}}(\Delta t_b, 0)}} \nonumber \\
	&\times \sqrt{\frac{(2\pi)^{2(i_{\text{max}}-1)}}{\det(P_{\text{r}}[\cov_{\text{vac}}^{-1} + \cov_{\text{d}}^{-1}]P_{\text{r}})}} \exp[\frac{1}{2} \vec{\zeta}_{\text{d}}^\T(x_a, p_b) \Omega P_{\text{r}} \{P_{\text{r}}(\cov_{\text{vac}}^{-1} + \cov_{\text{d}}^{-1}) P_{\text{r}}\}^{-1} P_{\text{r}} \Omega^\T \vec{\zeta}_{\text{d}}(x_a, p_b)]
,\end{align}
to express the probability distribution of the joint quadrature measurement of the Fock state as
\begin{align}
	p(x_a, p_b) &= 	N(x_a, p_b) \frac{(-1)^n}{\pi} \sqrt{\frac{2\pi}{\sigma_x}}\sqrt{\frac{2\pi}{\sigma_p}} e^{\frac{[\vec{\zeta}_{\text{d,ph}}^\T(x_a, p_b) \vec{e}_x]^2}{2\sigma_x}} e^{\frac{[\vec{\zeta}_{\text{d,ph}}^\T(x_a, p_b) \vec{e}_p]^2}{2\sigma_p}} \nonumber \\
	&\times \sum_{i=0}^n \left(1 - \frac{4}{\sigma_x}\right)^{i} \left(1 - \frac{4}{\sigma_p}\right)^{n-i} L_{i}^{(-\frac{1}{2})}\left[2\frac{\{\vec{\zeta}_{\text{d,ph}}^\T(x_a, p_b) \vec{e}_x\}^2}{\sigma_x^2-4\sigma_x}\right] L_{n-i}^{(-\frac{1}{2})}\left[2\frac{\{\vec{\zeta}_{\text{d,ph}}^\T(x_a, p_b) \vec{e}_p\}^2}{\sigma_p^2-4\sigma_p}\right]
.\end{align}




\begin{acknowledgments}
	We acknowledge funding by the Deutsche Forschungsgemeinschaft (DFG) - Project No. 425217212 - SFB 1432.
\end{acknowledgments}

\begin{flushleft}
\textbf{\large Supplementary Information}
\end{flushleft}

\section{Subcycle mode basis}
Multicycle dynamics is usually well captured by a Hermite-Gauss function basis defining a set of mode operators $\hat{a}_i = \int_0^\infty f_i(\omega) \hat{a}_\omega \dd \omega$.
However, while the local oscillator is multicycle, the nonlinear interaction in electro-optic sampling involves some subcycle dynamics, thus requiring a mode basis able to capture subcycle as well as multicycle dynamics.
Therefore, we introduce the subcycle mode basis, based on the generalised Laguerre-Gauss modes.
The first mode is the two parameter, $\sigma_0, k_0 > 0$, positive frequency, $\omega \geq 0$, distribution
\begin{equation}\label{eq:fundamental_mode_function_si}
	f_0(\omega) = \left[\frac{2 / \sigma_0}{\Gamma(k_0 + 1 / 2)}\right]^{\frac{1}{2}} \left(\frac{\omega}{\sigma_0}\right)^{k_0} \exp(-\frac{\omega^2}{2\sigma_0^2})
.\end{equation}
All higher modes can be calculated using
\begin{equation}\label{eq:polynomial_si}
	\tilde{L}_i(\omega) = \left[\frac{\Gamma(i + 1) \Gamma(k_0 + 1 / 2)}{\Gamma(i + k_0 + 1 / 2)}\right]^{\frac{1}{2}} \laguerre_i^{(k_0 - 1 / 2)}\left(\frac{\omega^2}{\sigma_0^2}\right)
,\end{equation}
with the generalized Laguerre polynomials, $\laguerre_i^{(\alpha)}\left(x\right)$, and are given by
\begin{equation}\label{eq:mode_functions_si}
	f_i(\omega) = f_0(\omega) \tilde{L}_i(\omega)
.\end{equation}
We can calculate the central frequency of the fundamental mode, $f_0(\omega)$, and approximate it for large $k_0$,
\begin{equation}\label{eq:average_frequency}
	\bar{\omega}_0 = \sigma_0 \frac{\Gamma(k_0 + 1) }{\Gamma(k_0 + 1 / 2)} \approx \sigma_0 \sqrt{k_0 + \frac{1}{\pi}}
.\end{equation}
Adding $1/\pi$ to $k_0$ ensures the correct value at $k_0 = 0$.
Similarly, we obtain the bandwidth spread of the distribution as
\begin{equation}\label{eq:bandwidth}
	\Delta \omega_0 = 2 \sqrt{2\ln(2)}\sigma_0 \sqrt{k_0 + \frac{1}{2} - \frac{\Gamma^2(k_0 + 1)}{\Gamma^2(k_0 + 1 / 2)}} \approx \sqrt{2\ln(2)} \sigma_0
,\end{equation}
since the distribution $f_0(\omega)$ resembles a Gaussian for large $k_0$.
The ratio $\bar{\omega}_0/\Delta \omega_0$ between the central frequency and the spread depends solely on $k_0$ with a minimum of $\sqrt{2/(\pi - 2)}$ at $k_0 = 0$ and strictly monotonously increasing with $k_0$.
Thus, we can interpret $k_0$ as a parameter determining the number of optical cycles completed during the pulse duration.
To gain a better understanding of the cycle parameter $k_0$, we calculate the Fourier transform of the fundamental mode,
\begin{align}\label{eq:fundamental_wavefunction}
	f_0(t) = \frac{1}{\sqrt{2\pi}} \left[\frac{2}{\sigma_0} \frac{1}{\Gamma(k_0 + 1 / 2)}\right]^{\frac{1}{2}} 2^{\frac{k_0-1}{2}} \Bigg[&\Gamma\left(\frac{k_0+1}{2}\right)\qFp{1}{1}\left\{\frac{k_0+1}{2}; \frac{1}{2}; -\frac{\sigma_0^2}{2} t^2\right\} \nonumber\\
	&- \iu t \sqrt{2} \sigma_0 \Gamma\left(\frac{k_0}{2} + 1\right) \qFp{1}{1}\left\{\frac{k_0}{2} + 1; \frac{3}{2}; - \frac{\sigma_0^2}{2} t^2\right\}
		\Bigg]
,\end{align}
and can use this time-domain representation to define three regimes for the number of optical cycles listed in Tab.~\ref{tab:cyclness}.
\begin{table}
	\caption{Comparison of the three regimes for the fundamental mode function $f_0$ defined in equation~\eqref{eq:fundamental_mode_function} and \eqref{eq:fundamental_wavefunction}}
	\label{tab:cyclness}
	\begin{tabular}{l|c|c}
        Regime & number of roots in $\Im[f_0(t)]$ & Cycle parameter \\
        \hline \hline
		Subcycle & 1 & $0 < k_0 \leq 1$ \\
		Single cycle & 3 & $1 < k_0 \leq 3$ \\
		Multicycle & $>3$ & $3 < k_0$
	\end{tabular}
\end{table}

\begin{figure}
	\includegraphics[width=\textwidth]{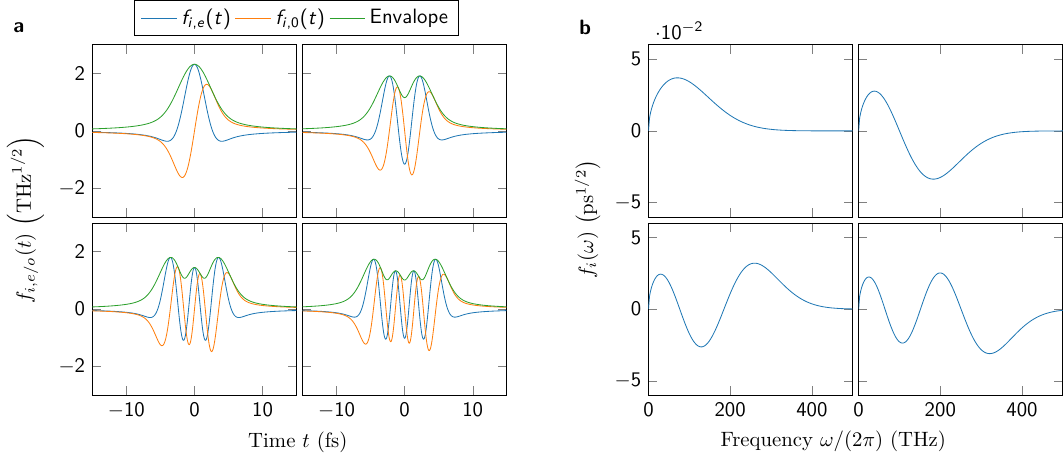}
	\caption{%
		\label{fig:subcyclebasis}
		\textbf{a}, The first four mode function in the time domain with parameters $k_0 = 0.5$ and $\sigma_0 / (2\pi) = \SI{100}{\tera\hertz}$.
        Thus, the fundamental mode is subcycle.
        \textbf{b}, The same four modes as in \textbf{a}, but in the (positive) frequency domain.}
\end{figure}

The orthogonal functions $f_i(\omega) = [f_{i, e}(\omega) + f_{i, o}(\omega)] / 2$ can be extended to negative frequencies by decomposing them into even and an odd functions $f_{i, e}(\omega)$, $f_{i, o}(\omega)$.
Thus, we can decompose the mode operator $\hat{a}_i = \frac{1}{\sqrt{2}}(\hat{x}_i + \iu \hat{p}_i)$ into the even quadrature $\hat{x}_i = \frac{1}{\sqrt{2}}\int_{-\infty}^\infty f_{i, e}(\omega) \hat{a}_\omega \dd \omega$ and the odd quadrature $\hat{p}_i = \frac{-\iu}{\sqrt{2}}\int_{-\infty}^\infty f_{i, o}(\omega) \hat{a}_\omega \dd \omega$, by deploying the convention $\hat{a}_\omega = \hat{a}_{-\omega}^\dagger$.
In the time domain we have $\hat{x}_i = \frac{1}{\sqrt{2}}\int_{-\infty}^\infty f_{i, e}(t) \hat{a}_t \dd t$ and $\hat{p}_i = \frac{1}{\sqrt{2}}\int_{-\infty}^\infty f_{i, o}(t) \hat{a}_t \dd t$.
An example for the subcycle mode basis can be seen in Fig.~\ref{fig:subcyclebasis} for a cycle parameter $k_0 = 0.5$ and bandwidth $\sigma_0 / (2\pi) = \SI{100}{\tera\hertz}$
In praxis, calculation with a countable infinite set of modes is not always feasible.
Thus, the set has to be truncated at an index capturing the dynamics of interest.
We call the maximal index $i_{\text{max}}$.

\section{Derivation of the measurement operator}\label{sm:measrument_operator}
The detected photon-count difference with quantum efficiency $\eta(\omega)$ is
\begin{equation}\label{eq:photon_difference_1}
	\hat{q} = \int_0^\infty \eta(\omega) \{[\hat{a}_s^{\prime\prime}(\omega)]^\dagger\hat{a}_s^{\prime\prime}(\omega) - \hat{a}_z^{\prime\prime}(\omega)]^\dagger\hat{a}_z^{\prime\prime}(\omega)\} \dd \omega
.\end{equation}
The action of the wave plate is independent of frequency, thus
\begin{align}
	\hat{a}_s^{\prime\prime}(\omega) &= \e^{-\iu \tilde{\Phi} / 2}\left[w_1\hat{a}_s^\prime(\omega) + w_2\hat{a}_z^\prime(\omega)\right] \\
	\hat{a}_z^{\prime\prime}(\omega) &= \e^{-\iu \tilde{\Phi} / 2}\left[w_2\hat{a}_s^\prime(\omega) + w_1^\ast\hat{a}_z^\prime(\omega)\right]
.\end{align}
with 
\begin{align}
	w_1 &= \cos(\Phi / 2) - \iu \sin(\Phi / 2)\cos(2\theta) \label{eq:wp1} \\
	w_2 &= -\iu \sin(\Phi / 2) \sin(2\theta) \label{eq:wp2}
.\end{align}
Substituting into equation~\eqref{eq:photon_difference_1} gives:
\begin{align}
	\hat{q} = \int_0^\infty \eta(\omega) \{[&\abs{w_1}^2 - \abs{w_2}^2] [\hat{a}_s^\prime(\omega)]^\dagger\hat{a}_s^\prime(\omega) + w_1^\ast[w_2 - w_2^\ast] [\hat{a}_s^\prime(\omega)]^\dagger\hat{a}_z^\prime(\omega) \\
		&+  w_1[w_2^\ast - w_2][\hat{a}_z^\prime(\omega)]^\dagger\hat{a}_s^\prime(\omega) + [\abs{w_2}^2 - \abs{w_1}^2] [\hat{a}_z^\prime(\omega)]^\dagger\hat{a}_z^\prime(\omega)\} \dd \omega
.\end{align}
Thus, to obtain an observable linear in $\hat{a}_s^\prime(\omega)$, we require $0 = \abs{w_1}^2 - \abs{w_2}^2$ or
\begin{equation}
	\theta = \frac{1}{2}\acos(\sqrt{\frac{1 - 2\cos^2(\Phi / 2)}{2\sin^2(\Phi / 2)}})
.\end{equation}
Inserted in equation~\eqref{eq:wp1} and \eqref{eq:wp2} and restricting $\Phi \in [\frac{1}{2}\pi, \frac{3}{2}\pi]$, we obtain $w_1 = \cos(\Phi / 2) - \iu \frac{1}{\sqrt{2}}\sqrt{1 - 2\cos^2(\Phi / 2)}$ and $w_2 = -\iu \frac{1}{\sqrt{2}}$. Furthermore, by defining
\begin{equation}
	\varphi^\prime = -\atan(\sqrt{\frac{2\cos^2(\Phi / 2)}{1 - 2\cos^2(\Phi / 2)}})
,\end{equation}
we can write $\abs{w_1}^2 - \abs{w_2}^2 = 0$, $w_1^\ast[w_2 - w_2^\ast] = \e^{\iu \varphi^\prime}$, $w_1[w_2^\ast - w_2] = \e^{-\iu \varphi^\prime}$ and $\abs{w_2}^2 - \abs{w_1}^2 = 0$.
With this, the photon number difference simplifies to:
\begin{equation}
	\hat{q} = \int_0^\infty \eta(\omega) \left[\e^{\iu\varphi^\prime}\{\hat{a}_s^\prime(\omega)\}^\dagger\hat{a}_z^\prime(\omega) + \e^{-\iu\varphi^\prime}\{\hat{a}_z^\prime(\omega)\}^\dagger\hat{a}_s^\prime(\omega)\right] \dd \omega
.\end{equation}
For a half-wave plate, $\Phi = \pi$ and $\theta = \SI{22.5}{\degree}$, we have $\varphi^\prime = 0$, resulting in a $\hat{x}$-quadrature measurement, while for a quarter-wave plate, $\Phi = \pi / 2$ and $\theta = \SI{45}{\degree}$, we have $\varphi^\prime = -\pi / 2$, resulting in a $\hat{p}$-quadrature measurement, as we will show next.
Before the wave plate, a coherent probe is introduced described by the displacement operator, $\hat{D}_z[\alpha_{\text{LO}}(\omega)]$, acting on the continuous-frequency mode operator as
\begin{equation}
	\hat{a}_z^\prime(\omega) = \hat{D}_z^\dagger[\alpha_{\text{LO}}(\omega)] \hat{a}_z(\omega) \hat{D}_z[\alpha_{\text{LO}}(\omega)] = \hat{a}_z(\omega) + \alpha_{\text{LO}}(\omega)
.\end{equation}
Neglecting all contributions below the first order in $\alpha_{\text{LO}}(\omega)$ approximates the photon-count difference by a quadrature, 
\begin{equation}\label{eq:msnt_q_1_si}
	\hat{q} \approx \int_0^\infty \eta(\omega) \left[\e^{\iu\varphi^\prime}\alpha_{\text{LO}}(\omega)\{\hat{a}_s^\prime(\omega)\}^\dagger + \e^{-\iu\varphi^\prime}\alpha_{\text{LO}}^\ast(\omega)\hat{a}_s^\prime(\omega)\right] \dd \omega
,\end{equation}
justifying the name $\hat{q}$ for the measurement operator.
Inserting into equation~\eqref{eq:msnt_q_1_si} the relation $\hat{a}^\prime_s(\omega) = \sum_i f_i^\ast(\omega) \hat{a}^\prime_i$,
\begin{equation}
	\hat{q} \approx \sum_i\e^{\iu\varphi^\prime}\int_0^\infty \eta(\omega) \alpha_{\text{LO}}(\omega)f_i(\omega) \dd \omega \{\hat{a}_{i}^\prime\}^\dagger + \sum_i\e^{-\iu\varphi^\prime}\int_0^\infty \eta(\omega)\alpha_{\text{LO}}^\ast(\omega) f_i^\ast(\omega) \dd \omega \hat{a}_{i}^\prime
,\end{equation}
and by defining
\begin{equation}
	\vec{\zeta}_{\text{LO,c}} = \left(\int_0^\infty \eta(\omega) \alpha_{\text{LO}}(\omega)f_0(\omega) \dd \omega, \ldots, \int_0^\infty \eta(\omega)\alpha_{\text{LO}}^\ast(\omega) f_0^\ast(\omega) \dd \omega, \ldots\right)^\T
,\end{equation}
as well as $R_{\text{c}}(\varphi^\prime) = \diag(\e^{\iu \varphi^\prime}, \e^{\iu \varphi^\prime}, \ldots, \e^{-\iu \varphi^\prime}, \e^{-\iu \varphi^\prime}, \ldots)$ and $\hat{\vec{\zeta}}_{\text{LO,c}} = (\hat{a}_0, \hat{a}_1, \ldots, \hat{a}^\dagger_0, \hat{a}^\dagger_1, \ldots)^\T$, we can rewrite the measurement quadrature operator as
\begin{equation}
	\hat{q} = \left(R_{\text{c}}^\T(\varphi^\prime) \vec{\zeta}_{\text{LO,c}}\right)^\dagger \hat{\vec{\zeta}}_{\text{LO,c}}
.\end{equation}
We can change to real phase space using the transformation matrix
\begin{equation}
	T = \frac{1}{\sqrt{2}}\begin{pmatrix}
		\one & \iu \one \\
		\one & -\iu \one
	\end{pmatrix}
,\end{equation}
with $\one$ being the identity matrix, resulting in the local oscillator state vector
\begin{equation}
	\vec{\zeta}_{\text{LO}} = T^\dagger\vec{\zeta}_{\text{LO,c}} = (\Re{[\vec{\zeta}_{\text{LO,c}}]_0}, \ldots, \Im{[\vec{\zeta}_{\text{LO,c}}]_0}, \ldots)^\T
,\end{equation}
The action of the wave plate corresponds to a rotation in the $x_i$-$p_i$-area of the phase space,
\begin{equation}
	R(\varphi^\prime) = T^\dagger R_{\text{c}}(\varphi^\prime) T = \begin{pmatrix} 
		\cos(\varphi^\prime) \one & -\sin(\varphi^\prime) \one \\
		\sin(\varphi^\prime) \one & \cos(\varphi^\prime) \one \\
	\end{pmatrix}
.\end{equation}
Together, we can express the measurement operator expressed in the real phase space,
\begin{equation}
	\hat{q} = \left(R_{\text{c}}^\T(\varphi^\prime)T T^\dagger \vec{\zeta}_{\text{LO,c}}\right)^\dagger T T^\dagger \hat{\vec{\zeta}}_{\text{LO,c}} = \left(R^\T(\varphi^\prime)\vec{\zeta}_{\text{LO}}\right)^\T \hat{\vec{\zeta}}_{\text{LO}}
.\end{equation}
Since the local oscillator pulse can be time-delayed against the quantum pulse, the expansion of the local oscillator pulse in the mode basis would have to be done for each time delay.
However, it is computationally more effective to calculate the relation between the mode basis of the time-delayed local oscillator and the mode basis in which the local oscillator is not time-delayed. 
The two bases are related by the free time-evolution of the electromagnetic field,
\begin{equation}
	\hat{U}_{\Delta t} = \exp(-\iu \frac{1}{2} \Delta t \int_0^\infty 
    \left(\omega \hat{a}_\omega^\dagger \hat{a}_\omega + \omega \hat{a}_\omega \hat{a}_\omega^\dagger \right) \dd \omega)
.\end{equation}
Exploiting again the relation $\hat{a}^\prime_s(\omega) = \sum_i f_i^\ast(\omega) \hat{a}^\prime_i$ to calculate $\tilde{\omega}_{ij} = \int_0^\infty \omega f_i(\omega) f_j(\omega) \dd \omega$ and defining
\begin{equation}
	G_{\Delta t} = \Delta t \begin{pmatrix} 
		\Im(\tilde{\omega}) & \Re(\tilde{\omega}) \\
		-\Re(\tilde{\omega}) & \Im(\tilde{\omega})
	\end{pmatrix}, \qquad \Omega = \begin{pmatrix} 
		0 & \one \\
		-\one & 0
		\end{pmatrix}
,\end{equation}
the free time evolution can be rewritten as
\begin{equation}
	\hat{U}_{\Delta t} = \exp(-\hat{\vec{\zeta}}_{\arm}^\T \Omega G_{\Delta t} \hat{\vec{\zeta}}_{\arm})
.\end{equation}
Following \cite{Adesso2014}, we can express the unitary (free) time evolution using a symplectic matrix, 
\begin{equation}
	M(\Delta t) = \exp(G_{\Delta t})
,\end{equation}
to relate the quadrature operators in the two bases,
\begin{equation}
	\hat{\vec{\zeta}}_{\text{LO}} = \hat{U}_{\Delta t}^\dagger \hat{\vec{\zeta}}_{\arm} \hat{U}_{\Delta t} = M(\Delta t) \hat{\vec{\zeta}}_{\arm}
.\end{equation}
A matrix $M$ is symplectic iff it fulfils the relation $M^T \Omega M = \Omega$.
Thus, we can further rewrite the expression for the measurement quadrature operator,
\begin{equation}\label{eq:msnt_quadrature_homodyne}
	\hat{q} = \left[M^\T(\Delta t) R^\T(\varphi^\prime)\vec{\zeta}_{\text{LO}}\right]^\T \hat{\vec{\zeta}}_\arm
.\end{equation}
By introducing the abbreviation for the transformed local oscillator state,
\begin{equation}\label{eq:lo_homodyne}
	\vec{\zeta}_{\text{LO}}(\Delta t, \varphi^\prime) = M^\T(\Delta t) R^\T(\varphi^\prime)\vec{\zeta}_{\text{LO}}
,\end{equation}
we can express the measurement quadrature operator as the projection of the vector of operators $\hat{\vec{\zeta}}_\arm$ onto the vector of the transformed local oscillator,
\begin{equation}
	\hat{q} = \vec{\zeta}_{\text{LO}}^\T(\Delta t, \varphi^\prime)\hat{\vec{\zeta}}_\arm
.\end{equation}
As a last step, we have to implement the simultaneous measurement using the beam splitter.
We denote the quadrature operators at the empty port of the beam splitter with $\hat{\vec{\zeta}}_{\text{vac}}$, the input of the quantum state $\hat{\vec{\zeta}}$ and at the two output ports with $\hat{\vec{\zeta}}_{a}$, $\hat{\vec{\zeta}}_{b}$, respectively.
These operators are related to each other by the following equations,
\begin{align}
	\hat{\vec{\zeta}}_a &= \frac{1}{\sqrt{2}}\left(-\Omega \hat{\vec{\zeta}}_{\text{vac}} + \hat{\vec{\zeta}}\right) \label{eq:beamsplitter_a}\\
	\hat{\vec{\zeta}}_b &= \frac{1}{\sqrt{2}}\left(\hat{\vec{\zeta}}_{\text{vac}} -\Omega \hat{\vec{\zeta}}\right) \label{eq:beamsplitter_b}\\
.\end{align}
The simultaneous measurement consists of two copies of the quadrature measurement described above, performed on the two output ports of the beam splitter.
Therefore, we can identify, the operators $\hat{\vec{\zeta}}_a$ and $\hat{\vec{\zeta}}_b$ with $\hat{\vec{\zeta}}_\arm$.
From here on, we will refer to $a$ and $b$ using $x=a,b$.
Two measurements means two measurement operators $\hat{q}_\arm = \vec{\zeta}_{\text{LO}}^\T(\Delta t_\arm, \varphi_\arm^\prime)\hat{\vec{\zeta}}_\arm$ with $\arm=a,b$.
Exploiting $\Omega R^\T(\varphi^\prime) = R^\T(\varphi^\prime - \pi / 2)$, and inserting equation~\eqref{eq:beamsplitter_a} and \eqref{eq:beamsplitter_b}, the two quadrature measurement operators are
\begin{align}
	\hat{q}_a &= \vec{\zeta}_{\text{LO}}^\T(\Delta t_a, \varphi_a^\prime) \hat{\vec{\zeta}} + \vec{\zeta}_{\text{LO}}^\T(\Delta t_a, \varphi_a^\prime - \pi / 2) \hat{\vec{\zeta}}_{\text{vac}} \\
	\hat{q}_b &= \vec{\zeta}_{\text{LO}}^\T(\Delta t_b, \varphi_b^\prime - \pi / 2) \hat{\vec{\zeta}} + \vec{\zeta}_{\text{LO}}^\T(\Delta t_b, \varphi_b^\prime) \hat{\vec{\zeta}}_{\text{vac}}
.\end{align}
By defining $\varphi_a = \varphi_a^\prime$ and $\varphi_b = \varphi_b^\prime - \pi / 2$, the two equations above can be written in a more compact form
\begin{equation}\label{eq:quadrature_homodyne}
	\hat{q}_\arm(\Delta t_\arm, \varphi_\arm) = \hat{q}_\arm = \vec{\zeta}_{\text{LO}}^\T(\Delta t_\arm, \varphi_\arm) \hat{\vec{\zeta}} + \vec{\zeta}_{\text{LO}}^\T(\Delta t_\arm, \varphi_\arm + \Delta \varphi_\arm) \hat{\vec{\zeta}}_{\text{vac}}
,\end{equation}
with $\Delta \varphi_a = -\pi / 2$ and $\Delta \varphi_a = \pi / 2$.
equation~\eqref{eq:quadrature_homodyne} describes the measurement operator for the implementation of eight-port homodyne detection described in the main text.
The only difference to equation~(2) is the missing nonlinear interaction, necessary to implement electro-optic sampling.
This will be introduced in the following two sections.

\section{Discretizing the nonlinear interaction}\label{sm:nl}
The second-order nonlinear interaction is mediated by a zinc-blende type nonlinear crystal and the paraxial approximation is applied. The interaction is of first order in the $z$-polarized electric field, $\hat{E}_z(x,t)$, and second order in the $s$-polarized field, $\hat{E}_s(x,t)$ \cite{Moskalenko2015}.
The second-order nonlinear interaction along the propagation direction is assumed to be given by $\lambda(x) = \lambda f_{\lambda}(x)$ with $\lambda = A \varepsilon_0 d / 2$ defined in terms of the beam waist area $A = \pi (\SI{3}{\micro\meter})^2$, the vacuum permittivity $\varepsilon_0$ and the interaction parameter $d = - n^4(\bar{\omega}_{\text{p}}) r_{41}$ which depends on the refractive index at the central frequency of the probe pulse $\bar{\omega}_{\text{p}}$ and the electro-optic coefficient of zinc tellurid, $r_{41} = \SI{4}{\pico\meter\per\volt}$ \cite{Boyd2019}.
The refractive index is modelled by the Sellmeier equation,
\begin{equation}\label{eq:refractive_index}
    n(\omega) = \left(a + b \frac{[2 \pi c]^2}{[2\pi c]^2 - \omega^2 \gamma} \right)^{\frac{1}{2}},
\end{equation}
with $a = 4.27$, $b = 3.01$, $\gamma = \SI{0.142}{\cdot 10^{-12} \meter^2}$ and $c$ the speed of light in vacuum \cite{Marple1964}.
For free space optics, we will assume a rectangular profile of the crystal with length $L$, i.e., $f_{\lambda}(x) = \rect(x / L)$.
The Hamiltonian,
\begin{equation}\label{eq:nl_hamiltonian_sm}
	\hat{H}_{\text{NL}}(t) = \int_{-\infty}^\infty \lambda(x) \hat{E}_z(x,t) \hat{E}_s^2(x,t)
,\end{equation}
describing the nonlinear interaction results in a time evolution described by the unitary operator,
\begin{equation}
	\hat{U}_{\text{NL}} = \mathcal{T} \exp(-\frac{\iu}{\hbar}\int_{-\infty}^\infty \hat{H}_{\text{NL}}(t) \dd t)
,\end{equation}
with $\mathcal{T}$ accounting for the time ordering protocol.
To avoid complications with the time ordering, we can use a Magnus expansion
\begin{equation}
	\hat{U}_{\text{NL}} = \exp(\sum_{k=1}^\infty \hat{\Xi}_k)
,\end{equation}
with
\begin{align}
	\hat{\Xi}_1 &= -\frac{\iu}{\hbar} \int_{-\infty}^\infty \hat{H}_{\text{NL}}(t) \dd t \\
	\hat{\Xi}_2 &= \frac{1}{2\hbar^2}\int_{-\infty}^\infty \int_{-\infty}^t [\hat{H}_{\text{NL}}(t), \hat{H}_{\text{NL}}(t^\prime)]\dd t^\prime \dd t \label{eq:second_order_magnus}
.\end{align}
First, we will have a closer look at the first order, $\hat{\Xi}_1$.
The electric field of polarization, $\mu=s,z$, is given as $\hat{E}_{\mu}(x, t) = \int_{0}^\infty E(\Omega) e^{\iu(k_\Omega x - \Omega t)} \hat{a}_{\mu}(\Omega) \dd \Omega + \hc$ with $E(\Omega) = -\iu \sqrt{\hbar \Omega / (4\pi \varepsilon_0 c n(\Omega) A)}$.
The dispersion relation is $c k_\Omega = \Omega n(\Omega)$.
The state of the $z$-polarized field is a strong coherent pulse, which can be described using the displacement operator $\hat{D}[\alpha(\omega)] = \exp(\int_0^\infty \alpha(\omega) \hat{a}_z^\dagger(\omega) \dd \omega - \hc)$,
\begin{align}
	\hat{E}_{z}(x, t) &\to \hat{D}^\dagger[\alpha(\omega)]\hat{E}_{z}(x, t) \hat{D}[\alpha(\omega)] = \hat{E}_{z}(x, t) + E_{\text{p}}(x, t) \approx E_{p}(x, t), \\
	E_{\text{p}}(x, t) &= \int_{0}^\infty E(\Omega) e^{\iu(k_\Omega x - \Omega t)} \alpha(\Omega) \dd \Omega + \hc
.\end{align}
To keep the result general, we introduce the Fourier transforms
\begin{align}
	\widehat{\lambda}(k) &= \frac{1}{\sqrt{2\pi}}\int_{-\infty}^\infty \lambda(x) e^{\iu k x} \dd x \\
	\delta(\omega) &= \frac{1}{2\pi}\int_{-\infty}^\infty e^{-\iu \omega t} \dd t
.\end{align}
In case of the rectangular profile of the crystal, the Fourier transform of the crystal is $\widehat{\lambda}(k) = \lambda\frac{L}{\sqrt{2\pi}}\sinc(k L / 2)$.
The definitions $\Delta k(\Omega, \omega) = k_{\Omega + \omega} - k_{\Omega} - k_{\omega}$, $E_\text{p}(\Omega) = E(\Omega)\alpha(\Omega)\Theta(\Omega)$, with the Heaviside function $\Theta(\Omega)$, $\alpha^2 = \int_0^{\infty} \abs{\alpha(\omega)}^2 \dd \omega$, $f_{\alpha}(\omega) = \alpha(\omega) / \alpha$ as well as
\begin{align}
	S(\Omega, \omega) &= \frac{1}{\hbar} (2\pi)^{3/2}\left(\frac{\hbar}{4\pi\varepsilon_0 c A}\right)^{3/2} \sqrt{\frac{\omega + \Omega}{n(\omega + \Omega)}} f_{\alpha}(\omega + \Omega) \sqrt{\frac{\omega\Omega}{n(\omega)n(\Omega}} \widehat{\lambda}[\Delta k(\Omega, \omega)] \label{eq:squeezing_nl_si} \\
	B(\Omega, \omega) &= \frac{1}{\hbar} (2\pi)^{3/2}\left(\frac{\hbar}{4\pi\varepsilon_0 c A}\right)^{3/2} \sqrt{\frac{\omega + \Omega}{n(\omega + \Omega)}} \left[f_{\alpha}^\ast(\omega - \Omega) + f_{\alpha}(\Omega - \omega)\right] \sqrt{\frac{\omega\Omega}{n(\omega)n(\Omega}} \widehat{\lambda}[\Delta k(\Omega, \omega)] \label{eq:beam_splitting_nl_si}
,\end{align}
allow us to express the unitary with only the first order of the Magnus expansion as
\begin{equation}\label{eq:nl_unitary_first_order}
	\hat{U}_{\text{NL}} \approx \exp(\abs{\alpha} \iint_0^\infty S[\Omega, \omega] \hat{a}_\omega \hat{a}_\Omega \dd \omega \dd \Omega - \hc + 2\abs{\alpha} \iint_0^\infty B[\Omega, \omega] \hat{a}_\omega^\dagger \hat{a}_\Omega \dd \omega \dd \Omega)
.\end{equation}
Scrutinizing equation~\eqref{eq:squeezing_nl_si} and \eqref{eq:beam_splitting_nl_si} we can differentiate two cases:
Case one simultaneously creates or annihilates a photon at $\omega$ and $\bar{\omega}_{\text{p}} - \Omega$.
In the context of electro-optic sampling we will call this difference-frequency generation (DFG).
Case two creates/annihilates a photon at $\omega$, while annihilating/creating a photon at $\omega + \bar{\omega}_{\text{p}}$.
In electro-optic sampling, we will call this sum-frequency generation (SFG).

Inserting the relation $\hat{a}_\omega = \sum_{i=0}^\infty f_i^\ast(\omega) \hat{a}_i$ into equation~\eqref{eq:nl_unitary_first_order} and defining the Matrix $S$ and $B$ by
\begin{align}
	S_{ij} &= 2 \iint_0^\infty S(\Omega, \omega) f_i^\ast(\Omega) f_j^\ast(\omega) \dd \Omega \dd \omega \\
	B_{ij} &= 2 \iint_0^\infty B(\Omega, \omega) f_i(\Omega) f_j^\ast(\omega) \dd \Omega \dd \omega
,\end{align}
allows to rewrite
\begin{align}
	\hat{U}_{\text{NL}} &= \exp(\frac{1}{2} \abs{\alpha}\sum_{i,j=0}^\infty S_{ij}\hat{a}_i \hat{a}_j - \hc + \abs{\alpha}\sum_{i,j=0}^\infty B_{ij} \hat{a}_i^\dagger \hat{a}_j) \\
		&= \exp(-\frac{\iu}{2}\abs{\alpha}\hat{\vec{\zeta}}_c^\T \begin{pmatrix} 
		\iu B & (\iu S)^\ast \\
		\iu S & (\iu B)^\ast
	\end{pmatrix}\hat{\vec{\zeta}}_c)
.\end{align}
Defining the matrix
\begin{equation}\label{eq:nl_hamiltonian_matrix_si}
	G_{\text{NL}} = \Omega T^\dagger \begin{pmatrix} 
		\iu B & (\iu S)^\ast \\
		\iu S & (\iu B)^\ast
	\end{pmatrix} T = \begin{pmatrix} 
		-\Re(S - B) & \Im(S - B) \\
		\Im(S + B) & \Re(S + B)
		\end{pmatrix}
,\end{equation}
allows us to express the unitary of the nonlinear interaction as
\begin{equation}
	\hat{U}_{\text{NL}} = \exp(-\abs{\alpha} \hat{\vec{\zeta}}^\T \Omega G_{\text{NL}} \hat{\vec{\zeta}})
.\end{equation}
Since the Hamiltonian of the nonlinear interaction is quadratic in the annihilation and creation operators, it is sufficient to look at the transformation of the quadrature operators and the unitary evolution.
By introducing the symplectic matrix
\begin{equation}\label{eq:nl_symplectic_matrix}
	M_{\text{NL}}(\alpha) = \exp[\abs{\alpha} G_{\text{NL}}]
,\end{equation}
we can express the evolution of the quadrature operators in a compact form, $\hat{\vec{\zeta}}^\prime = \hat{U}_{\text{NL}}^\dagger\hat{\vec{\zeta}}\hat{U}_{\text{NL}} = M_{\text{NL}}\hat{\vec{\zeta}}$.

\section{Second order Magnus expansion}
\begin{figure}
	\centering
	\includegraphics[width=0.5\textwidth]{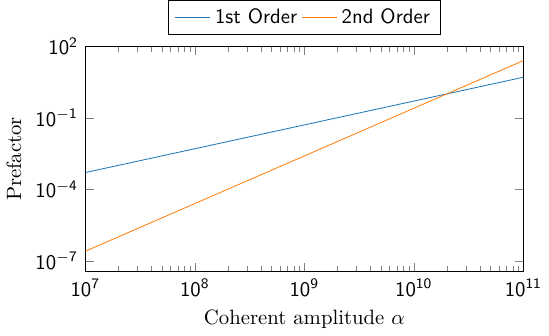}
	\caption{Comparison between the prefactors in the first order and second order Magnus expansion as a function of the coherent amplitude $\alpha$ of the $z$-polarized classical pulse in the nonlinear interaction defined by equation~\eqref{eq:nl_hamiltonian_sm}.}\label{fig:magnus}
\end{figure}

Evaluation of the second order in the Magnus expansion in equation~\eqref{eq:second_order_magnus} leads to
\begin{align}
    \hat{\Xi}_2 &= \int_{-\infty}^\infty \int_{-\infty}^{t_1} \iint_{-\infty}^\infty \lambda(x_1)\lambda(x_2) E_z(x_1, t_1)E_z(x_2, t_2) \nonumber \\
    &[\hat{E}_s(x_1, t_1), \hat{E}_s(x_2, t_2)]\{\hat{E}_s(x_1, t_1), \hat{E}_s(x_2, t_2) \} \dd x_1 \dd x_2 \dd t_1 \dd t_2 
.\end{align}
with the commutator $[\hat{A}, \hat{B}] = \hat{A}\hat{B} - \hat{B}\hat{A}$ and the anticommutator $\{\hat{A}, \hat{B}\} = \hat{A}\hat{B} + \hat{B}\hat{A}$.
The response function with $\tau = \frac{n(\Omega)}{c}(x_2 - x_1) - (t_2 - t_1)$,
\begin{align}
    \mathcal{R}(\tau) &= \frac{\iu}{\hbar}[\hat{E}_s(x_1, t_1), \hat{E}_s(x_2, t_2)] \\
    &= -\frac{\iu}{\hbar}\int_{-\infty}^{\infty} E^2(\Omega) e^{\iu \Omega \tau} \dd \Omega \\
    &= \frac{-\iu}{4\pi\varepsilon_0 c A} \int_{-\infty}^{\infty} \frac{\Omega}{n(\Omega)} e^{\iu \Omega \tau} \dd \Omega
,\end{align}
restricts the interaction of the fields at $(c t_1, x_1)$ and $(c t_2, x_2)$ to the light cone \cite{Lindel2024}.
This can be seen if we neglect dispersion, $n(\Omega) = n$.
In this case, the response function can be evaluated to 
\begin{equation}
    \mathcal{R}(\tau) = \frac{-\iu}{4\pi\varepsilon_0 c A n}  \int_{-\infty}^{\infty} \Omega e^{i \Omega \tau}\dd \Omega = \frac{1}{2\varepsilon_0 c A n} \frac{\dd }{\dd \tilde{\tau}} \delta(\tilde{\tau}) |_{\tilde{\tau} = \tau}
.\end{equation}
To obtain an estimate for the contribution of the second order in the Magnus expansion to the nonlinear interaction, it is sufficient to compare the prefactor of $\lambda(x_1)\lambda(x_2) E_z(x_1, t_1)E_z(x_2, t_2)$ to the equivalent prefactor of $\lambda(x) E_z(x, t)$ corresponding to the first order. The prefactor for the second-order Magnus expansion, $\alpha^2 \frac{\lambda^2 \hbar}{4\pi \varepsilon_0 c A}$, and the first order $\alpha \lambda\sqrt{\frac{\hbar}{4\pi \varepsilon_0 c A}}$ are compared in Fig.~\ref{fig:magnus}.

The two contribute the same at a coherent amplitude of $\alpha = \frac{1}{\lambda}\sqrt{\frac{4\pi \varepsilon_0 c A}{\hbar}} \approx 2 \cdot 10^{10}$.
Therefore, it is sufficient to consider only the first order since we assume the pulse power to stay way below the threshold defined by our estimate. 

\section{Electro-optic sampling}\label{sm:eos}
Electro-optic sampling is, to a large part, equivalent to the homodyne setup described in Sec.~\ref{sm:measrument_operator}.
Yet, after the quantum pulse and the local oscillator combine in a polarizing beam splitter, they enter a nonlinear crystal, of the type described in the previous Sec.~\ref{sm:nl}.
Therefore, we have to modify equation~\eqref{eq:lo_homodyne} to 
\begin{equation}
	\vec{\zeta}_{\text{LO}}(\Delta t, \alpha, \varphi) = M^\T(\Delta t) M_{\text{NL}}^\T(\alpha) R^\T(\varphi) \vec{\zeta}_{\text{LO}}
,\end{equation}
resulting in equation~(2),
\begin{equation}
	\hat{q}_\arm(\Delta t_\arm, \alpha_\arm, \varphi_\arm) = \vec{\zeta}_{\text{LO}}^\T(\Delta t_\arm, \alpha_\arm, \varphi_\arm) \hat{\vec{\zeta}} + \vec{\zeta}_{\text{LO}}^\T(\Delta t_\arm, \alpha_\arm, \varphi_\arm + \Delta \varphi_\arm) \hat{\vec{\zeta}}_{\text{vac}}
,\end{equation}
describing the measurement operators of simultaneous electro-optic sampling.

\begin{figure}[h]
	\centering
	\includegraphics[width=\textwidth]{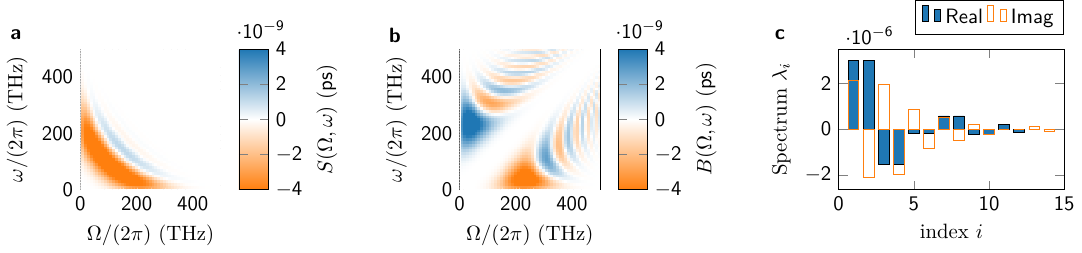}
	\caption{\textbf{a}, The kernel $S(\Omega, \omega)$ describing the squeezing in the nonlinear interaction between the frequencies $\Omega$ and $\omega$.
    The detection crystal is assumed to be made from zinc telluride, as described in Sec.~\ref{sm:nl}, placed in free space and of length $L = \SI{20}{\micro\meter}$.
    The probe parameters are $k_{\text{DX}} = 4$, $\sigma_{\text{DX}} = \SI{100}{\tera\hertz}$.
    \textbf{b}, The kernel $B(\Omega, \omega)$ describing the beam-splitter interaction between the frequencies $\Omega$ and $\omega$.
    The crystal and probe properties are the same as in \textbf{a}.
    \textbf{c}, Complex eigenvalues of the matrix $G_{\text{NL}}$, defined in equation~\eqref{eq:nl_hamiltonian_matrix_si}, showing the multimode character of the interaction.
    The eigenvalues are complex due to the simultaneous presence of the sum- and difference frequency processes.}\label{fig:eos_jsa}
\end{figure}

Predicting the effect of the symplectic matrix in equation~\eqref{eq:nl_symplectic_matrix} on the quadratures is not a straightforward endeavour.
While perturbation theory simplifies the interaction in the weak probe regime, stronger probe amplitudes reduce shot noise \cite{Guedes2023, Hubenschmid2024}.
If the nonlinear interaction contains only squeezing, i.e., $B(\Omega, \omega) = 0$, one can simplify the interaction to a few principle modes using a Schmidt decomposition \cite{Law2000}.
This is not possible for the nonlinear action underlying electro-optic sampling, since the simultaneous Schmidt decomposition of $S(\Omega, \omega)$ and $B(\Omega, \omega)$ is, in general, not possible.
An example for the two kernels $S(\Omega, \omega)$ and $B(\Omega, \omega)$ for a nonlinear crystal as described in Sec.~\ref{sm:nl} with a rectangular profile of length $L = \SI{20}{\micro\meter}$ and a probe with $k_{\text{DX}} = 4$, $\sigma_{\text{DX}} = \SI{100}{\tera\hertz}$ is shown in Fig.~\ref{fig:eos_jsa} \textbf{a} and \textbf{b}.
The (complex) spectrum $\lambda_i \in \mathbb{C}$ obtained from the corresponding matrix $G_{\text{NL}}$ defined in equation~\eqref{eq:nl_hamiltonian_matrix_si} can be taken from Fig.~\ref{fig:eos_jsa} \textbf{c}.
The first four modes contribute the most to the spectrum, implying an effective description of the nonlinear interaction based on four modes.
Yet, the question remains how to obtain this effective description.

\begin{figure}[h]
	\centering
	\includegraphics[width=\textwidth]{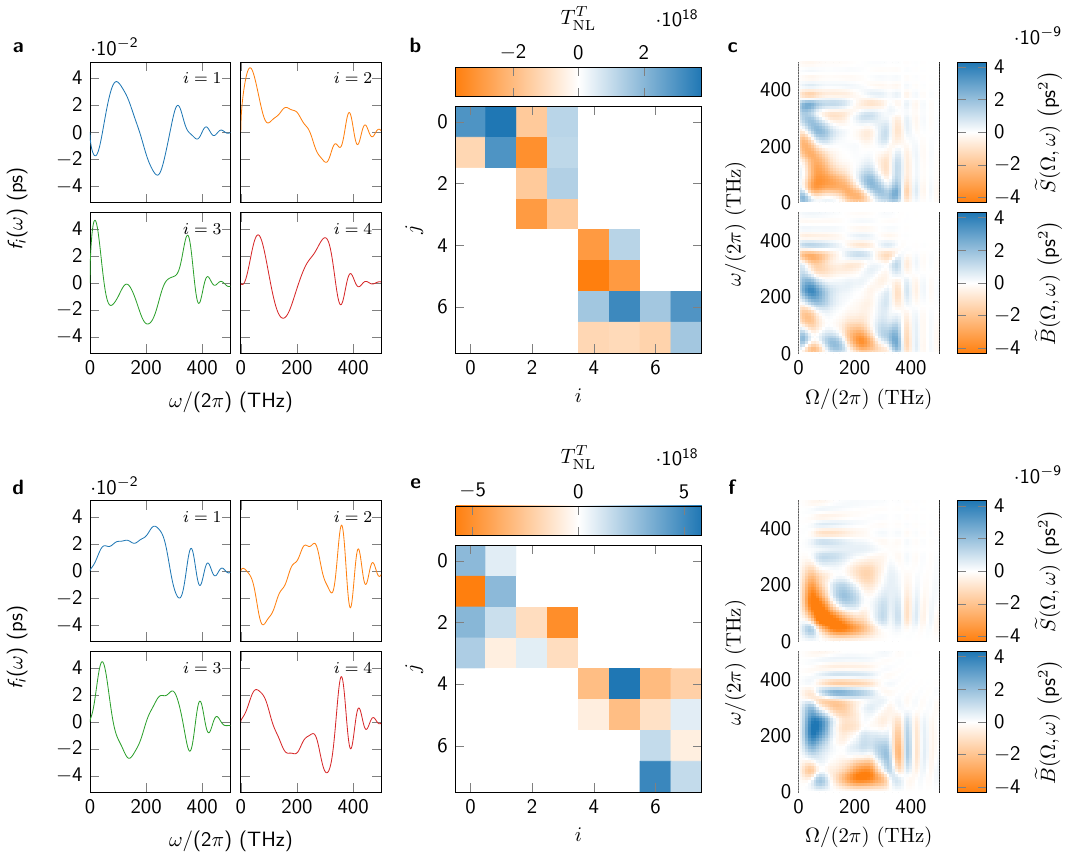}
	\caption{\textbf{a}, First four principle modes of the real Schur decomposition applied to the $\hat{p}$-quadrature block of $G_{\text{NL}}$, defined in equation~\eqref{eq:nl_hamiltonian_matrix_si}. The nonlinear interaction corresponds to the one shown in Fig.~\ref{fig:eos_jsa}.
    \textbf{b}, The triangular matrix resulting from the real Schur decomposition truncated to the first four principle modes.
    If the spectrum in Fig.~\ref{fig:eos_jsa} contains two eigenvalues $\lambda_i$ and $\lambda_i^\ast$, the triangular matrix has a corresponding $2 \times 2$-block on the diagonal.
    \textbf{c}, the kernels $\widetilde{S}(\Omega, \omega)$, $\widetilde{B}(\Omega, \omega)$ reconstructed from the truncated triangular matrix and first four principle modes.
    \textbf{d}-\textbf{f}, similar to \textbf{a}-\textbf{c}, but the Schur decomposition is applied to the $p$-quadrature block of $G_{\text{NL}}$.
    }\label{fig:eos_effective}
\end{figure}

Diagonalization of the matrix $G_{\text{NL}}$ defined in equation~\eqref{eq:nl_hamiltonian_matrix} with orthogonal transformation matrices is only possible if $G_{\text{NL}}$ is a normal matrix.
Here we will restrict our discussion to $S, B \in \mathbb{R}$, which is achieve by restricting the phase of the probe to be integer multiples of $\pi$, i.e., $\alpha \in \mathbb{R}$.
With $S = S^\T$ being symmetric and $B = -B^\T$ antisymmetric we can calculate the commutator
\begin{equation}
	\left[\begin{pmatrix} 
		-S + B & 0 \\
		0 & S + B
	\end{pmatrix},
	\begin{pmatrix} 
		-S + B & 0 \\
		0 & S + B
	\end{pmatrix}^\T\right]
	= 2
	\begin{pmatrix} 
		[S,B] & 0 \\
		0 & -[S,B]
	\end{pmatrix}
,\end{equation}
which implies that $G_{\text{NL}}$ is only normal if $S$ and $B$ commute, i.e., $[S,B] = 0$.
Since this is not the case in general, diagonalization with orthogonal transformation matrices is not possible.
Yet, we can still use the real Schur decomposition to transform $G_{\text{NL}}$ to a (block) triagonal matrix $T_{\text{NL}}$ using a orthogonal transformation matrix $O$, i.e., $G_{\text{NL}} = O^\T T_{\text{NL}} O$.
To transform the $\hat{x}_i$ and $\hat{p}_i$ quadratures with the same transformation, we apply the Schur decomposition only to the upper (lower) $N \times N$ block of $G_{\text{NL}}$ and transform the lower (upper) block accordingly.
The triagonalized $N \times N$ block of $G_{\text{NL}}$ is thus an upper triangular matrix with the real part of the eigenvalues, $\Re(\lambda_i)$, on the diagonal and the lower $N \times N$ block a lower triangular matrix with eigenvalues $-\Re(\lambda_i)$.
Complex eigenvalues of the upper block of $G_{\text{NL}}$ will lead to two by two blocks with the diagonal $\diag(\Re(\lambda_i), \Re(\lambda_i))$.
The Schur decomposition, truncated to the first four Schur modes of the upper $\hat{x}$-quadrature and lower $\hat{p}$-quadrature $N \times N$ block of $G_{\text{NL}}$ can be seen in Fig.~\ref{fig:eos_effective} \textbf{a}, \textbf{b} and \textbf{d}, \textbf{e} respectively.
Here we consider the transpose, since the vector describing the local oscillator $\vec{\zeta}_{\text{LO}}(\Delta t_\arm, \varphi_\arm, 0)$ transforms with the transpose of the symplectic matrix, $\vec{\zeta}_{\text{LO}}(\Delta t_\arm, \varphi_\arm, \alpha) = M_{\text{NL}}^\T(\alpha) \vec{\zeta}_{\text{LO}}(\Delta t_\arm, \varphi_\arm, 0)$.

\begin{figure}
	\centering
	\includegraphics[width=\textwidth]{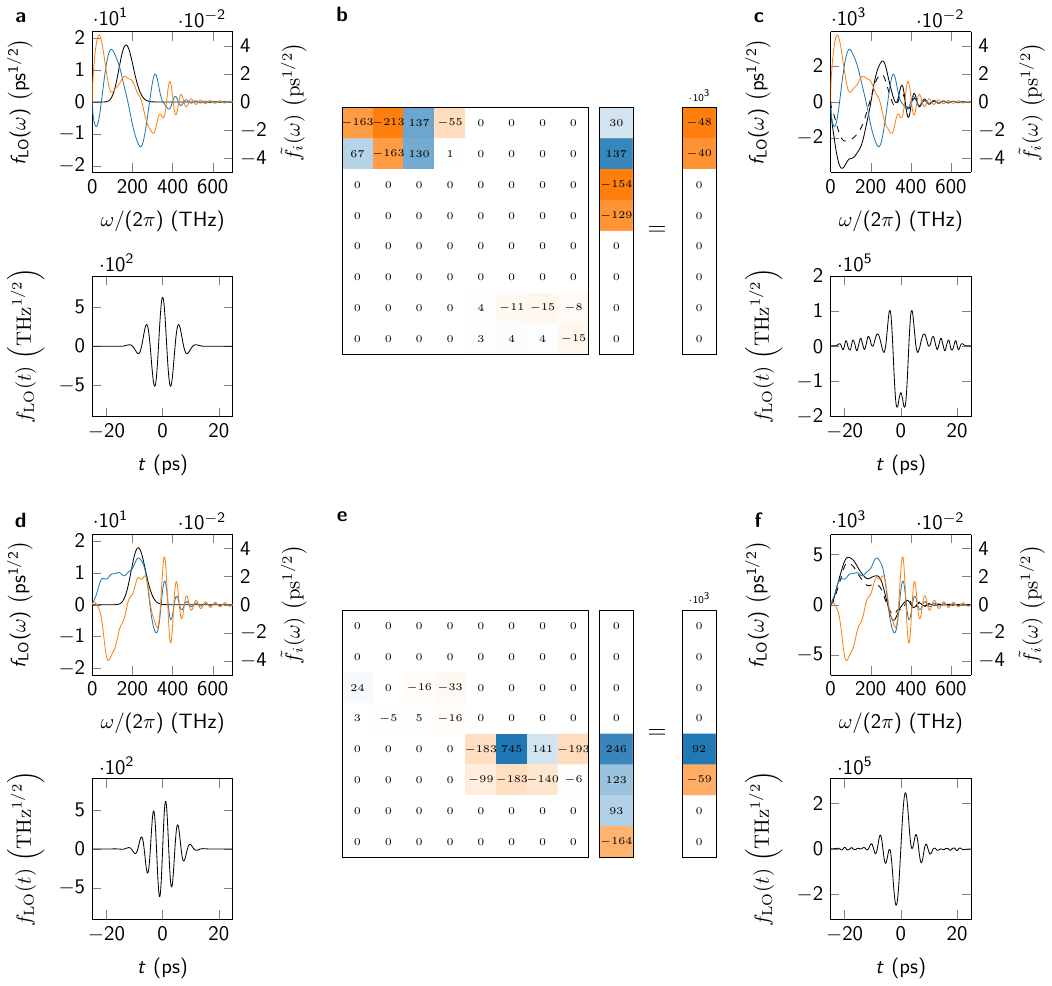}
	\caption{\textbf{a}, First two principle modes of the Schur decomposition in the frequency domain (solid blue and orange) together with the input local oscillator, with $\sigma_{\text{LO}} / (2\pi) = \SI{230}{\tera\hertz}$ and $k_{\text{LO}} = 21.16$, in the frequency and time domain (black) for the Schur decomposition of the $\hat{x}$-block of $G_{\text{NL}}$.
    \textbf{b}, Corresponding matrix input output relation $\tilde{M}_{\text{NL}, trunc}^\T\vec{\tilde{\zeta}}_{\text{LO}}(0, 0, 0) = \vec{\tilde{\zeta}}_{\text{LO}}(0, \alpha, 0)$ truncated to four principle modes, describing the transformation of the local oscillator due to the nonlinear interaction. The coherent amplitude of the probe pulse $\alpha = 1,78 \cdot 10^6$ is determined by the optimisation procedure described in the text.
    \textbf{c}, Local oscillator after transformation with $M_{\text{NL}}$ in frequency and time domain (black).
    The dashed line describes the resulting mode if the truncated matrix, $\tilde{M}_{\text{NL}, trunc}$, is used.
    \textbf{d}-\textbf{f}, Similar to \textbf{a}-\textbf{c}, , but the Schur decomposition is applied to the $\hat{p}$-quadrature block of $G_{\text{NL}}$. 
    The coherent amplitude of the probe pulse $\alpha = -1.94 \cdot 10^6$.}\label{fig:eos_lo}
\end{figure}

The triagonal form of $\tilde{M}_{\text{NL}}^\T(\alpha) = O^\T M_{\text{NL}}^\T(\alpha) O$ defines a sequence of $\hat{q}=\hat{x}$ or $\hat{q}=\hat{p}$ subspaces $V_{\hat{q}}^{(n)}$ closed under $\tilde{M}_{\text{NL}}^\T(\alpha)$, spanned by the Schur modes of the first $n$ blocks, i.e. $V_{\hat{q}}^{(1)} \subset V_{\hat{q}}^{(2)} \subset \ldots \mathbb{R}^{2 N}$.
To better understand the resulting effective nonlinear interaction, we will restrict the further discussion to the first block in the Schur decomposition, i.e., the first two Schur modes.
Since the first two pairs of eigenvalues of the Matrix $M_{\text{NL}}$ have different signs (c.f. Fig.~\ref{fig:eos_jsa} \textbf{c}), one contributes as squeezing while the other manifests as antisqueezing, which is why the local oscillator after transformation with $M_{\text{NL}}^T$ is well described by the first two modes (see Fig.~\ref{fig:eos_lo}).
However, a realistic local oscillator prior to transformation with $M_{\text{NL}}^\T$ can have additional contributions from higher-order principle modes, leading to divinations between the truncated and complete transformation.
The truncated triagonal matrix takes the form [$a = \Re(\lambda_1)$]
\begin{equation}\label{eq:triagonal_trunc}
	T_{\text{NL,trunc}} = \begin{pmatrix} 
		a & -b & 0 & 0 \\
		c & a & 0 & 0 \\
		0 & 0 & -a & -c \\
		0 & 0 & b & -a \\
	\end{pmatrix}
.\end{equation}
Defining $\delta = \alpha\sqrt{bc}$ and $\gamma = \sqrt{b / c}$, the exponentiation of the above truncated triagonal matrix, $\tilde{M}_{\text{NL, trunc}}^\T = \exp(\alpha T_{\text{NL, trunc}})^\T$, can be calculated analytically \cite{Bernstein1993},
\begin{equation}\label{eq:truncated_nl}
	\tilde{M}_{\text{NL, trunc}}^\T = \begin{pmatrix} 
		\e^{-\alpha a} & 0 & 0 & 0 \\
		0 & \e^{-\alpha a} & 0 & 0 \\
		0 & 0 & \e^{\alpha a} & 0 \\
		0 & 0 & 0 & \e^{\alpha a} \\
	\end{pmatrix}\begin{pmatrix} 
		\cos(\delta) & 1 / \gamma \sin(\delta) & 0 & 0 \\
		-\gamma \sin(\delta) & \cos(\delta) & 0 & 0 \\
		0 & 0 & \cos(\delta) & \gamma\sin(\delta) \\
		0 & 0 & -1 / \gamma \sin(\delta) & \cos(\delta) \\
	\end{pmatrix}
.\end{equation}
The truncation to a two-mode interaction in equation~\eqref{eq:truncated_nl} can capture the time evolution of the ($q$-quadrature) local oscillator, if it is contained well in the subspace $V_{\hat{q}}^{(1)}$.
It even becomes exact if $\vec{\tilde{\zeta}}_{\text{LO}}(\Delta t_\arm, 0, \varphi_\arm) = O^\T \vec{\zeta}_{\text{LO}}(\Delta t_\arm, 0, \varphi_\arm) \in V_{\hat{q}}^{(1)}$ and since the first subspace corresponds to the eigenvalues with largest real part, the squeezing interaction is strongest for these modes [e.g., the squeezing in equation~\eqref{eq:truncated_nl} is given by $\alpha a = \alpha \Re(\lambda_1)$].
The goal of electro-optic sampling is to correlate a low frequency range with a higher one to obtain information about the quantum state in the lower end of the spectrum by measuring in the higher frequencies.
In the case of a $x$-quadrature local oscillator, $\vec{\zeta}_{\text{LO}}(\Delta t_\arm, 0, \pi /2)$, the high-frequency contribution to the first two principle modes destructively interfere, if the two modes are added.
Therefore, if we assume the local oscillator prior to the nonlinear symplectic transformation is $\vec{\zeta}_{\text{LO}}(\Delta t_x, 0, 0) = (x_{\text{LO}, 1}, x_{\text{LO}, 2}, \ldots)^\T$, we require $x_{\text{LO}, 1}\cos(\delta) + x_{\text{LO}, 2}\frac{1}{\gamma} \sin(\delta) = -x_{\text{LO}, 1}\gamma \sin(\delta) + x_{\text{LO}, 2} \cos(\delta)$ or
\begin{equation}\label{eq:optimal_probe_strength_p}
	\delta_x = \atan(\sqrt{bc}\frac{x_{\text{LO},1} - x_{\text{LO},2}}{x_{\text{LO},1}c + x_{\text{LO},2}b}) \mod \pi 
.\end{equation}
The transformation of a $x$-quadrature local oscillator with $\sigma_{\text{LO}} / (2\pi) = \SI{170}{\tera\hertz}$ and $k_{\text{LO}} = 11.24$ by a nonlinear interaction with $\alpha = \delta_x / \sqrt{bc} = 1.78 \cdot 10^6$ is shown in Fig.~\ref{fig:eos_lo} \textbf{a} - \textbf{b}.
For the case of a $\hat{p}$-quadrature local oscillator, $\vec{\zeta}_{\text{LO}}(\Delta t_\arm, 0, \pi /2)$, the high frequency part of the first two Schur modes can be suppressed while simultaneously amplifying the lower frequencies by subtracting them.
For $\vec{\zeta}_{\text{LO}}(\Delta t_\arm, 0, \pi /2) = (0, \ldots, 0, p_{\text{LO}, 1}, p_{\text{LO}, 2}, \ldots)^\T$,subtraction leads to the requirement $p_{\text{LO}, 1}\cos(\delta) + p_{\text{LO}, 2}\gamma \sin(\delta) = p_{\text{LO}, 1}\frac{1}{\gamma} \sin(\delta) - p_{\text{LO}, 2} \cos(\delta)$ or
\begin{equation}\label{eq:optimal_probe_strength_p_si}
	\delta_p = \atan(\sqrt{bc}\frac{p_{\text{LO},1} + p_{\text{LO},2}}{p_{\text{LO},1}c - p_{\text{LO},2}b}) \mod \pi 
.\end{equation}
The transformation of a $p$-quadrature local oscillator with $\sigma_{\text{LO}} / (2\pi) = \SI{230}{\tera\hertz}$ and $k_{\text{LO}} = 21.16$ is visualized in Fig.~\ref{fig:eos_lo} \textbf{d} - \textbf{f}.
The optimal probe strength $\alpha = \delta_p / \sqrt{bc} = -1.94 \cdot 10^{6}$ corresponds to the second solution of equation~\eqref{eq:optimal_probe_strength_p_si}.

\section{Free space multimode squeezed vacuum}\label{sm:multimode}
\begin{figure}
	\centering
	\includegraphics[width=\textwidth]{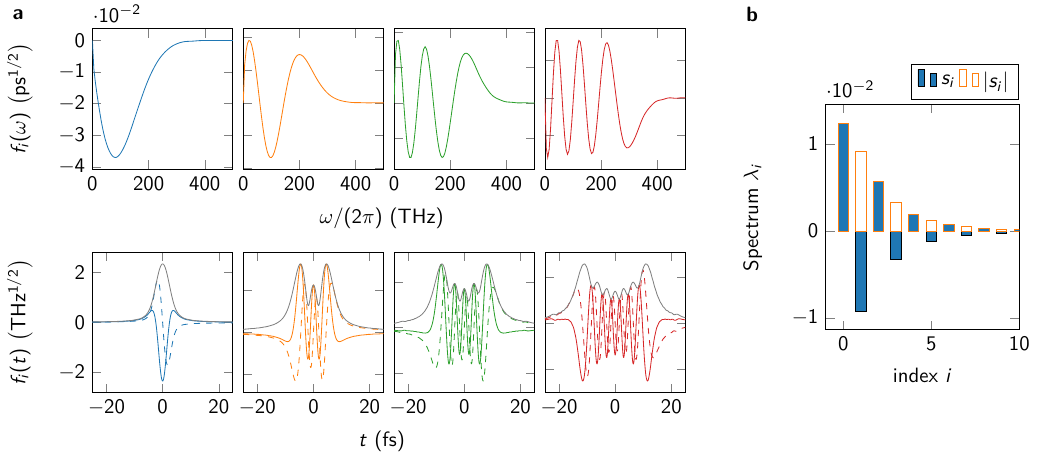}
	\caption{\textbf{a}, Four principle modes of the Bloch-Messiah decomposition in the frequency and time domain.
    Solid lines correspond to the $q$ quadrature, dashed lines to the $p$ quadratures and black lines to the envelope.
    \textbf{b}, Singular values of the Bloch-Messiah decomposition. The parameters are the same as in Fig.~\ref{fig:eos_jsa}}\label{fig:mmsv}
\end{figure}

In the case of the squeezed vacuum we can use the Bloch-Messiah decomposition to simplify the multimode quantum state.
Different to the previous section we diagonalize the symplectic matrix $M_{\text{NL}}(\alpha_{\text{GX}})$ using a singular value decomposition (SVD), $M_{\text{NL}}(\alpha_{\text{GX}}) = U D_{\text{NL}} V^\T$, where $D_{\text{NL}} = \diag(e^{2s_0}, e^{2s_1}, \ldots, e^{-2s_0}, e^{-2s_1}, \ldots)$ and in general $U \neq V$.
Thus, the covariance matrix of the Gaussian state after squeezing the vacuum is ($\vec{\tilde{\zeta}}_{x,i} = V_{\text{NL}}\vec{\zeta}_{x,i}$)
\begin{align}
	\cov_{mmsv} &= M_{\text{NL}}^\T(\alpha_{\text{GX}}) \cov_{\text{vac}} M_{\text{NL}}(\alpha_{\text{GX}}) = \frac{1}{2} V_{\text{NL}} D_{\text{NL}}^2 V^\T_{\text{NL}} \\
		&= \frac{1}{2} \sum_{i=1}^{\infty} e^{2 s_i} \vec{\tilde{\zeta}}_{x,i} \vec{\tilde{\zeta}}_{x,i}^\T + \frac{1}{2} \sum_{i=1}^{\infty} e^{-2 s_i} \vec{\tilde{\zeta}}_{p,i} \vec{\tilde{\zeta}}_{p,i}^\T \label{eq:mmsv}
.\end{align}
Therefore, as a result from the Bloch-Messiah decomposition, we can express the multimode squeezed vacuum as independent single-mode squeezed states.
The respective principle modes and squeezing parameters are shown in Fig.~\ref{fig:mmsv} \textbf{a} and \textbf{b}.

Since the state is clearly not described by a single temporal mode, a time local measurement, i.e., a measurement with the same time delay on each arm, $\Delta t_a = \Delta t_b = \Delta t$, has manifold of modes contributing to the time local correlator
\begin{align}
	g(\Delta t, \Delta t, \varphi_a, \varphi_b) &= \frac{1}{2}\{\ev{\hat{q}_a(\Delta t, \alpha_{\text{DX}}, \varphi_a), \hat{q}_b(\Delta t, \alpha_{\text{DX}}, \varphi_b)\}} \nonumber \\
		&= \frac{1}{2} \sum_{i=1}^{\infty} e^{2 s_i} \abs{\vec{\zeta}_{\text{LO}}(\Delta t, \alpha_{\text{DX}}, \varphi) \vec{\tilde{\zeta}}_{x,i}}^2 + \frac{1}{2} \sum_{i=1}^{\infty} e^{-2 s_i} \abs{\vec{\zeta}_{\text{LO}}(\Delta t, \alpha_{\text{DX}}, \varphi) \vec{\tilde{\zeta}}_{p,i}}^2 \nonumber\\
        &+ \frac{1}{2} \norm{\vec{\zeta}_{\text{LO}}(\Delta t, \alpha_{\text{DX}}, \varphi)}^2
.\end{align}
Since to local oscillator overlaps with all principle modes to some degree, it is not possible to obtain all significant squeezing parameters and corresponding principle modes by measuring $g(\Delta t, \Delta t, \varphi_a, \varphi_b)$ for different time delays $\Delta t$, as pointed out by Yang \textit{et al.} \cite{Yang2023}.

\section{Dispersion engineering}
The number of principle modes contributing to the squeezed state in equation~\eqref{eq:mmsv} can be manipulated by engineering the nonlinear interaction.
One possibility is to periodically pole the nonlinear crystal, i.e., to change $\lambda(x) \to \lambda(x) \sign[\cos(2\pi x / \Lambda)]$ \cite{Boyd2019}.
Yet, following \cite{Boyd2019} will lead to a non-Hermitian Hamiltonian.
Instead, we will approximate the periodic poling by
\begin{equation}
	\lambda(x) \to \lambda(x) \cos(2\pi x / \Lambda)
.\end{equation}
The Fourier transform of the crystal profile for this case is
\begin{equation}
	\frac{1}{\sqrt{2\pi}}\int_{-\infty}^\infty \lambda(x) \cos(2\pi x / \Lambda) e^{\iu k x} \dd x = \frac{1}{2}\left[\widehat{\lambda}(k + \frac{2\pi}{\Lambda}) + \widehat{\lambda}(k - \frac{2\pi}{\Lambda})\right]
.\end{equation}
Thus, the periodic poling leads to an additional wave vector $\pm 2\pi / \Lambda$.
Another possibility to manipulate the wave-vector mismatch is to have birefringence in the material of the nonlinear crystal.
This means the refractive index $n_z(\omega)$ for $z$ polarized field is different to the refractive index $n_s(\omega)$ for $s$-polarized field, without the presence of a finite electric field.
The wave vector mismatch has to be changed to $c\Delta k = (\Omega + \omega)n_z(\Omega + \omega) - \Omega n_s(\Omega) - \omega n_s(\omega)$.
Follow Horoshko \textit{et al.}\cite{Horoshko2024} (see also \cite{RomanRodriguez2021}), we introduce $\xi = \frac{1}{2}(\omega + \Omega)$, $\upsilon = \frac{1}{2}(\omega - \Omega)$ and use a Taylor expansion around $(\xi_0, \upsilon_0)$,
\begin{align}
	\Delta k(\xi,\upsilon) \approx \Delta k(\xi_0, \upsilon_0) &+ \frac{\partial \Delta k}{\partial\xi}\Big\vert_{(\xi_0, \upsilon_0)} (\xi - \xi_0) + \frac{\partial \Delta k}{\partial\upsilon}\Big\vert_{(\xi_0, \upsilon_0)} (\upsilon - \upsilon_0) + \frac{\partial^2 \Delta k}{\partial\upsilon\partial\xi}\Big\vert_{(\xi_0, \upsilon_0)} (\xi - \xi_0)(\upsilon - \upsilon_0) \nonumber \\
	&+ \frac{\partial^2 \Delta k}{\partial\xi^2}\Big\vert_{(\xi_0, \upsilon_0)} (\xi - \xi_0)^2 + \frac{\partial^2 \Delta k}{\partial\upsilon^2}\Big\vert_{(\xi_0, \upsilon_0)} (\upsilon - \upsilon_0)^2
,\end{align}
the wave-vector mismatch can be simplified.
Using the group refractive index $n_{g,\mu}(\omega) = \frac{\dd \omega n_{\mu}(\omega)}{\dd \omega}$ and the dispersion parameter $D_{\mu}(\omega) = \frac{\dd^2 \omega n_{\mu}(\omega)}{\dd \omega^2}$, we can write the derivations as
\begin{align}
	\frac{\partial \Delta k}{\partial\xi}\Big\vert_{(\xi_0, \upsilon_0)} &= \frac{1}{c}\left[2n_{g,z}(2\xi_0) - n_{g,s}(\xi_0 + \upsilon_0) - n_{g,s}(\xi_0 - \upsilon_0)\right] \\
	\frac{\partial \Delta k}{\partial\upsilon}\Big\vert_{(\xi_0, \upsilon_0)} &= \frac{1}{c}\left[- n_{g,s}(\xi_0 + \upsilon_0) - n_{g,s}(\xi_0 - \upsilon_0)\right]\\
	\frac{\partial^2 \Delta k}{\partial\upsilon\partial\xi}\Big\vert_{(\xi_0, \upsilon_0)} &= \frac{1}{c}\left[- D_{s}(\xi_0 + \upsilon_0) - D_{s}(\xi_0 - \upsilon_0)\right]\\
	\frac{\partial^2 \Delta k}{\partial\xi^2}\Big\vert_{(\xi_0, \upsilon_0)} &= \frac{1}{c}\left[4D_{z}(2\xi_0) - D_{s}(\xi_0 + \upsilon_0) - D_{s}(\xi_0 - \upsilon_0)\right]\\
	\frac{\partial^2 \Delta k}{\partial\upsilon^2}\Big\vert_{(\xi_0, \upsilon_0)} &= -\frac{1}{c}\left[D_{s}(\xi_0 + \upsilon_0) + D_{s}(\xi_0 - \upsilon_0)\right]
.\end{align}
For the choice $\xi_0 = \frac{1}{2}\bar{\omega}_{\text{p}}$ and $\upsilon_0 = 0$, where the maximum of the probes contribution is expected, and using $r_{\pm} = \Delta k(\frac{1}{2}\bar{\omega}_{\text{p}}, 0) \pm \frac{2\pi}{\Lambda}$, $s = \frac{1}{c}[n_{g,z}(\bar{\omega}_{\text{p}}) - n_{g,s}(\bar{\omega}_{\text{p}} / 2)]$, $t = \frac{1}{4c}[2D_{z}(\bar{\omega}_{\text{p}}) - D_{s}(\bar{\omega}_{\text{p}} / 2)]$, $u = \frac{1}{4c}D_{s}(\bar{\omega}_{\text{p}} / 2)$ we find from the condition of no wave vector mismatch,
\begin{align}
	0 &= \Delta k(\xi, \upsilon) \pm \frac{2\pi}{\Lambda} \\
	&= r + s(2\xi - \bar{\omega}_{\text{p}}) + t(2\xi - \bar{\omega}_{\text{p}})^2 - 4u\upsilon^2
,\end{align}
with the solution
\begin{equation}\label{eq:qpm_solution}
	2\upsilon = \pm \frac{1}{\sqrt{u}}\left[r + s(2\xi - \bar{\omega}_{\text{p}}) + t(2\xi - \bar{\omega}_{\text{p}})^2\right]^{\frac{1}{2}}
.\end{equation}

\section{Time-local sampling of single (temporal) mode squeezed light}\label{a:single_mode}

\begin{figure}
\centering
	\includegraphics[width=\textwidth]{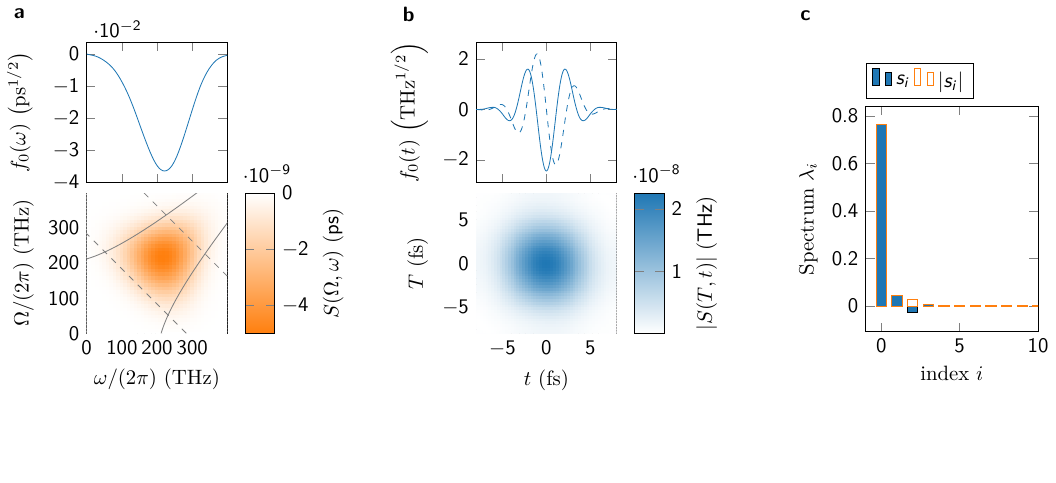}
	\caption{\textbf{a}, The kernel of the dispersion engineered single mode squeezing in the frequency domain, $S(\Omega, \omega)$, and time domain, $S(T, t)$, as well as the first principle mode in the frequency and time domain (real part as solid line and imaginary as dashed).
    The profile of the crystal is assumed to be exponential, $\widehat{f}_{\lambda}(k) = \exp(-| L k / 2|)$, with length $L = \SI{14}{\micro\meter}$, possibly achievable in waveguides.
    The parameter of the pump are assumed to be $k_{GX} = 8.2$ and $\sigma_{GX} / (2\pi) = \SI{140}{\tera\hertz}$.
    \textbf{b}, eigenvalues of the matrix $M_{\text{NL}}$ for the dispersion engineered single mode squeezing.
    The first eigenvalue contributes significantly more to  the nonlinear interaction than the others, justifying the assumption of single mode squeezing.
    }\label{fig:smsv}
\end{figure}

There is a special case which requires only time local measurement, i.e., measurements with $\Delta t_a = \Delta t_b$.
However, we require single mode (broadband) squeezing, $\sigma_x \neq \sigma_p$,
\begin{equation}\label{eq:single_mode_squeezed_state}
	\cov_{\text{smsv}} = (\sigma_x - \frac{1}{2}) \vec{\tilde{\zeta}}_x \vec{\tilde{\zeta}}_x^\T + (\sigma_p - \frac{1}{2}) \vec{\tilde{\zeta}}_p \vec{\tilde{\zeta}}_p^\T + \frac{1}{2} \one
.\end{equation}
For a single temporal mode nonlinear interaction, the function of no wave-vector mismatch, equation~\eqref{eq:qpm_solution}, should be orthogonal to the function of the probe $f_{\alpha}(\omega + \Omega) = f_{\alpha}(2\xi)$, i.e., along the line $\omega = \Omega$.
This can be achieved by choosing $n_z(\bar{\omega}_{\text{p}}) = n_s(\bar{\omega}_{\text{p}} / 2)$.
This choice results in a approximate wave-vector mismatch of $\Delta k(\Omega,\omega) = -\frac{1}{4c}D(\bar{\omega}_{\text{p}} / 2)(\omega - \Omega)^2$.
If the longitudinal profile of the crystal is smooth, e.g., 
\begin{equation}
	\widehat{\lambda}(\Delta k) \approx \frac{\lambda}{\sqrt{2\pi}} \exp[-\frac{L}{c}\abs{D(\bar{\omega}_{\text{p}} / 2)}(\omega - \Omega)^2]
,\end{equation}
which may be achievable in waveguides, and we choose the spectrum of the pump to be Gaussian,
\begin{equation}
	f_{\alpha, \text{smsv}}(\omega) = (\sqrt{2\pi}\sigma)^{-1/2}\exp[-\frac{(\omega - \bar{\omega}_{\text{p}})^2}{(2 \sigma)^2}]
,\end{equation}
we obtain (broadband) single mode squeezing, if $L = \frac{c}{\sigma}\abs{D(\bar{\omega}_{\text{p}} / 2)}^{-1}$.
The squeezing kernel $S(\Omega, \omega)$ of the single mode squeezing interaction, as well as the first principle mode are shown in Fig.~\ref{fig:smsv} \textbf{a}.
The distribution of the singular values of the covariance matrix corresponding to the single (broadband) mode squeezed vacuum in Fig.~\ref{fig:smsv} \textbf{b} displays the dominant contribution of the first principle mode to the squeezed vacuum, confirming the single (broadband) mode nature of the nonlinear interaction.

The correlation function, measured with the setup proposed in the main text, of the single mode squeezed state takes the form
\begin{equation}\label{eq:single_mode_time}
	g(\Delta t, \Delta t, \varphi, \varphi) = (\sigma_x - \frac{1}{2}) (\vec{\zeta}_{\text{LO}}^\T(\Delta t, \varphi) \vec{\tilde{\zeta}}_x)^2 + (\sigma_p - \frac{1}{2}) (\vec{\zeta}_{\text{LO}}^\T(\Delta t, \varphi) \vec{\tilde{\zeta}}_p)^2 + \frac{1}{2} \norm{\vec{\zeta}_{\text{LO}}(\Delta t, \varphi)}^2
.\end{equation}
Using equation~\eqref{eq:single_mode_time}, we can reconstruct the single mode squeezed state using the following algorithm.
\begin{enumerate}
	\item Using $\vec{\zeta}_{\text{LO}}^\T(0, \pi / 2) \vec{\tilde{\zeta}}_x = 0$ and $\vec{\zeta}_{\text{LO}}^\T(0, 0) \vec{\tilde{\zeta}}_p = 0$, as well as assuming $\norm{\vec{\zeta}_{\text{LO}}(\Delta t, \varphi)} = 1$, we can calculate the squeezing ratio
	\begin{equation}\label{eq:squeezing_ratio}
		r = \frac{g(0, 0, 0, 0) - 1 / 2}{g(0, 0, \pi / 2, \pi / 2) - 1 / 2} = \frac{\sigma_x - 1 / 2}{\sigma_p - 1 / 2}
	,\end{equation}
	independent of the overlap between the local oscillator and the state.

	\item Now we can calculate the scaled modulus of the overlap of the local oscillator with one of the broadband quadratures
	\begin{equation}
		\sqrt{[g(\Delta t, \Delta t, 0, 0) - \frac{1}{2}] - r [g(\Delta t, \Delta t, \pi / 2, \pi / 2) - \frac{1}{2}]} = \pm \kappa \vec{\zeta}_{\text{LO}}^\T(\Delta t, 0) \vec{\zeta}_p
	,\end{equation}
	with $\kappa = \sqrt{[1 - (\frac{\sigma_x - 1 / 2}{\sigma_p - 1 / 2})^2] (\sigma_p - \frac{1}{2})}$.
    The sign has to be chosen accordingly.
    
	\item Making measurements for different $\Delta t_i$ and collecting them in a matrix $X_{\text{LO}} = (\vec{\zeta}_{\text{LO}}^\T(\Delta t_1, 0), \vec{\zeta}_{\text{LO}}^\T(\Delta t_2, 0), \ldots)^\T$ and $\vec{\sigma} = (g(\Delta t_1, \Delta t_1, 0, 0) - 1 / 2, g(\Delta t_2, \Delta t_2, 0, 0) - 1 / 2, \ldots)^\T$, we can obtain a scaled projection, $P = X_{\text{LO}}^+ X_{\text{LO}}$, of the broadband quadrature
	\begin{equation}\label{eq:quadrature_projection}
		\kappa P \vec{\zeta}_p = \PI{X_{\text{LO}}} \vec{\sigma}
	.\end{equation}
	
	\item If $X_{\text{LO}}$ is of full rank, we can calculate $\vec{\zeta}_p$ by normalizing $\PI{X_{\text{LO}}} \vec{\sigma}$.
    
	\item Now we can calculate $\sigma_x$, $\sigma_p$ and in a similar manner to $\vec{\tilde{\zeta}}_x$.
\end{enumerate}

\begin{figure}
	\centering
	\includegraphics[width=\textwidth]{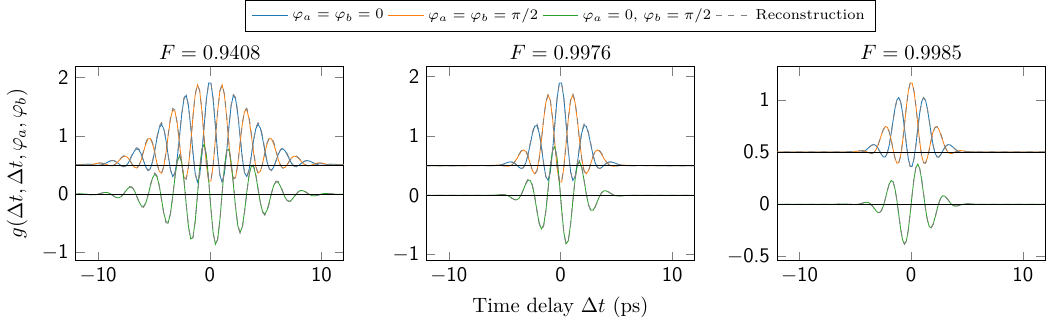}
	\caption{Time local sampled variances $g(\Delta t, \Delta t, 0,  0)$, $g(\Delta t, \Delta t, \pi / 2,  \pi / 2)$ and covariance $g(\Delta t, \Delta t, 0,  \pi / 2)$ for a single mode squeezed state shown in Fig.~\ref{fig:smsv}.
    The central frequency of the local oscillator ist fixed at $\bar{\omega}_{\text{LO}}/(2\pi) = \SI{230}{\tera\hertz}$ and the bandwidth is $\Delta \omega_{\text{LO}}/(2\pi) = \SI{50}{\tera\hertz}$ in \textbf{a}, $\Delta \omega_{\text{LO}}/(2\pi) = \SI{200}{\tera\hertz}$ in \textbf{b}, and $\Delta \omega_{\text{LO}}/(2\pi) = \SI{400}{\tera\hertz}$ in \textbf{c}.
    The dashed line corresponds to the signal, which would be obtained from sampling the quantum state reconstructed with the algorithm described in Section~\ref{a:single_mode}.}\label{fig:smsv_signal}
\end{figure}

Fig.~\ref{fig:smsv_signal} shows an example of time local sampling of the single mode squeezed states as well as the signal obtained from the reconstructed state using the above algorithm, showing good agreement of the two signals.

Squeezing is necessary since otherwise (i.e., $\sigma_x = \sigma_p = \sigma$) we get a signal
\begin{equation}\label{eq:time_local_therm}
	g(\Delta t, \Delta t, \varphi, \varphi) = (\sigma - \frac{1}{2}) [(\vec{\zeta}_{\text{LO}}^\T(\Delta t, \varphi) \vec{\tilde{\zeta}}_x)^2 + (\vec{\zeta}_{\text{LO}}^\T(\Delta t, \varphi) \vec{\tilde{\zeta}}_p)^2] + \frac{1}{2} \norm{\vec{\zeta}_{\text{LO}}(\Delta t, \varphi)}^2
,\end{equation}
which only contains information about the envelope, corresponding to a thermal state as discussed in the following section.

\section{Correlations in thermal states}
\begin{figure}
	\centering
	\includegraphics[width=\textwidth]{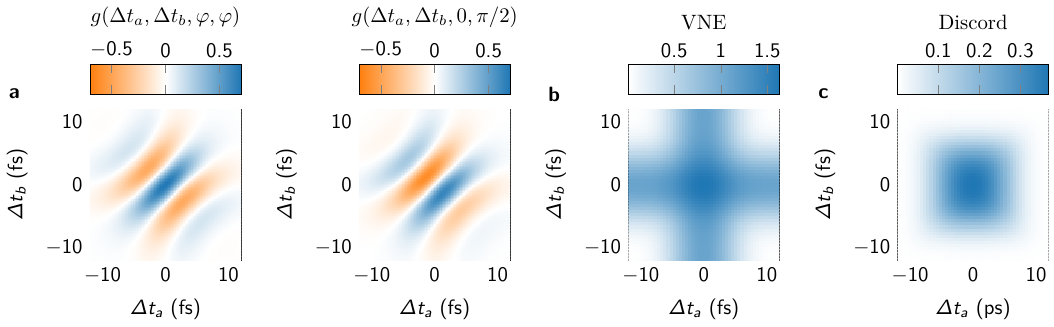}
	\caption{The time domain correlated signal $g(\Delta t_a, \Delta t_b, \varphi_a, \varphi_b)$ for a $\hat{x}\hat{x}$-, $\hat{p}\hat{p}$- and $\hat{x}\hat{p}$-quadrature measurement as a function of the two time delays $\Delta t_a$ and $\Delta t_b$ of a thermal state with temperature $T =\SI{1000}{\kelvin}$ and the local oscillator is centered at $\bar{\omega}_{\text{LO}} = \SI{230}{\tera\hertz}$ and $\Delta \omega_{\text{LO}} = \SI{50}{\tera\hertz}$.}\label{fig:corr_therm}
\end{figure}

Let us assume the pulsed quantum state is thermal radiation with a temperature $T$ and average photon number distribution
\begin{equation}
	n_{\text{th}}(\omega) = \left(\exp[\frac{\hbar\omega}{k_{B}T}] - 1\right)^{-1}
.\end{equation}
Expressing the average photon number distribution $n_{\text{th}}(\omega) = n_{\text{th}} g_0(\omega)$ using the average photon number $n_{\text{th}}$ and a normalized function $g_1(\omega)$, we can complete $g_1(\omega)$ to a mode basis $\{g_0, g_1, \ldots\}$ and define a corresponding phase-space basis $\{\vec{\tilde{\zeta}}_{x,0}, \vec{\tilde{\zeta}}_{x,1}, \ldots, \vec{\tilde{\zeta}}_{p,0}, \vec{\tilde{\zeta}}_{p,1}, \ldots\}$. The thermal state expressed in the discrete basis is thus, 
\begin{equation}\label{eq:thermal_state}
	\cov_{\text{th}} = n_{\text{th}} (\vec{\tilde{\zeta}}_{x,0} \vec{\tilde{\zeta}}_{x,0}^\T + \vec{\tilde{\zeta}}_{p,0} \vec{\tilde{\zeta}}_{p,0}^\T) + \frac{1}{2} \one
.\end{equation}
We can change to the calculation basis using $g_0(\omega) = \sum_{i=0}^{\infty} f_i(\omega) \braket{f_i}{g_0}$.
The correlated signal
\begin{equation}\label{eq:corr_therm}
	g(\Delta t_a, \Delta t_b, \varphi_a, \varphi_b) = \vec{\zeta}_{\text{LO}}^\T(\Delta t_a, 0, \varphi_a) \cov_{\text{th}} \vec{\zeta}_{\text{LO}}(\Delta t_b, 0, \varphi_b)
,\end{equation}
is shown in Fig.~\ref{fig:corr_therm} for a thermal state with temperature $T = \SI{1000}{\kelvin}$.
Along the diagonal $\Delta t_a = \Delta t_b$ of the time local signal, no oscillation and thus no phase information about the state is visible.
Only along the time non-local axis $\Delta t_a = - \Delta t_b$, the oscillations become apparent.
Time local measurements are thus not sufficient to fully reconstruct thermal state of light.
Yet, since the thermal states are Gaussian, we can use correlation measurements and the algorithm presented in Sec.~II, to fully reconstruct the thermal states.

\section{Non-Gaussian states}

Going beyond Gaussian states requires the consideration of higher order moments.
One possibility to capture the complete statistics is by means of the characteristic function $\chi$, defined using the displacement operator $\hat{D}(\tilde{x},\tilde{p}) = \exp[\iu(\tilde{x}\hat{p} - \tilde{p}\hat{x})]$.
The characteristic function of the quantum state at the input of the measurement setup,
\begin{equation}
	\chi_{\hat{\rho}}(\tilde{x}_a,\tilde{p}_a,\tilde{x}_b,\tilde{p}_b) = \tr[\hat{D}(\tilde{x}_a,\tilde{p}_a)\hat{D}(\tilde{x}_b,\tilde{p}_b) \hat{\rho} \otimes \ket{0}\bra{0}]
,\end{equation}
is calculated from the quantum state composed of the states at the two beam splitter inputs, i.e., the input quantum state $\hat{\rho}$ and the vacuum $\ket{0}\bra{0}$ at the empty port.
The characteristic function of the measurement operator $\ket{x_a}\bra{x_a}\otimes\ket{p_b}\bra{p_b}$, corresponding to a $\hat{x}$-quadrature measurement at detection arm $a$ and a $\hat{p}$-quadrature measurement at the detection arm $b$, is
\begin{equation}
	\chi_{x_a, p_b}(\tilde{x}_a,\tilde{p}_a,\tilde{x}_b,\tilde{p}_b) = \tr[\hat{D}(\tilde{x}_a,\tilde{p}_a)\hat{D}(\tilde{x}_b,\tilde{p}_b)\ket{x_a}\bra{x_a}\otimes\ket{p_b}\bra{p_b}]
.\end{equation}
We use $x_a = x_a(\Delta t_a, \varphi_a=0)$ and $p_b = p_b(\Delta t_b, \varphi_a=\pi / 2)$ as an abbreviation.
The probability to obtain the result $(x_a, p_b)$ from the simultaneous $\hat{x}$- and $\hat{p}$-quadrature measurement, can thus be expressed using the characteristic functions, 
\begin{equation}
	p[x_a, p_b] = \tr(\hat{\rho}\ket{x_a}\bra{x_a}\otimes\ket{p_b}\bra{p_b}) = \frac{1}{2\pi}\iiiint \chi_{\hat{\rho}}(\tilde{x}_a,\tilde{p}_a,\tilde{x}_b,\tilde{p}_b) \chi_{x_a, p_b}(\tilde{x}_a,\tilde{p}_a,\tilde{x}_b,\tilde{p}_b) \dd\tilde{x}_a\dd\tilde{p}_a\dd\tilde{x}_b\dd\tilde{p}_b
.\end{equation}
Exploiting
\begin{equation}
	\chi_{x_a, x_b}(\tilde{x}_a,\tilde{p}_a,\tilde{x}_b,\tilde{p}_b) = \begin{cases}
		\delta(\tilde{x}_a)\delta(\tilde{x}_b)\exp[-\iu(\tilde{p}_a x_a + \tilde{p}_b x_b)], & \varphi_a = 0, \varphi_b = 0 \\
		\delta(\tilde{x}_a)\delta(\tilde{p}_b)\exp[-\iu(\tilde{p}_a x_a - \tilde{x}_b p_b)], & \varphi_a = 0, \varphi_b = \pi / 2 \\
		\delta(\tilde{p}_a)\delta(\tilde{x}_b)\exp[\iu(\tilde{x}_a p_a - \tilde{p}_b x_b)], & \varphi_a = \pi / 2, \varphi_b = 0 \\
		\delta(\tilde{p}_a)\delta(\tilde{p}_b)\exp[\iu(\tilde{p}_a x_a + \tilde{p}_b x_b)], & \varphi_a = \pi / 2, \varphi_b = \pi / 2
	\end{cases}
,\end{equation}
for the case of a $\hat{x}$-quadrature measurement in detection arm $a$ and a $\hat{p}$-quadrature measurement in detection arm $b$, i.e. $\varphi_a = 0$ and $\varphi_b = \pi / 2$, we can express the probability distribution,
\begin{equation}
	p[x_a, p_b] = \frac{1}{2\pi}\iint \exp[-\iu(\tilde{p}_a x_a - \tilde{x}_b p_b)] \chi_{\hat{\rho}}(0,\tilde{p}_a,\tilde{x}_b,0) \dd\tilde{p}_a\dd\tilde{x}_b
,\end{equation}
as the Fourier transform of the characteristic function of the quantum state. This is possible since the measurement operators at the two output ports of the beam splitter commute.
Thus, the characteristic function corresponds to an actual probability distribution.
Moving to the discretised basis again using [we use $\vec{\zeta}_{\text{LO}}(\Delta t_\arm, \alpha, \varphi_\arm) = \vec{\zeta}_{\text{LO}}(\Delta t_\arm, \varphi_\arm)$ as shorthand]
\begin{align}
	\vec{\zeta}(\tilde{x}_a, \tilde{p}_a, \tilde{x}_b, \tilde{p}_b) &= \tilde{x}_a \vec{\zeta}_{\text{LO}}(\Delta t_a, 0) + \tilde{p}_a \vec{\zeta}_{\text{LO}}(\Delta t_a, \pi / 2) \\ 
	&+ \tilde{x}_b \vec{\zeta}_{\text{LO}}(\Delta t_b, 0) + \tilde{p}_b \vec{\zeta}_{\text{LO}}(\Delta t_b, \pi / 2) \\
	\vec{\zeta}_{\text{vac}}(\tilde{x}_a, \tilde{p}_a, \tilde{x}_b, \tilde{p}_b) &= \tilde{x}_a \vec{\zeta}_{\text{LO}}(\Delta t_a, 0) - \tilde{p}_a \vec{\zeta}_{\text{LO}}(\Delta t_a, -\pi / 2) \\ 
	&+ \tilde{x}_b \vec{\zeta}_{\text{LO}}(\Delta t_b, \pi) - \tilde{p}_b \vec{\zeta}_{\text{LO}}(\Delta t_b, \pi / 2)
,\end{align}
and
\begin{equation}
	\hat{D}(\vec{\zeta}) = \exp[\iu \vec{\zeta}^\T \Omega \hat{\vec{\zeta}}]
,\end{equation}
we can express the characteristic function of the quantum state as 
\begin{equation}
	\chi_{\hat{\rho}}(\tilde{x}_a,\tilde{p}_a,\tilde{x}_b,\tilde{p}_b) = \tr{\hat{D}[\vec{\zeta}(\tilde{x}_a, \tilde{p}_a, \tilde{x}_b, \tilde{p}_b)]\hat{\rho}} \bra{0}\hat{D}[\vec{\zeta}_{\text{vac}}(\tilde{x}_a, \tilde{p}_a, \tilde{x}_b, \tilde{p}_b)]\ket{0}
.\end{equation}
Inserting the characteristic function of the vacuum,
\begin{equation}
	\bra{0}\hat{D}[\vec{\zeta}_{\text{vac}}(\tilde{x}_a, \tilde{p}_a, \tilde{x}_b, \tilde{p}_b)]\ket{0} = \exp[- \frac{1}{4}\norm{\vec{\zeta}_{\text{vac}}(\tilde{x}_a, \tilde{p}_a, \tilde{x}_b, \tilde{p}_b)}^2]
,\end{equation}
and rewriting the multimode characteristic function as the symplectic Fourier transform of the multimode Wignerfunction,
\begin{equation}
	\chi(\vec{\tilde{\zeta}}) = \int\ldots\int \e^{-\iu \vec{\tilde{\zeta}}^\T \Omega \vec{\zeta}} W(\vec{\zeta}) \dd^{2 i_{\text{max}}} \zeta
,\end{equation}
we can motivate the definition of the integration kernel
\begin{equation}\label{eq:kernel_definition}
	K(x_a, p_b | \vec{\zeta}) = \frac{1}{2\pi} \iint \exp{-\iu(\tilde{p}_a x_a - \tilde{x}_b p_b) - \iu [\vec{\zeta}(0, \tilde{p}_a, \tilde{x}_b, 0)]^\T \Omega \vec{\zeta} - \frac{1}{4}\norm{\vec{\zeta}_{\text{vac}}(0, \tilde{p}_a, \tilde{x}_b, 0)}^2} \dd\tilde{p}_a\dd\tilde{x}_b
.\end{equation}
The above definition allows us to write the joint probability distribution,
\begin{equation}\label{eq:joint_prob_dist}
	p(x_a, p_b) = \int \ldots \int K(x_a, p_b | \vec{\zeta}) W(\vec{\zeta}) \dd^{2 i_{\text{max}}} \zeta
,\end{equation}
in a compact form. Furthermore, we can restrict to the special case,
\begin{align}
	\vec{\zeta}(0, \tilde{p}_a, \tilde{x}_b, 0) &= \tilde{p}_a \vec{\zeta}_{\text{LO}}(\Delta t_a, \pi / 2) + \tilde{x}_b \vec{\zeta}_{\text{LO}}(\Delta t_b, 0) \\
	\vec{\zeta}_{\text{vac}}(0, \tilde{p}_a, \tilde{x}_b, 0) &= - \tilde{p}_a \vec{\zeta}_{\text{LO}}(\Delta t_a, -\pi / 2) + \tilde{x}_b \vec{\zeta}_{\text{LO}}(\Delta t_b, \pi)
.\end{align}
The integral in the definition of the Kernel, equation~\eqref{eq:kernel_definition} can be solved for.
To allow for a compact representation of the result we introduce the LO-matrix
\begin{equation}
	P_{\text{LO}} = \vec{\zeta}_{\text{LO}}(\Delta t_a, \pi / 2)[\vec{\zeta}_{\text{LO}}(\Delta t_a, \pi / 2)]^\T + \vec{\zeta}_{\text{LO}}(\Delta t_b, 0)[\vec{\zeta}_{\text{LO}}(\Delta t_b, 0)]^\T
,\end{equation}
which is proportional to a projector for $\Delta t_a = \Delta t_b$, and detectable phase space plane spanned by
\begin{align}
	\vec{\zeta}_{\text{d}}(x_a, p_b) = 4\frac{P_{\text{LO}} [x_a\vec{\zeta}_{\text{LO}}(\Delta t_a, \pi / 2) - p_b \vec{\zeta}_{\text{LO}}(\Delta t_b, 0)]}{\norm{\vec{\zeta}_{\text{LO}}(\Delta t_a, \pi / 2) - \vec{\zeta}_{\text{LO}}(\Delta t_b, 0)}^2}
,\end{align}
as well as the covariance matrix
\begin{align}
	\cov_d^{-1} &= 4\frac{\Omega^\T P_{\text{LO}}^2 \Omega}{\norm{\vec{\zeta}_{\text{LO}}(\Delta t_a, \pi / 2) - \vec{\zeta}_{\text{LO}}(\Delta t_b, 0)}^2}
.\end{align}
With these definitions, the kernel takes the form
\begin{align}\label{eq:kernel_si}
	K(x_a, p_b | \vec{\zeta}) &= \frac{2\sqrt{2}\exp[-2\frac{\norm{x_a\vec{\zeta}_{\text{LO}}(\Delta t_a, \pi / 2) - p_b \vec{\zeta}_{\text{LO}}(\Delta t_b, 0)}^2}{\norm{\vec{\zeta}_{\text{LO}}(\Delta t_a, \pi / 2) - \vec{\zeta}_{\text{LO}}(\Delta t_b, 0)}^2}]}{\norm{\vec{\zeta}_{\text{LO}}(\Delta t_a, \pi / 2) - \vec{\zeta}_{\text{LO}}(\Delta t_b, 0)}}  \exp[\vec{\zeta}_{\text{d}}^\T(x_a, p_b) \vec{\zeta} - \frac{1}{2}\vec{\zeta}^\T \cov_d^{-1} \vec{\zeta}]
.\end{align}
Further calculation require additional assumptions about the quantum state.
For now we will focus on the special case of a single mode Fock state with multimode Wignerfunction
\begin{equation}
	W_{\text{smfs}}(\vec{\zeta}) = (-1)^n L_n\left(2\norm{\vec{\zeta}_{\text{ph}}}^2\right) W_{\text{vac}}(\vec{\zeta})
,\end{equation}
with $\vec{\zeta}_{\text{ph}}$ in the two dimensional phase space plain of the mode corresponding to the Fock state and $W_{\text{vac}}(\vec{\zeta})$ the Gaussian Wignerfunction of the vacuum with covariance matrix $\cov_{\text{vac}} = \frac{1}{2}\one$.
Inserting equation~\eqref{eq:kernel} for the kernel and the multimode Wignerfunction of the Fock state into the expression of the joint probability distribution in equation~\eqref{eq:joint_prob_dist}, we obtain
\begin{align}
	p(x_a, p_b) &= \frac{2\sqrt{2}\exp[-2\frac{\norm{x_a\vec{\zeta}_{\text{LO}}(\Delta t_a, \pi / 2) - p_b \vec{\zeta}_{\text{LO}}(\Delta t_b, 0)}^2}{\norm{\vec{\zeta}_{\text{LO}}(\Delta t_a, \pi / 2) - \vec{\zeta}_{\text{LO}}(\Delta t_b, 0)}^2}]}{\norm{\vec{\zeta}_{\text{LO}}(\Delta t_a, \pi / 2) - \vec{\zeta}_{\text{LO}}(\Delta t_b, 0)}} \nonumber \\
	&\times \frac{(-1)^n}{\pi} \int L_n\left(2\norm{\vec{\zeta}_{\text{ph}}}^2\right) \int \ldots \int \exp[\vec{\zeta}_{\text{d}}^\T(x_a, p_b) \vec{\zeta} - \frac{1}{2}\vec{\zeta}^\T (\cov_{\text{vac}}^{-1} + \cov_d^{-1}) \vec{\zeta}] \dd^{2 i_{\text{max}}} \zeta \label{eq:prob_dist_1}
.\end{align}
Using the projector $P_{\text{ph}}$, projecting into the phase space plane of the Fock state, and $P_{\text{r}} = \one - P_{\text{ph}}$, the covariance matrix can be seperated into blocks. We define the Schur complement,
\begin{equation}
	\cov_{\text{schur}}^{-1} = P_{\text{ph}}(\cov_{\text{vac}}^{-1} + \cov_d^{-1}) P_{\text{ph}} - P_{\text{ph}} \cov_d^{-1} P_{\text{r}} [P_{\text{r}}(\cov_{\text{vac}}^{-1} + \cov_d^{-1})P_{\text{r}}]^{-1} P_{\text{r}} \cov_d^{-1} P_{\text{ph}}
,\end{equation}
and the detected quadrature projected into the Fock states mode,
\begin{equation}
	\vec{\zeta}_{\text{d,ph}}(x_a, p_b) = P_{\text{ph}}[\one - \cov_d^{-1} P_{\text{r}} \{ P_{\text{r}} (\cov_{\text{vac}}^{-1} + \cov_d^{-1}) P_{\text{r}} \}^{-1} P_{\text{r}}] \vec{\zeta}_d(x_a, p_b)
.\end{equation}
The Gaussian part of the multidimensional integral in equation~\eqref{eq:prob_dist_1} can be solved. Together with the definition
\begin{align}
	&N(x_a, p_a) = \frac{2\sqrt{2}\exp[-2\frac{\norm{x_a\vec{\zeta}_{\text{LO}}(\Delta t_a, \pi / 2) - p_b \vec{\zeta}_{\text{LO}}(\Delta t_b, 0)}^2}{\norm{\vec{\zeta}_{\text{LO}}(\Delta t_a, \pi / 2) - \vec{\zeta}_{\text{LO}}(\Delta t_b, 0)}^2}]}{\norm{\vec{\zeta}_{\text{LO}}(\Delta t_a, \pi / 2) - \vec{\zeta}_{\text{LO}}(\Delta t_b, 0)}} \nonumber \\
	&\times \sqrt{\frac{(2\pi)^{2(i_{\text{max}}-1)}}{\det(P_{\text{r}}[\cov_{\text{vac}}^{-1} + \cov_d^{-1}]P_{\text{r}})}} \exp[\frac{1}{2} \vec{\zeta}_{\text{d}}^\T(x_a, p_b) \Omega P_{\text{r}} \{P_{\text{r}}(\cov_{\text{vac}}^{-1} + \cov_{\text{d}}^{-1}) P_{\text{r}}\}^{-1} P_{\text{r}} \Omega^\T \vec{\zeta}_{\text{d}}(x_a, p_b)]
,\end{align}
the joint probability distribution can be expressed as
\begin{equation}\label{eq:prob_dist_2}
	p(x_a, p_b) =  N(x_a, p_b) \frac{(-1)^n}{\pi} \int L_n\left(2\norm{\vec{\zeta}_{\text{ph}}}^2\right) \exp[ - \frac{1}{2}\vec{\zeta}_{\text{ph}}^\T \cov_{\text{schur}}^{-1} \vec{\zeta}_{\text{ph}} + \vec{\zeta}_{\text{d,ph}}^\T(x_a, p_b) \vec{\zeta}_{\text{ph}}] \dd^{2} \zeta_{\text{ph}}
.\end{equation}
The last two integrals can be solved using the singular value decomposition $\cov_{\text{schur}}^{-1} = \sigma_x \vec{e}_x\vec{e}_x^\T + \sigma_p \vec{e}_p\vec{e}_p^\T$ or for the inverse, $\cov_{\text{schur}} = \frac{1}{\sigma_x} \vec{e}_x\vec{e}_x^\T + \frac{1}{\sigma_p} \vec{e}_p\vec{e}_p^\T$.
The integration is taken along the two axis defined by the vectors $\vec{e}_x$ and $\vec{e}_p$, i.e., $\vec{\zeta}_{\text{ph}} = x_{\text{ph}} \vec{e}_x + p_{\text{ph}} \vec{e}_p$.
With the result of the integral,
\begin{align}
	&\int L_n\left(2\norm{\vec{\zeta}_{\text{ph}}}^2\right) \exp[ - \frac{1}{2}\vec{\zeta}_{\text{ph}}^\T \cov_{\text{schur}}^{-1} \vec{\zeta}_{\text{ph}} + \vec{\zeta}_{\text{d,ph}}^\T(x_a, p_b) \vec{\zeta}_{\text{ph}}] \dd^{2} \zeta_{\text{ph}} \nonumber \\
	&= \sqrt{\frac{2\pi}{\sigma_x}}\sqrt{\frac{2\pi}{\sigma_p}} \e^{\frac{[\vec{\zeta}_{\text{d,ph}}^\T(x_a, p_b) \vec{e}_x]^2}{2\sigma_x}} \e^{\frac{[\vec{\zeta}_{\text{d,ph}}^\T(x_a, p_b) \vec{e}_p]^2}{2\sigma_p}} \nonumber \\
	&\times \sum_{i=0}^n \left(1 - \frac{4}{\sigma_x}\right)^{i} \left(1 - \frac{4}{\sigma_p}\right)^{n-i} L_{i}^{(-\frac{1}{2})}\left[2\frac{\{\vec{\zeta}_{\text{d,ph}}^\T(x_a, p_b) \vec{e}_x\}^2}{\sigma_x^2-4\sigma_x}\right] L_{n-i}^{(-\frac{1}{2})}\left[2\frac{\{\vec{\zeta}_{\text{d,ph}}^\T(x_a, p_b) \vec{e}_p\}^2}{\sigma_p^2-4\sigma_p}\right]
,\end{align}
The joint probability distribution of the simultaneous quadrature measurement of a single mode Fock state is given by
\begin{align}
	p(x_a, p_b) &= 	N(x_a, p_b) \frac{(-1)^n}{\pi} \sqrt{\frac{2\pi}{\sigma_x}}\sqrt{\frac{2\pi}{\sigma_p}} \e^{\frac{[\vec{\zeta}_{\text{d,ph}}^\T(x_a, p_b) \vec{e}_x]^2}{2\sigma_x}} \e^{\frac{[\vec{\zeta}_{\text{d,ph}}^\T(x_a, p_b) \vec{e}_p]^2}{2\sigma_p}} \nonumber \\
	&\times \sum_{i=0}^n \left(1 - \frac{4}{\sigma_x}\right)^{i} \left(1 - \frac{4}{\sigma_p}\right)^{n-i} L_{i}^{(-\frac{1}{2})}\left[2\frac{\{\vec{\zeta}_{\text{d,ph}}^\T(x_a, p_b) \vec{e}_x\}^2}{\sigma_x^2-4\sigma_x}\right] L_{n-i}^{(-\frac{1}{2})}\left[2\frac{\{\vec{\zeta}_{\text{d,ph}}^\T(x_a, p_b) \vec{e}_p\}^2}{\sigma_p^2-4\sigma_p}\right]
.\end{align}
The joint probability distribution allows us to scrutinize two special cases in which the two well known statistics of a Fock state are recovered: The Poissonian distribution of the Husimi function (photodetection) and the quadrature distribution of a Fock state following Hermite-Gauss functions.
\begin{enumerate}
	\item Let us consider the case with $\sigma_x = \sigma_p = \sigma$. This occurs for example for time local detection, i.e., $\Delta t_a = \Delta t_b$. In this case we can define the detection probability $p = \sigma_x / 4$, the binomial distribution $\text{bin}(k;p,n) = \binom{n}{k} (1 - p)^{n-k} p^k$ and the Husimi function of a Fock state,
\begin{equation}
	Q_n(x,p) = \left(\frac{x^2 + p^2}{2}\right)^n \frac{1}{\pi n!}\exp(-\frac{1}{2}\frac{x^2 + p^2}{2})
.\end{equation}
With these definitions the joint probability distribution in the case of equal singular values, $\sigma_x = \sigma_p = \sigma$, can be expressed as a statistical mixture of Fock states:
	\begin{align}
		&p(x_a, p_b) = N(x_a, p_b) \frac{2\pi}{\sigma} \e^{\frac{[\vec{\zeta}_{\text{d,ph}}^\T(x_a, p_b) \vec{e}_x]^2}{2\sigma}} \e^{\frac{[\vec{\zeta}_{\text{d,ph}}^\T(x_a, p_b) \vec{e}_p]^2}{2\sigma}} \e^{\frac{[\vec{\zeta}_{\text{d,ph}}^\T(x_a, p_b) \vec{e}_x]^2}{4\sigma^2}} \e^{\frac{[\vec{\zeta}_{\text{d,ph}}^\T(x_a, p_b) \vec{e}_p]^2}{4\sigma^2}} \frac{1}{p^n} \nonumber \\
		&\times \sum_{i=0}^n \text{bin}(k; p, n) Q_k\left(\frac{\vec{\zeta}_{\text{d,ph}}^\T(x_a, p_b)\vec{e}_x}{\sigma/2}, \frac{\vec{\zeta}_{\text{d,ph}}^\T(x_a, p_b)\vec{e}_p}{\sigma/2}\right) \label{eq:prob_dist_case1}
	.\end{align}
	Thus, the state appears to the detection as a $k$-photon Fock state with probability $\text{bin}(k;p,n)$. For complete mode function overlap between local oscillator and Fock state, i.e. $\sigma \to 4$ or $p \to 1$, we have $\text{bin}(k; p, n) = \delta_{k,n}$ and $p(x_a, p_b) = \tilde{N}(x_a, p_b) Q_n[2\vec{\zeta}_{\text{d,ph}}^\T(x_a, p_b)\vec{e}_x, 2\vec{\zeta}_{\text{d,ph}}^\T(x_a, p_b)\vec{e}_p]$, with $\tilde{N}(x_a, p_b)$ being an appropriate normalization envalope, and the statistics corresponds to a pure $n$-photon Fock state as expected from a single mode optical homodyne detection \cite{Freyberger1993}.
    
    \item The second case occurs for correlations measurements with $\Delta t_a \neq \Delta t_b$ if $\sigma_x \approx 2$. In this case the probability distribution along the phase-space line defined by $\vec{\zeta}_{\text{d,ph}}^\T(x_a, p_b)\vec{e}_p = 0$ can be written as
\begin{equation}\label{eq:prob_dist_case3}
	p(x_a, p_b) =  N(x_a, p_b) \frac{2^{-n}}{\sqrt{2} \pi n!} \int \abs{H_n(x)}^2 \exp[x\vec{\zeta}_{\text{d,ph}}^\T(x_a, p_b)\vec{e}_x - \frac{\sigma_x}{2}x^2] \dd x
.\end{equation}
There are two cases with $\sigma_p \approx 2$.
\begin{enumerate}
    \item The intermediate case, in which $\sigma_x$ becomes large. This occurs, if the time delay on one arm, e.g., $\Delta t_b$, corresponds to a uneven multiple of a quater cycle of the quantum pulse while the other time delay is fixed, e.g., $\Delta t_a = \SI{0}{\femto\second}$. In this case, both arms measure the same quadrature of the Fock state. By defining  $\vec{\tilde{\zeta}}_{d,0}(x_a, p_b) = \lim_{\sigma_x \to \infty} \vec{\zeta}_{\text{d,ph}}(x_a, p_b) / \sigma_x$, the probability distribution of the measurement outcomes,
    \begin{equation}
	\lim_{\sigma_x \to \infty} p(x_a, p_b) =  \tilde{N}(x_a, p_b) \abs{H_n[\vec{\tilde{\zeta}}_{d,0}(x_a, p_b)]}^2
,\end{equation}
 in the limit of infinitely large $\sigma_x$ along the line defined by $\vec{\zeta}_{\text{d,ph}}^\T(x_a, p_b)\vec{e}_p = 0$ follows the quadrature distribution of a $n$-photon Fock state.

\item Let us consider $\Delta t_a = 0$ and $\Delta t_b \gg 0$. This means $\vec{\zeta}_{\text{d}}^\T(x_a, p_b)\vec{e}_p \approx 0$ and $\sigma_p \to 2$. In this case only detection arm $a$ measures the Fock state, while the other detects the vacuum. The measurement probability distribution therefore follows the statistics of a single quadrature measurement admixed with vacuum noise from the second arm resulting in a smoothed out version of the Hermite-Gauss functions given by equation~\eqref{eq:prob_dist_case3}.

\end{enumerate}
\end{enumerate}


\bibliography{correlation_tomography}

\begin{thebibliography}{93}%
\makeatletter
\providecommand \@ifxundefined [1]{%
 \@ifx{#1\undefined}
}%
\providecommand \@ifnum [1]{%
 \ifnum #1\expandafter \@firstoftwo
 \else \expandafter \@secondoftwo
 \fi
}%
\providecommand \@ifx [1]{%
 \ifx #1\expandafter \@firstoftwo
 \else \expandafter \@secondoftwo
 \fi
}%
\providecommand \natexlab [1]{#1}%
\providecommand \enquote  [1]{``#1''}%
\providecommand \bibnamefont  [1]{#1}%
\providecommand \bibfnamefont [1]{#1}%
\providecommand \citenamefont [1]{#1}%
\providecommand \href@noop [0]{\@secondoftwo}%
\providecommand \href [0]{\begingroup \@sanitize@url \@href}%
\providecommand \@href[1]{\@@startlink{#1}\@@href}%
\providecommand \@@href[1]{\endgroup#1\@@endlink}%
\providecommand \@sanitize@url [0]{\catcode `\\12\catcode `\$12\catcode
  `\&12\catcode `\#12\catcode `\^12\catcode `\_12\catcode `\%12\relax}%
\providecommand \@@startlink[1]{}%
\providecommand \@@endlink[0]{}%
\providecommand \url  [0]{\begingroup\@sanitize@url \@url }%
\providecommand \@url [1]{\endgroup\@href {#1}{\urlprefix }}%
\providecommand \urlprefix  [0]{URL }%
\providecommand \Eprint [0]{\href }%
\providecommand \doibase [0]{https://doi.org/}%
\providecommand \selectlanguage [0]{\@gobble}%
\providecommand \bibinfo  [0]{\@secondoftwo}%
\providecommand \bibfield  [0]{\@secondoftwo}%
\providecommand \translation [1]{[#1]}%
\providecommand \BibitemOpen [0]{}%
\providecommand \bibitemStop [0]{}%
\providecommand \bibitemNoStop [0]{.\EOS\space}%
\providecommand \EOS [0]{\spacefactor3000\relax}%
\providecommand \BibitemShut  [1]{\csname bibitem#1\endcsname}%
\let\auto@bib@innerbib\@empty
\bibitem [{\citenamefont {Gulla}\ \emph {et~al.}(2021)\citenamefont {Gulla},
  \citenamefont {Ryen},\ and\ \citenamefont {Skaar}}]{Gulla2021}%
  \BibitemOpen
  \bibfield  {author} {\bibinfo {author} {\bibfnamefont {J.}~\bibnamefont
  {Gulla}}, \bibinfo {author} {\bibfnamefont {K.}~\bibnamefont {Ryen}},\ and\
  \bibinfo {author} {\bibfnamefont {J.}~\bibnamefont {Skaar}},\ }\href@noop {}
  {\bibinfo {title} {Limits for realizing single photons}} (\bibinfo {year}
  {2021}),\ \Eprint {https://arxiv.org/abs/2109.06472} {arXiv:2109.06472
  [quant-ph]} \BibitemShut {NoStop}%
\bibitem [{\citenamefont {Yanagimoto}\ \emph {et~al.}(2024)\citenamefont
  {Yanagimoto}, \citenamefont {Ng}, \citenamefont {Jankowski}, \citenamefont
  {Nehra}, \citenamefont {McKenna}, \citenamefont {Onodera}, \citenamefont
  {Wright}, \citenamefont {Hamerly}, \citenamefont {Marandi}, \citenamefont
  {Fejer},\ and\ \citenamefont {Mabuchi}}]{Yanagimoto2024}%
  \BibitemOpen
  \bibfield  {author} {\bibinfo {author} {\bibfnamefont {R.}~\bibnamefont
  {Yanagimoto}}, \bibinfo {author} {\bibfnamefont {E.}~\bibnamefont {Ng}},
  \bibinfo {author} {\bibfnamefont {M.}~\bibnamefont {Jankowski}}, \bibinfo
  {author} {\bibfnamefont {R.}~\bibnamefont {Nehra}}, \bibinfo {author}
  {\bibfnamefont {T.~P.}\ \bibnamefont {McKenna}}, \bibinfo {author}
  {\bibfnamefont {T.}~\bibnamefont {Onodera}}, \bibinfo {author} {\bibfnamefont
  {L.~G.}\ \bibnamefont {Wright}}, \bibinfo {author} {\bibfnamefont
  {R.}~\bibnamefont {Hamerly}}, \bibinfo {author} {\bibfnamefont
  {A.}~\bibnamefont {Marandi}}, \bibinfo {author} {\bibfnamefont {M.~M.}\
  \bibnamefont {Fejer}},\ and\ \bibinfo {author} {\bibfnamefont
  {H.}~\bibnamefont {Mabuchi}},\ }\bibfield  {title} {\bibinfo {title}
  {Mesoscopic ultrafast nonlinear optics—the emergence of multimode quantum
  non-gaussian physics},\ }\href {https://doi.org/10.1364/optica.514075}
  {\bibfield  {journal} {\bibinfo  {journal} {Optica}\ }\textbf {\bibinfo
  {volume} {11}},\ \bibinfo {pages} {896} (\bibinfo {year} {2024})}\BibitemShut
  {NoStop}%
\bibitem [{\citenamefont {Weedbrook}\ \emph {et~al.}(2012)\citenamefont
  {Weedbrook}, \citenamefont {Pirandola}, \citenamefont {García-Patrón},
  \citenamefont {Cerf}, \citenamefont {Ralph}, \citenamefont {Shapiro},\ and\
  \citenamefont {Lloyd}}]{Weedbrook2012}%
  \BibitemOpen
  \bibfield  {author} {\bibinfo {author} {\bibfnamefont {C.}~\bibnamefont
  {Weedbrook}}, \bibinfo {author} {\bibfnamefont {S.}~\bibnamefont
  {Pirandola}}, \bibinfo {author} {\bibfnamefont {R.}~\bibnamefont
  {García-Patrón}}, \bibinfo {author} {\bibfnamefont {N.~J.}\ \bibnamefont
  {Cerf}}, \bibinfo {author} {\bibfnamefont {T.~C.}\ \bibnamefont {Ralph}},
  \bibinfo {author} {\bibfnamefont {J.~H.}\ \bibnamefont {Shapiro}},\ and\
  \bibinfo {author} {\bibfnamefont {S.}~\bibnamefont {Lloyd}},\ }\bibfield
  {title} {\bibinfo {title} {Gaussian quantum information},\ }\href
  {https://doi.org/10.1103/revmodphys.84.621} {\bibfield  {journal} {\bibinfo
  {journal} {Reviews of Modern Physics}\ }\textbf {\bibinfo {volume} {84}},\
  \bibinfo {pages} {621} (\bibinfo {year} {2012})}\BibitemShut {NoStop}%
\bibitem [{\citenamefont {Braunstein}\ and\ \citenamefont {van
  Loock}(2005)}]{Braunstein2005}%
  \BibitemOpen
  \bibfield  {author} {\bibinfo {author} {\bibfnamefont {S.~L.}\ \bibnamefont
  {Braunstein}}\ and\ \bibinfo {author} {\bibfnamefont {P.}~\bibnamefont {van
  Loock}},\ }\bibfield  {title} {\bibinfo {title} {Quantum information with
  continuous variables},\ }\href {https://doi.org/10.1103/revmodphys.77.513}
  {\bibfield  {journal} {\bibinfo  {journal} {Rev. Mod. Phys.}\ }\textbf
  {\bibinfo {volume} {77}},\ \bibinfo {pages} {513} (\bibinfo {year}
  {2005})}\BibitemShut {NoStop}%
\bibitem [{\citenamefont {Weedbrook}\ \emph {et~al.}(2004)\citenamefont
  {Weedbrook}, \citenamefont {Lance}, \citenamefont {Bowen}, \citenamefont
  {Symul}, \citenamefont {Ralph},\ and\ \citenamefont {Lam}}]{Weedbrook2004}%
  \BibitemOpen
  \bibfield  {author} {\bibinfo {author} {\bibfnamefont {C.}~\bibnamefont
  {Weedbrook}}, \bibinfo {author} {\bibfnamefont {A.~M.}\ \bibnamefont
  {Lance}}, \bibinfo {author} {\bibfnamefont {W.~P.}\ \bibnamefont {Bowen}},
  \bibinfo {author} {\bibfnamefont {T.}~\bibnamefont {Symul}}, \bibinfo
  {author} {\bibfnamefont {T.~C.}\ \bibnamefont {Ralph}},\ and\ \bibinfo
  {author} {\bibfnamefont {P.~K.}\ \bibnamefont {Lam}},\ }\bibfield  {title}
  {\bibinfo {title} {Quantum cryptography without switching},\ }\href
  {https://link.aps.org/doi/10.1103/PhysRevLett.93.170504} {\bibfield
  {journal} {\bibinfo  {journal} {Phys. Rev. Lett.}\ }\textbf {\bibinfo
  {volume} {93}},\ \bibinfo {pages} {170504} (\bibinfo {year}
  {2004})}\BibitemShut {NoStop}%
\bibitem [{\citenamefont {Lance}\ \emph {et~al.}(2005)\citenamefont {Lance},
  \citenamefont {Symul}, \citenamefont {Sharma}, \citenamefont {Weedbrook},
  \citenamefont {Ralph},\ and\ \citenamefont {Lam}}]{Lance2005}%
  \BibitemOpen
  \bibfield  {author} {\bibinfo {author} {\bibfnamefont {A.~M.}\ \bibnamefont
  {Lance}}, \bibinfo {author} {\bibfnamefont {T.}~\bibnamefont {Symul}},
  \bibinfo {author} {\bibfnamefont {V.}~\bibnamefont {Sharma}}, \bibinfo
  {author} {\bibfnamefont {C.}~\bibnamefont {Weedbrook}}, \bibinfo {author}
  {\bibfnamefont {T.~C.}\ \bibnamefont {Ralph}},\ and\ \bibinfo {author}
  {\bibfnamefont {P.~K.}\ \bibnamefont {Lam}},\ }\bibfield  {title} {\bibinfo
  {title} {No-switching quantum key distribution using broadband modulated
  coherent light},\ }\href
  {https://link.aps.org/doi/10.1103/PhysRevLett.95.180503} {\bibfield
  {journal} {\bibinfo  {journal} {Phys. Rev. Lett.}\ }\textbf {\bibinfo
  {volume} {95}},\ \bibinfo {pages} {180503} (\bibinfo {year}
  {2005})}\BibitemShut {NoStop}%
\bibitem [{\citenamefont {Madsen}\ \emph {et~al.}(2012)\citenamefont {Madsen},
  \citenamefont {Usenko}, \citenamefont {Lassen}, \citenamefont {Filip},\ and\
  \citenamefont {Andersen}}]{Madsen2012}%
  \BibitemOpen
  \bibfield  {author} {\bibinfo {author} {\bibfnamefont {L.~S.}\ \bibnamefont
  {Madsen}}, \bibinfo {author} {\bibfnamefont {V.~C.}\ \bibnamefont {Usenko}},
  \bibinfo {author} {\bibfnamefont {M.}~\bibnamefont {Lassen}}, \bibinfo
  {author} {\bibfnamefont {R.}~\bibnamefont {Filip}},\ and\ \bibinfo {author}
  {\bibfnamefont {U.~L.}\ \bibnamefont {Andersen}},\ }\bibfield  {title}
  {\bibinfo {title} {Continuous variable quantum key distribution with
  modulated entangled states},\ }\bibfield  {journal} {\bibinfo  {journal}
  {Nat. Commun.}\ }\textbf {\bibinfo {volume} {3}},\ \href
  {https://doi.org/10.1038/ncomms2097} {10.1038/ncomms2097} (\bibinfo {year}
  {2012})\BibitemShut {NoStop}%
\bibitem [{\citenamefont {Usenko}\ and\ \citenamefont
  {Grosshans}(2015)}]{Usenko2015}%
  \BibitemOpen
  \bibfield  {author} {\bibinfo {author} {\bibfnamefont {V.~C.}\ \bibnamefont
  {Usenko}}\ and\ \bibinfo {author} {\bibfnamefont {F.}~\bibnamefont
  {Grosshans}},\ }\bibfield  {title} {\bibinfo {title} {Unidimensional
  continuous-variable quantum key distribution},\ }\href
  {https://doi.org/10.1103/physreva.92.062337} {\bibfield  {journal} {\bibinfo
  {journal} {Phys. Rev. A}\ }\textbf {\bibinfo {volume} {92}},\ \bibinfo
  {pages} {062337} (\bibinfo {year} {2015})}\BibitemShut {NoStop}%
\bibitem [{\citenamefont {Diamanti}\ and\ \citenamefont
  {Leverrier}(2015)}]{Diamanti2015}%
  \BibitemOpen
  \bibfield  {author} {\bibinfo {author} {\bibfnamefont {E.}~\bibnamefont
  {Diamanti}}\ and\ \bibinfo {author} {\bibfnamefont {A.}~\bibnamefont
  {Leverrier}},\ }\bibfield  {title} {\bibinfo {title} {Distributing secret
  keys with quantum continuous variables: Principle, security and
  implementations},\ }\href {https://doi.org/10.3390/e17096072} {\bibfield
  {journal} {\bibinfo  {journal} {Entropy}\ }\textbf {\bibinfo {volume} {17}},\
  \bibinfo {pages} {6072} (\bibinfo {year} {2015})}\BibitemShut {NoStop}%
\bibitem [{\citenamefont {Hosseinidehaj}\ \emph {et~al.}(2019)\citenamefont
  {Hosseinidehaj}, \citenamefont {Babar}, \citenamefont {Malaney},
  \citenamefont {Ng},\ and\ \citenamefont {Hanzo}}]{Hosseinidehaj2019}%
  \BibitemOpen
  \bibfield  {author} {\bibinfo {author} {\bibfnamefont {N.}~\bibnamefont
  {Hosseinidehaj}}, \bibinfo {author} {\bibfnamefont {Z.}~\bibnamefont
  {Babar}}, \bibinfo {author} {\bibfnamefont {R.}~\bibnamefont {Malaney}},
  \bibinfo {author} {\bibfnamefont {S.~X.}\ \bibnamefont {Ng}},\ and\ \bibinfo
  {author} {\bibfnamefont {L.}~\bibnamefont {Hanzo}},\ }\bibfield  {title}
  {\bibinfo {title} {Satellite-based continuous-variable quantum
  communications: State-of-the-art and a predictive outlook},\ }\href
  {https://doi.org/10.1109/comst.2018.2864557} {\bibfield  {journal} {\bibinfo
  {journal} {IEEE Commun. Surv. Tutor.}\ }\textbf {\bibinfo {volume} {21}},\
  \bibinfo {pages} {881} (\bibinfo {year} {2019})}\BibitemShut {NoStop}%
\bibitem [{\citenamefont {Silberhorn}\ \emph {et~al.}(2002)\citenamefont
  {Silberhorn}, \citenamefont {Ralph}, \citenamefont {Lütkenhaus},\ and\
  \citenamefont {Leuchs}}]{Silberhorn2002}%
  \BibitemOpen
  \bibfield  {author} {\bibinfo {author} {\bibfnamefont {C.}~\bibnamefont
  {Silberhorn}}, \bibinfo {author} {\bibfnamefont {T.~C.}\ \bibnamefont
  {Ralph}}, \bibinfo {author} {\bibfnamefont {N.}~\bibnamefont {Lütkenhaus}},\
  and\ \bibinfo {author} {\bibfnamefont {G.}~\bibnamefont {Leuchs}},\
  }\bibfield  {title} {\bibinfo {title} {Continuous variable quantum
  cryptography: Beating the 3 db loss limit},\ }\href
  {https://doi.org/10.1103/physrevlett.89.167901} {\bibfield  {journal}
  {\bibinfo  {journal} {Physical Review Letters}\ }\textbf {\bibinfo {volume}
  {89}},\ \bibinfo {pages} {167901} (\bibinfo {year} {2002})}\BibitemShut
  {NoStop}%
\bibitem [{\citenamefont {Hillery}(2000)}]{Hillery2000}%
  \BibitemOpen
  \bibfield  {author} {\bibinfo {author} {\bibfnamefont {M.}~\bibnamefont
  {Hillery}},\ }\bibfield  {title} {\bibinfo {title} {Quantum cryptography with
  squeezed states},\ }\href {https://doi.org/10.1103/physreva.61.022309}
  {\bibfield  {journal} {\bibinfo  {journal} {Physical Review A}\ }\textbf
  {\bibinfo {volume} {61}},\ \bibinfo {pages} {022309} (\bibinfo {year}
  {2000})}\BibitemShut {NoStop}%
\bibitem [{\citenamefont {Slusher}\ \emph {et~al.}(1987)\citenamefont
  {Slusher}, \citenamefont {Grangier}, \citenamefont {LaPorta}, \citenamefont
  {Yurke},\ and\ \citenamefont {Potasek}}]{Slusher1987}%
  \BibitemOpen
  \bibfield  {author} {\bibinfo {author} {\bibfnamefont {R.~E.}\ \bibnamefont
  {Slusher}}, \bibinfo {author} {\bibfnamefont {P.}~\bibnamefont {Grangier}},
  \bibinfo {author} {\bibfnamefont {A.}~\bibnamefont {LaPorta}}, \bibinfo
  {author} {\bibfnamefont {B.}~\bibnamefont {Yurke}},\ and\ \bibinfo {author}
  {\bibfnamefont {M.~J.}\ \bibnamefont {Potasek}},\ }\bibfield  {title}
  {\bibinfo {title} {Pulsed squeezed light},\ }\href
  {https://doi.org/10.1103/physrevlett.59.2566} {\bibfield  {journal} {\bibinfo
   {journal} {Phys. Rev. Lett.}\ }\textbf {\bibinfo {volume} {59}},\ \bibinfo
  {pages} {2566} (\bibinfo {year} {1987})}\BibitemShut {NoStop}%
\bibitem [{\citenamefont {Hirano}\ and\ \citenamefont
  {Matsuoka}(1990)}]{Hirano1990}%
  \BibitemOpen
  \bibfield  {author} {\bibinfo {author} {\bibfnamefont {T.}~\bibnamefont
  {Hirano}}\ and\ \bibinfo {author} {\bibfnamefont {M.}~\bibnamefont
  {Matsuoka}},\ }\bibfield  {title} {\bibinfo {title} {Broadband squeezing of
  light by pulse excitation},\ }\href {https://doi.org/10.1364/ol.15.001153}
  {\bibfield  {journal} {\bibinfo  {journal} {Opt. Lett.}\ }\textbf {\bibinfo
  {volume} {15}},\ \bibinfo {pages} {1153} (\bibinfo {year}
  {1990})}\BibitemShut {NoStop}%
\bibitem [{\citenamefont {Smithey}\ \emph {et~al.}(1992)\citenamefont
  {Smithey}, \citenamefont {Beck}, \citenamefont {Belsley},\ and\ \citenamefont
  {Raymer}}]{Smithey1992}%
  \BibitemOpen
  \bibfield  {author} {\bibinfo {author} {\bibfnamefont {D.~T.}\ \bibnamefont
  {Smithey}}, \bibinfo {author} {\bibfnamefont {M.}~\bibnamefont {Beck}},
  \bibinfo {author} {\bibfnamefont {M.}~\bibnamefont {Belsley}},\ and\ \bibinfo
  {author} {\bibfnamefont {M.~G.}\ \bibnamefont {Raymer}},\ }\bibfield  {title}
  {\bibinfo {title} {Sub-shot-noise correlation of total photon number using
  macroscopic twin pulses of light},\ }\href
  {https://doi.org/10.1103/physrevlett.69.2650} {\bibfield  {journal} {\bibinfo
   {journal} {Phys. Rev. Lett.}\ }\textbf {\bibinfo {volume} {69}},\ \bibinfo
  {pages} {2650} (\bibinfo {year} {1992})}\BibitemShut {NoStop}%
\bibitem [{\citenamefont {Smithey}\ \emph
  {et~al.}(1993{\natexlab{a}})\citenamefont {Smithey}, \citenamefont {Beck},
  \citenamefont {Cooper},\ and\ \citenamefont {Raymer}}]{Smithey1993a}%
  \BibitemOpen
  \bibfield  {author} {\bibinfo {author} {\bibfnamefont {D.~T.}\ \bibnamefont
  {Smithey}}, \bibinfo {author} {\bibfnamefont {M.}~\bibnamefont {Beck}},
  \bibinfo {author} {\bibfnamefont {J.}~\bibnamefont {Cooper}},\ and\ \bibinfo
  {author} {\bibfnamefont {M.~G.}\ \bibnamefont {Raymer}},\ }\bibfield  {title}
  {\bibinfo {title} {Measurement of number-phase uncertainty relations of
  optical fields},\ }\href {https://doi.org/10.1103/physreva.48.3159}
  {\bibfield  {journal} {\bibinfo  {journal} {Phys. Rev. A}\ }\textbf {\bibinfo
  {volume} {48}},\ \bibinfo {pages} {3159} (\bibinfo {year}
  {1993}{\natexlab{a}})}\BibitemShut {NoStop}%
\bibitem [{\citenamefont {Smithey}\ \emph
  {et~al.}(1993{\natexlab{b}})\citenamefont {Smithey}, \citenamefont {Beck},
  \citenamefont {Raymer},\ and\ \citenamefont {Faridani}}]{Smithey1993b}%
  \BibitemOpen
  \bibfield  {author} {\bibinfo {author} {\bibfnamefont {D.~T.}\ \bibnamefont
  {Smithey}}, \bibinfo {author} {\bibfnamefont {M.}~\bibnamefont {Beck}},
  \bibinfo {author} {\bibfnamefont {M.~G.}\ \bibnamefont {Raymer}},\ and\
  \bibinfo {author} {\bibfnamefont {A.}~\bibnamefont {Faridani}},\ }\bibfield
  {title} {\bibinfo {title} {Measurement of the {W}igner distribution and the
  density matrix of a light mode using optical homodyne tomography: Application
  to squeezed states and the vacuum},\ }\href
  {https://doi.org/10.1103/physrevlett.70.1244} {\bibfield  {journal} {\bibinfo
   {journal} {Phys. Rev. Lett.}\ }\textbf {\bibinfo {volume} {70}},\ \bibinfo
  {pages} {1244} (\bibinfo {year} {1993}{\natexlab{b}})}\BibitemShut {NoStop}%
\bibitem [{\citenamefont {Zavatta}\ \emph {et~al.}(2002)\citenamefont
  {Zavatta}, \citenamefont {Bellini}, \citenamefont {Ramazza}, \citenamefont
  {Marin},\ and\ \citenamefont {Arecchi}}]{Zavatta2002}%
  \BibitemOpen
  \bibfield  {author} {\bibinfo {author} {\bibfnamefont {A.}~\bibnamefont
  {Zavatta}}, \bibinfo {author} {\bibfnamefont {M.}~\bibnamefont {Bellini}},
  \bibinfo {author} {\bibfnamefont {P.~L.}\ \bibnamefont {Ramazza}}, \bibinfo
  {author} {\bibfnamefont {F.}~\bibnamefont {Marin}},\ and\ \bibinfo {author}
  {\bibfnamefont {F.~T.}\ \bibnamefont {Arecchi}},\ }\bibfield  {title}
  {\bibinfo {title} {Time-domain analysis of quantum states of light: noise
  characterization and homodyne tomography},\ }\href
  {https://doi.org/10.1364/josab.19.001189} {\bibfield  {journal} {\bibinfo
  {journal} {J. Opt. Soc. Am. B.}\ }\textbf {\bibinfo {volume} {19}},\ \bibinfo
  {pages} {1189} (\bibinfo {year} {2002})}\BibitemShut {NoStop}%
\bibitem [{\citenamefont {Zavatta}\ \emph {et~al.}(2005)\citenamefont
  {Zavatta}, \citenamefont {Viciani},\ and\ \citenamefont
  {Bellini}}]{Zavatta2005}%
  \BibitemOpen
  \bibfield  {author} {\bibinfo {author} {\bibfnamefont {A.}~\bibnamefont
  {Zavatta}}, \bibinfo {author} {\bibfnamefont {S.}~\bibnamefont {Viciani}},\
  and\ \bibinfo {author} {\bibfnamefont {M.}~\bibnamefont {Bellini}},\
  }\bibfield  {title} {\bibinfo {title} {Non-classical field characterization
  by high-frequency, time-domain quantum homodyne tomography},\ }\href
  {https://doi.org/10.1002/lapl.200510060} {\bibfield  {journal} {\bibinfo
  {journal} {Laser Phys. Lett.}\ }\textbf {\bibinfo {volume} {3}},\ \bibinfo
  {pages} {3} (\bibinfo {year} {2005})}\BibitemShut {NoStop}%
\bibitem [{\citenamefont {Haderka}\ \emph {et~al.}(2009)\citenamefont
  {Haderka}, \citenamefont {Mich{\'{a}}lek}, \citenamefont
  {Urb{\'{a}}{\v{s}}ek},\ and\ \citenamefont {Je{\v{z}}ek}}]{Haderka2009}%
  \BibitemOpen
  \bibfield  {author} {\bibinfo {author} {\bibfnamefont {O.}~\bibnamefont
  {Haderka}}, \bibinfo {author} {\bibfnamefont {V.}~\bibnamefont
  {Mich{\'{a}}lek}}, \bibinfo {author} {\bibfnamefont {V.}~\bibnamefont
  {Urb{\'{a}}{\v{s}}ek}},\ and\ \bibinfo {author} {\bibfnamefont
  {M.}~\bibnamefont {Je{\v{z}}ek}},\ }\bibfield  {title} {\bibinfo {title}
  {Fast time-domain balanced homodyne detection of light},\ }\href
  {https://doi.org/10.1364/ao.48.002884} {\bibfield  {journal} {\bibinfo
  {journal} {Appl. Optics}\ }\textbf {\bibinfo {volume} {48}},\ \bibinfo
  {pages} {2884} (\bibinfo {year} {2009})}\BibitemShut {NoStop}%
\bibitem [{\citenamefont {Okubo}\ \emph {et~al.}(2008)\citenamefont {Okubo},
  \citenamefont {Hirano}, \citenamefont {Zhang},\ and\ \citenamefont
  {Hirano}}]{Okubo2008}%
  \BibitemOpen
  \bibfield  {author} {\bibinfo {author} {\bibfnamefont {R.}~\bibnamefont
  {Okubo}}, \bibinfo {author} {\bibfnamefont {M.}~\bibnamefont {Hirano}},
  \bibinfo {author} {\bibfnamefont {Y.}~\bibnamefont {Zhang}},\ and\ \bibinfo
  {author} {\bibfnamefont {T.}~\bibnamefont {Hirano}},\ }\bibfield  {title}
  {\bibinfo {title} {Pulse-resolved measurement of quadrature phase amplitudes
  of squeezed pulse trains at a repetition rate of 76 {MHz}},\ }\href
  {https://doi.org/10.1364/ol.33.001458} {\bibfield  {journal} {\bibinfo
  {journal} {Opt. Lett.}\ }\textbf {\bibinfo {volume} {33}},\ \bibinfo {pages}
  {1458} (\bibinfo {year} {2008})}\BibitemShut {NoStop}%
\bibitem [{\citenamefont {Ansari}\ \emph {et~al.}(2017)\citenamefont {Ansari},
  \citenamefont {Harder}, \citenamefont {Allgaier}, \citenamefont {Brecht},\
  and\ \citenamefont {Silberhorn}}]{Ansari2017}%
  \BibitemOpen
  \bibfield  {author} {\bibinfo {author} {\bibfnamefont {V.}~\bibnamefont
  {Ansari}}, \bibinfo {author} {\bibfnamefont {G.}~\bibnamefont {Harder}},
  \bibinfo {author} {\bibfnamefont {M.}~\bibnamefont {Allgaier}}, \bibinfo
  {author} {\bibfnamefont {B.}~\bibnamefont {Brecht}},\ and\ \bibinfo {author}
  {\bibfnamefont {C.}~\bibnamefont {Silberhorn}},\ }\bibfield  {title}
  {\bibinfo {title} {Temporal-mode measurement tomography of a quantum pulse
  gate},\ }\href {https://doi.org/10.1103/physreva.96.063817} {\bibfield
  {journal} {\bibinfo  {journal} {Phys. Rev. A}\ }\textbf {\bibinfo {volume}
  {96}},\ \bibinfo {pages} {063817} (\bibinfo {year} {2017})}\BibitemShut
  {NoStop}%
\bibitem [{\citenamefont {Tiedau}\ \emph {et~al.}(2018)\citenamefont {Tiedau},
  \citenamefont {Shchesnovich}, \citenamefont {Mogilevtsev}, \citenamefont
  {Ansari}, \citenamefont {Harder}, \citenamefont {Bartley}, \citenamefont
  {Korolkova},\ and\ \citenamefont {Silberhorn}}]{Tiedau2018}%
  \BibitemOpen
  \bibfield  {author} {\bibinfo {author} {\bibfnamefont {J.}~\bibnamefont
  {Tiedau}}, \bibinfo {author} {\bibfnamefont {V.~S.}\ \bibnamefont
  {Shchesnovich}}, \bibinfo {author} {\bibfnamefont {D.}~\bibnamefont
  {Mogilevtsev}}, \bibinfo {author} {\bibfnamefont {V.}~\bibnamefont {Ansari}},
  \bibinfo {author} {\bibfnamefont {G.}~\bibnamefont {Harder}}, \bibinfo
  {author} {\bibfnamefont {T.~J.}\ \bibnamefont {Bartley}}, \bibinfo {author}
  {\bibfnamefont {N.}~\bibnamefont {Korolkova}},\ and\ \bibinfo {author}
  {\bibfnamefont {C.}~\bibnamefont {Silberhorn}},\ }\bibfield  {title}
  {\bibinfo {title} {Quantum state and mode profile tomography by the
  overlap},\ }\href {https://doi.org/10.1088/1367-2630/aaad8a} {\bibfield
  {journal} {\bibinfo  {journal} {New J. Phys.}\ }\textbf {\bibinfo {volume}
  {20}},\ \bibinfo {pages} {033003} (\bibinfo {year} {2018})}\BibitemShut
  {NoStop}%
\bibitem [{\citenamefont {Ansari}\ \emph {et~al.}(2018)\citenamefont {Ansari},
  \citenamefont {Donohue}, \citenamefont {Allgaier}, \citenamefont {Sansoni},
  \citenamefont {Brecht}, \citenamefont {Roslund}, \citenamefont {Treps},
  \citenamefont {Harder},\ and\ \citenamefont {Silberhorn}}]{Ansari2018}%
  \BibitemOpen
  \bibfield  {author} {\bibinfo {author} {\bibfnamefont {V.}~\bibnamefont
  {Ansari}}, \bibinfo {author} {\bibfnamefont {J.~M.}\ \bibnamefont {Donohue}},
  \bibinfo {author} {\bibfnamefont {M.}~\bibnamefont {Allgaier}}, \bibinfo
  {author} {\bibfnamefont {L.}~\bibnamefont {Sansoni}}, \bibinfo {author}
  {\bibfnamefont {B.}~\bibnamefont {Brecht}}, \bibinfo {author} {\bibfnamefont
  {J.}~\bibnamefont {Roslund}}, \bibinfo {author} {\bibfnamefont
  {N.}~\bibnamefont {Treps}}, \bibinfo {author} {\bibfnamefont
  {G.}~\bibnamefont {Harder}},\ and\ \bibinfo {author} {\bibfnamefont
  {C.}~\bibnamefont {Silberhorn}},\ }\bibfield  {title} {\bibinfo {title}
  {Tomography and purification of the temporal-mode structure of quantum
  light},\ }\href {https://doi.org/10.1103/physrevlett.120.213601} {\bibfield
  {journal} {\bibinfo  {journal} {Phys. Rev. Lett.}\ }\textbf {\bibinfo
  {volume} {120}},\ \bibinfo {pages} {213601} (\bibinfo {year}
  {2018})}\BibitemShut {NoStop}%
\bibitem [{\citenamefont {Gil-Lopez}\ \emph {et~al.}(2021)\citenamefont
  {Gil-Lopez}, \citenamefont {Teo}, \citenamefont {De}, \citenamefont {Brecht},
  \citenamefont {Jeong}, \citenamefont {Silberhorn},\ and\ \citenamefont
  {S{\'{a}}nchez-Soto}}]{GilLopez2021}%
  \BibitemOpen
  \bibfield  {author} {\bibinfo {author} {\bibfnamefont {J.}~\bibnamefont
  {Gil-Lopez}}, \bibinfo {author} {\bibfnamefont {Y.~S.}\ \bibnamefont {Teo}},
  \bibinfo {author} {\bibfnamefont {S.}~\bibnamefont {De}}, \bibinfo {author}
  {\bibfnamefont {B.}~\bibnamefont {Brecht}}, \bibinfo {author} {\bibfnamefont
  {H.}~\bibnamefont {Jeong}}, \bibinfo {author} {\bibfnamefont
  {C.}~\bibnamefont {Silberhorn}},\ and\ \bibinfo {author} {\bibfnamefont
  {L.~L.}\ \bibnamefont {S{\'{a}}nchez-Soto}},\ }\bibfield  {title} {\bibinfo
  {title} {Universal compressive tomography in the time-frequency domain},\
  }\href {https://doi.org/10.1364/optica.427645} {\bibfield  {journal}
  {\bibinfo  {journal} {Optica}\ }\textbf {\bibinfo {volume} {8}},\ \bibinfo
  {pages} {1296} (\bibinfo {year} {2021})}\BibitemShut {NoStop}%
\bibitem [{\citenamefont {Kalash}\ and\ \citenamefont
  {Chekhova}(2023)}]{Kalash2023}%
  \BibitemOpen
  \bibfield  {author} {\bibinfo {author} {\bibfnamefont {M.}~\bibnamefont
  {Kalash}}\ and\ \bibinfo {author} {\bibfnamefont {M.~V.}\ \bibnamefont
  {Chekhova}},\ }\bibfield  {title} {\bibinfo {title} {Wigner function
  tomography via optical parametric amplification},\ }\href
  {https://doi.org/10.1364/optica.488697} {\bibfield  {journal} {\bibinfo
  {journal} {Optica}\ }\textbf {\bibinfo {volume} {10}},\ \bibinfo {pages}
  {1142} (\bibinfo {year} {2023})}\BibitemShut {NoStop}%
\bibitem [{\citenamefont {Serino}\ \emph {et~al.}(2023)\citenamefont {Serino},
  \citenamefont {Gil-Lopez}, \citenamefont {Stefszky}, \citenamefont {Ricken},
  \citenamefont {Eigner}, \citenamefont {Brecht},\ and\ \citenamefont
  {Silberhorn}}]{Serino2023}%
  \BibitemOpen
  \bibfield  {author} {\bibinfo {author} {\bibfnamefont {L.}~\bibnamefont
  {Serino}}, \bibinfo {author} {\bibfnamefont {J.}~\bibnamefont {Gil-Lopez}},
  \bibinfo {author} {\bibfnamefont {M.}~\bibnamefont {Stefszky}}, \bibinfo
  {author} {\bibfnamefont {R.}~\bibnamefont {Ricken}}, \bibinfo {author}
  {\bibfnamefont {C.}~\bibnamefont {Eigner}}, \bibinfo {author} {\bibfnamefont
  {B.}~\bibnamefont {Brecht}},\ and\ \bibinfo {author} {\bibfnamefont
  {C.}~\bibnamefont {Silberhorn}},\ }\bibfield  {title} {\bibinfo {title}
  {Realization of a multi-output quantum pulse gate for decoding
  high-dimensional temporal modes of single-photon states},\ }\href
  {https://doi.org/10.1103/prxquantum.4.020306} {\bibfield  {journal} {\bibinfo
   {journal} {PRX Quantum}\ }\textbf {\bibinfo {volume} {4}},\ \bibinfo {pages}
  {020306} (\bibinfo {year} {2023})}\BibitemShut {NoStop}%
\bibitem [{\citenamefont {Mrówczyński}\ and\ \citenamefont
  {Müller}(1994)}]{Mrowczynski1994}%
  \BibitemOpen
  \bibfield  {author} {\bibinfo {author} {\bibfnamefont {S.}~\bibnamefont
  {Mrówczyński}}\ and\ \bibinfo {author} {\bibfnamefont {B.}~\bibnamefont
  {Müller}},\ }\bibfield  {title} {\bibinfo {title} {Wigner functional
  approach to quantum field dynamics},\ }\href
  {https://doi.org/10.1103/physrevd.50.7542} {\bibfield  {journal} {\bibinfo
  {journal} {Physical Review D}\ }\textbf {\bibinfo {volume} {50}},\ \bibinfo
  {pages} {7542} (\bibinfo {year} {1994})}\BibitemShut {NoStop}%
\bibitem [{\citenamefont {Roux}\ and\ \citenamefont {Fabre}(2019)}]{Roux2019}%
  \BibitemOpen
  \bibfield  {author} {\bibinfo {author} {\bibfnamefont {F.~S.}\ \bibnamefont
  {Roux}}\ and\ \bibinfo {author} {\bibfnamefont {N.}~\bibnamefont {Fabre}},\
  }\href {https://doi.org/10.48550/ARXIV.1901.07782} {\bibinfo {title} {Wigner
  functional theory for quantum optics}} (\bibinfo {year} {2019})\BibitemShut
  {NoStop}%
\bibitem [{\citenamefont {Virally}\ and\ \citenamefont
  {Reulet}(2019)}]{Virally2019}%
  \BibitemOpen
  \bibfield  {author} {\bibinfo {author} {\bibfnamefont {S.}~\bibnamefont
  {Virally}}\ and\ \bibinfo {author} {\bibfnamefont {B.}~\bibnamefont
  {Reulet}},\ }\bibfield  {title} {\bibinfo {title} {Unidimensional time-domain
  quantum optics},\ }\href {https://doi.org/10.1103/physreva.100.023833}
  {\bibfield  {journal} {\bibinfo  {journal} {Physical Review A}\ }\textbf
  {\bibinfo {volume} {100}},\ \bibinfo {pages} {023833} (\bibinfo {year}
  {2019})}\BibitemShut {NoStop}%
\bibitem [{\citenamefont {Roux}(2020)}]{Roux2020}%
  \BibitemOpen
  \bibfield  {author} {\bibinfo {author} {\bibfnamefont {F.~S.}\ \bibnamefont
  {Roux}},\ }\bibfield  {title} {\bibinfo {title} {Erratum: Combining
  spatiotemporal and particle-number degrees of freedom [phys. rev. a 98,
  043841 (2018)]},\ }\href {https://doi.org/10.1103/physreva.101.019903}
  {\bibfield  {journal} {\bibinfo  {journal} {Physical Review A}\ }\textbf
  {\bibinfo {volume} {101}},\ \bibinfo {pages} {019903} (\bibinfo {year}
  {2020})}\BibitemShut {NoStop}%
\bibitem [{\citenamefont {Adesso}\ \emph {et~al.}(2014)\citenamefont {Adesso},
  \citenamefont {Ragy},\ and\ \citenamefont {Lee}}]{Adesso2014}%
  \BibitemOpen
  \bibfield  {author} {\bibinfo {author} {\bibfnamefont {G.}~\bibnamefont
  {Adesso}}, \bibinfo {author} {\bibfnamefont {S.}~\bibnamefont {Ragy}},\ and\
  \bibinfo {author} {\bibfnamefont {A.~R.}\ \bibnamefont {Lee}},\ }\bibfield
  {title} {\bibinfo {title} {Continuous variable quantum information: Gaussian
  states and beyond},\ }\bibfield  {journal} {\bibinfo  {journal} {Open Syst.
  Inf. Dyn. 21, 1440001 (2014)}\ }\href
  {https://doi.org/10.1142/S1230161214400010} {10.1142/S1230161214400010}
  (\bibinfo {year} {2014}),\ \Eprint {https://arxiv.org/abs/1401.4679}
  {arXiv:1401.4679 [quant-ph]} \BibitemShut {NoStop}%
\bibitem [{\citenamefont {Raymer}\ and\ \citenamefont
  {Walmsley}(2020)}]{Raymer2020}%
  \BibitemOpen
  \bibfield  {author} {\bibinfo {author} {\bibfnamefont {M.~G.}\ \bibnamefont
  {Raymer}}\ and\ \bibinfo {author} {\bibfnamefont {I.~A.}\ \bibnamefont
  {Walmsley}},\ }\bibfield  {title} {\bibinfo {title} {Temporal modes in
  quantum optics: then and now},\ }\href
  {https://doi.org/10.1088/1402-4896/ab6153} {\bibfield  {journal} {\bibinfo
  {journal} {Physica Scripta}\ }\textbf {\bibinfo {volume} {95}},\ \bibinfo
  {pages} {064002} (\bibinfo {year} {2020})}\BibitemShut {NoStop}%
\bibitem [{\citenamefont {Brecht}\ \emph {et~al.}(2015)\citenamefont {Brecht},
  \citenamefont {Reddy}, \citenamefont {Silberhorn},\ and\ \citenamefont
  {G.~Raymer}}]{Brecht2015}%
  \BibitemOpen
  \bibfield  {author} {\bibinfo {author} {\bibfnamefont {B.}~\bibnamefont
  {Brecht}}, \bibinfo {author} {\bibfnamefont {D.~V.}\ \bibnamefont {Reddy}},
  \bibinfo {author} {\bibfnamefont {C.}~\bibnamefont {Silberhorn}},\ and\
  \bibinfo {author} {\bibfnamefont {M.}~\bibnamefont {G.~Raymer}},\ }\bibfield
  {title} {\bibinfo {title} {Photon temporal modes: A complete framework for
  quantum information science},\ }\href
  {https://doi.org/10.1103/physrevx.5.041017} {\bibfield  {journal} {\bibinfo
  {journal} {Physical Review X}\ }\textbf {\bibinfo {volume} {5}},\ \bibinfo
  {pages} {041017} (\bibinfo {year} {2015})}\BibitemShut {NoStop}%
\bibitem [{\citenamefont {Walker}\ and\ \citenamefont
  {Carroll}(1986)}]{Walker1986}%
  \BibitemOpen
  \bibfield  {author} {\bibinfo {author} {\bibfnamefont {N.~G.}\ \bibnamefont
  {Walker}}\ and\ \bibinfo {author} {\bibfnamefont {J.~E.}\ \bibnamefont
  {Carroll}},\ }\bibfield  {title} {\bibinfo {title} {Multiport homodyne
  detection near the quantum noise limit},\ }\href
  {https://doi.org/10.1007/bf02032562} {\bibfield  {journal} {\bibinfo
  {journal} {Opt. Quant. Electron.}\ }\textbf {\bibinfo {volume} {18}},\
  \bibinfo {pages} {355} (\bibinfo {year} {1986})}\BibitemShut {NoStop}%
\bibitem [{\citenamefont {Freyberger}\ \emph {et~al.}(1993)\citenamefont
  {Freyberger}, \citenamefont {Vogel},\ and\ \citenamefont
  {Schleich}}]{Freyberger1993}%
  \BibitemOpen
  \bibfield  {author} {\bibinfo {author} {\bibfnamefont {M.}~\bibnamefont
  {Freyberger}}, \bibinfo {author} {\bibfnamefont {K.}~\bibnamefont {Vogel}},\
  and\ \bibinfo {author} {\bibfnamefont {W.~P.}\ \bibnamefont {Schleich}},\
  }\bibfield  {title} {\bibinfo {title} {From photon counts to quantum phase},\
  }\href {https://doi.org/10.1016/0375-9601(93)90313-o} {\bibfield  {journal}
  {\bibinfo  {journal} {Phys. Lett. A}\ }\textbf {\bibinfo {volume} {176}},\
  \bibinfo {pages} {41} (\bibinfo {year} {1993})}\BibitemShut {NoStop}%
\bibitem [{\citenamefont {Leonhardt}\ and\ \citenamefont
  {Paul}(1993)}]{Leonhardt1993}%
  \BibitemOpen
  \bibfield  {author} {\bibinfo {author} {\bibfnamefont {U.}~\bibnamefont
  {Leonhardt}}\ and\ \bibinfo {author} {\bibfnamefont {H.}~\bibnamefont
  {Paul}},\ }\bibfield  {title} {\bibinfo {title} {Realistic optical homodyne
  measurements and quasiprobability distributions},\ }\href
  {https://doi.org/10.1103/physreva.48.4598} {\bibfield  {journal} {\bibinfo
  {journal} {Phys. Rev. A}\ }\textbf {\bibinfo {volume} {48}},\ \bibinfo
  {pages} {4598} (\bibinfo {year} {1993})}\BibitemShut {NoStop}%
\bibitem [{\citenamefont {Zucchetti}\ \emph {et~al.}(1996)\citenamefont
  {Zucchetti}, \citenamefont {Vogel},\ and\ \citenamefont
  {Welsch}}]{Zucchetti1996}%
  \BibitemOpen
  \bibfield  {author} {\bibinfo {author} {\bibfnamefont {A.}~\bibnamefont
  {Zucchetti}}, \bibinfo {author} {\bibfnamefont {W.}~\bibnamefont {Vogel}},\
  and\ \bibinfo {author} {\bibfnamefont {D.-G.}\ \bibnamefont {Welsch}},\
  }\bibfield  {title} {\bibinfo {title} {Quantum-state homodyne measurement
  with vacuum ports},\ }\href {https://doi.org/10.1103/physreva.54.856}
  {\bibfield  {journal} {\bibinfo  {journal} {Phys. Rev. A}\ }\textbf {\bibinfo
  {volume} {54}},\ \bibinfo {pages} {856} (\bibinfo {year} {1996})}\BibitemShut
  {NoStop}%
\bibitem [{\citenamefont {{\v{R}}eh{\'{a}}{\v{c}}ek}\ \emph
  {et~al.}(2015)\citenamefont {{\v{R}}eh{\'{a}}{\v{c}}ek}, \citenamefont {T.},
  \citenamefont {Hradil},\ and\ \citenamefont {Wallentowitz}}]{Rehacek2015}%
  \BibitemOpen
  \bibfield  {author} {\bibinfo {author} {\bibfnamefont {J.}~\bibnamefont
  {{\v{R}}eh{\'{a}}{\v{c}}ek}}, \bibinfo {author} {\bibfnamefont {Y.~S.}\
  \bibnamefont {T.}}, \bibinfo {author} {\bibfnamefont {Z.}~\bibnamefont
  {Hradil}},\ and\ \bibinfo {author} {\bibfnamefont {S.}~\bibnamefont
  {Wallentowitz}},\ }\bibfield  {title} {\bibinfo {title} {Surmounting
  intrinsic quantum-measurement uncertainties in {G}aussian-state tomography
  with quadrature squeezing},\ }\bibfield  {journal} {\bibinfo  {journal} {Sci.
  Rep.}\ }\textbf {\bibinfo {volume} {5}},\ \href
  {https://doi.org/10.1038/srep12289} {10.1038/srep12289} (\bibinfo {year}
  {2015})\BibitemShut {NoStop}%
\bibitem [{\citenamefont {Hubenschmid}\ \emph {et~al.}(2022)\citenamefont
  {Hubenschmid}, \citenamefont {Guedes},\ and\ \citenamefont
  {Burkard}}]{Hubenschmid2022}%
  \BibitemOpen
  \bibfield  {author} {\bibinfo {author} {\bibfnamefont {E.}~\bibnamefont
  {Hubenschmid}}, \bibinfo {author} {\bibfnamefont {T.~L.~M.}\ \bibnamefont
  {Guedes}},\ and\ \bibinfo {author} {\bibfnamefont {G.}~\bibnamefont
  {Burkard}},\ }\bibfield  {title} {\bibinfo {title} {Complete positive
  operator-valued measure description of multichannel quantum electro-optic
  sampling with monochromatic field modes},\ }\href
  {https://doi.org/10.1103/physreva.106.043713} {\bibfield  {journal} {\bibinfo
   {journal} {Phys. Rev. A}\ }\textbf {\bibinfo {volume} {106}},\ \bibinfo
  {pages} {043713} (\bibinfo {year} {2022})}\BibitemShut {NoStop}%
\bibitem [{\citenamefont {Vogel}\ and\ \citenamefont
  {Risken}(1989)}]{Vogel1989}%
  \BibitemOpen
  \bibfield  {author} {\bibinfo {author} {\bibfnamefont {K.}~\bibnamefont
  {Vogel}}\ and\ \bibinfo {author} {\bibfnamefont {H.}~\bibnamefont {Risken}},\
  }\bibfield  {title} {\bibinfo {title} {Determination of quasiprobability
  distributions in terms of probability distributions for the rotated
  quadrature phase},\ }\href {https://doi.org/10.1103/physreva.40.2847}
  {\bibfield  {journal} {\bibinfo  {journal} {Phys. Rev. A}\ }\textbf {\bibinfo
  {volume} {40}},\ \bibinfo {pages} {2847} (\bibinfo {year}
  {1989})}\BibitemShut {NoStop}%
\bibitem [{\citenamefont {Leonhardt}\ and\ \citenamefont
  {Paul}(1994)}]{Leonhardt1994}%
  \BibitemOpen
  \bibfield  {author} {\bibinfo {author} {\bibfnamefont {U.}~\bibnamefont
  {Leonhardt}}\ and\ \bibinfo {author} {\bibfnamefont {H.}~\bibnamefont
  {Paul}},\ }\bibfield  {title} {\bibinfo {title} {High-accuracy optical
  homodyne detection with low-efficiency detectors: "preamplification" from
  antisqueezing},\ }\href {https://doi.org/10.1103/physrevlett.72.4086}
  {\bibfield  {journal} {\bibinfo  {journal} {Phys. Rev. Lett.}\ }\textbf
  {\bibinfo {volume} {72}},\ \bibinfo {pages} {4086} (\bibinfo {year}
  {1994})}\BibitemShut {NoStop}%
\bibitem [{\citenamefont {Wallentowitz}\ and\ \citenamefont
  {Vogel}(1996)}]{Wallentowitz1996}%
  \BibitemOpen
  \bibfield  {author} {\bibinfo {author} {\bibfnamefont {S.}~\bibnamefont
  {Wallentowitz}}\ and\ \bibinfo {author} {\bibfnamefont {W.}~\bibnamefont
  {Vogel}},\ }\bibfield  {title} {\bibinfo {title} {Unbalanced homodyning for
  quantum state measurements},\ }\href
  {https://doi.org/10.1103/physreva.53.4528} {\bibfield  {journal} {\bibinfo
  {journal} {Phys. Rev. A}\ }\textbf {\bibinfo {volume} {53}},\ \bibinfo
  {pages} {4528} (\bibinfo {year} {1996})}\BibitemShut {NoStop}%
\bibitem [{\citenamefont {Breitenbach}\ \emph {et~al.}(1997)\citenamefont
  {Breitenbach}, \citenamefont {Schiller},\ and\ \citenamefont
  {Mlynek}}]{Breitenbach1997}%
  \BibitemOpen
  \bibfield  {author} {\bibinfo {author} {\bibfnamefont {G.}~\bibnamefont
  {Breitenbach}}, \bibinfo {author} {\bibfnamefont {S.}~\bibnamefont
  {Schiller}},\ and\ \bibinfo {author} {\bibfnamefont {J.}~\bibnamefont
  {Mlynek}},\ }\bibfield  {title} {\bibinfo {title} {Measurement of the quantum
  states of squeezed light},\ }\href {https://doi.org/10.1038/387471a0}
  {\bibfield  {journal} {\bibinfo  {journal} {Nature}\ }\textbf {\bibinfo
  {volume} {387}},\ \bibinfo {pages} {471} (\bibinfo {year}
  {1997})}\BibitemShut {NoStop}%
\bibitem [{\citenamefont {Luis}\ \emph {et~al.}(2015)\citenamefont {Luis},
  \citenamefont {Sperling},\ and\ \citenamefont {Vogel}}]{Luis2015}%
  \BibitemOpen
  \bibfield  {author} {\bibinfo {author} {\bibfnamefont {A.}~\bibnamefont
  {Luis}}, \bibinfo {author} {\bibfnamefont {J.}~\bibnamefont {Sperling}},\
  and\ \bibinfo {author} {\bibfnamefont {W.}~\bibnamefont {Vogel}},\ }\bibfield
   {title} {\bibinfo {title} {Nonclassicality phase-space functions: More
  insight with fewer detectors},\ }\href
  {https://doi.org/10.1103/physrevlett.114.103602} {\bibfield  {journal}
  {\bibinfo  {journal} {Phys. Rev. Lett.}\ }\textbf {\bibinfo {volume} {114}},\
  \bibinfo {pages} {103602} (\bibinfo {year} {2015})}\BibitemShut {NoStop}%
\bibitem [{\citenamefont {Bohmann}\ \emph {et~al.}(2018)\citenamefont
  {Bohmann}, \citenamefont {Tiedau}, \citenamefont {Bartley}, \citenamefont
  {Sperling}, \citenamefont {Silberhorn},\ and\ \citenamefont
  {Vogel}}]{Bohmann2018}%
  \BibitemOpen
  \bibfield  {author} {\bibinfo {author} {\bibfnamefont {M.}~\bibnamefont
  {Bohmann}}, \bibinfo {author} {\bibfnamefont {J.}~\bibnamefont {Tiedau}},
  \bibinfo {author} {\bibfnamefont {T.}~\bibnamefont {Bartley}}, \bibinfo
  {author} {\bibfnamefont {J.}~\bibnamefont {Sperling}}, \bibinfo {author}
  {\bibfnamefont {C.}~\bibnamefont {Silberhorn}},\ and\ \bibinfo {author}
  {\bibfnamefont {W.}~\bibnamefont {Vogel}},\ }\bibfield  {title} {\bibinfo
  {title} {Incomplete detection of nonclassical phase-space distributions},\
  }\href {https://doi.org/10.1103/physrevlett.120.063607} {\bibfield  {journal}
  {\bibinfo  {journal} {Phys. Rev. Lett.}\ }\textbf {\bibinfo {volume} {120}},\
  \bibinfo {pages} {063607} (\bibinfo {year} {2018})}\BibitemShut {NoStop}%
\bibitem [{\citenamefont {Knyazev}\ \emph {et~al.}(2018)\citenamefont
  {Knyazev}, \citenamefont {Spasibko}, \citenamefont {Chekhova},\ and\
  \citenamefont {Khalili}}]{Knyazev2018}%
  \BibitemOpen
  \bibfield  {author} {\bibinfo {author} {\bibfnamefont {E.}~\bibnamefont
  {Knyazev}}, \bibinfo {author} {\bibfnamefont {K.~Y.}\ \bibnamefont
  {Spasibko}}, \bibinfo {author} {\bibfnamefont {M.~V.}\ \bibnamefont
  {Chekhova}},\ and\ \bibinfo {author} {\bibfnamefont {F.~Y.}\ \bibnamefont
  {Khalili}},\ }\bibfield  {title} {\bibinfo {title} {Quantum tomography
  enhanced through parametric amplification},\ }\href
  {https://doi.org/10.1088/1367-2630/aa99b4} {\bibfield  {journal} {\bibinfo
  {journal} {New J. Phys.}\ }\textbf {\bibinfo {volume} {20}},\ \bibinfo
  {pages} {013005} (\bibinfo {year} {2018})}\BibitemShut {NoStop}%
\bibitem [{\citenamefont {Olivares}\ \emph {et~al.}(2019)\citenamefont
  {Olivares}, \citenamefont {Allevi}, \citenamefont {Caiazzo}, \citenamefont
  {Paris},\ and\ \citenamefont {Bondani}}]{Olivares2019}%
  \BibitemOpen
  \bibfield  {author} {\bibinfo {author} {\bibfnamefont {S.}~\bibnamefont
  {Olivares}}, \bibinfo {author} {\bibfnamefont {A.}~\bibnamefont {Allevi}},
  \bibinfo {author} {\bibfnamefont {G.}~\bibnamefont {Caiazzo}}, \bibinfo
  {author} {\bibfnamefont {M.~G.~A.}\ \bibnamefont {Paris}},\ and\ \bibinfo
  {author} {\bibfnamefont {M.}~\bibnamefont {Bondani}},\ }\bibfield  {title}
  {\bibinfo {title} {Quantum tomography of light states by
  photon-number-resolving detectors},\ }\href
  {https://doi.org/10.1088/1367-2630/ab4afb} {\bibfield  {journal} {\bibinfo
  {journal} {New J. Phys.}\ }\textbf {\bibinfo {volume} {21}},\ \bibinfo
  {pages} {103045} (\bibinfo {year} {2019})}\BibitemShut {NoStop}%
\bibitem [{\citenamefont {Hubenschmid}\ \emph {et~al.}(2024)\citenamefont
  {Hubenschmid}, \citenamefont {Guedes},\ and\ \citenamefont
  {Burkard}}]{Hubenschmid2024}%
  \BibitemOpen
  \bibfield  {author} {\bibinfo {author} {\bibfnamefont {E.}~\bibnamefont
  {Hubenschmid}}, \bibinfo {author} {\bibfnamefont {T.~L.~M.}\ \bibnamefont
  {Guedes}},\ and\ \bibinfo {author} {\bibfnamefont {G.}~\bibnamefont
  {Burkard}},\ }\bibfield  {title} {\bibinfo {title} {Optical time-domain
  quantum state tomography on a subcycle scale},\ }\href
  {https://doi.org/10.1103/physrevx.14.041032} {\bibfield  {journal} {\bibinfo
  {journal} {Physical Review X}\ }\textbf {\bibinfo {volume} {14}},\ \bibinfo
  {pages} {041032} (\bibinfo {year} {2024})}\BibitemShut {NoStop}%
\bibitem [{\citenamefont {Yang}\ \emph {et~al.}(2023)\citenamefont {Yang},
  \citenamefont {Kizmann}, \citenamefont {Leitenstorfer},\ and\ \citenamefont
  {Moskalenko}}]{Yang2023}%
  \BibitemOpen
  \bibfield  {author} {\bibinfo {author} {\bibfnamefont {G.}~\bibnamefont
  {Yang}}, \bibinfo {author} {\bibfnamefont {M.}~\bibnamefont {Kizmann}},
  \bibinfo {author} {\bibfnamefont {A.}~\bibnamefont {Leitenstorfer}},\ and\
  \bibinfo {author} {\bibfnamefont {A.~S.}\ \bibnamefont {Moskalenko}},\
  }\href@noop {} {\bibinfo {title} {Subcycle tomography of quantum light}}
  (\bibinfo {year} {2023}),\ \Eprint {https://arxiv.org/abs/2307.12812}
  {arXiv:2307.12812 [quant-ph]} \BibitemShut {NoStop}%
\bibitem [{\citenamefont {Onoe}\ \emph {et~al.}(2023)\citenamefont {Onoe},
  \citenamefont {Virally},\ and\ \citenamefont {Seletskiy}}]{Onoe2023}%
  \BibitemOpen
  \bibfield  {author} {\bibinfo {author} {\bibfnamefont {S.}~\bibnamefont
  {Onoe}}, \bibinfo {author} {\bibfnamefont {S.}~\bibnamefont {Virally}},\ and\
  \bibinfo {author} {\bibfnamefont {D.~V.}\ \bibnamefont {Seletskiy}},\
  }\href@noop {} {\bibinfo {title} {Direct measurement of the husimi-q function
  of the electric-field in the time-domain}} (\bibinfo {year} {2023}),\ \Eprint
  {https://arxiv.org/abs/2307.13088} {arXiv:2307.13088 [quant-ph]} \BibitemShut
  {NoStop}%
\bibitem [{\citenamefont {Lordi}\ \emph {et~al.}(2024)\citenamefont {Lordi},
  \citenamefont {Tsao}, \citenamefont {Lind}, \citenamefont {Diddams},\ and\
  \citenamefont {Combes}}]{Lordi2024}%
  \BibitemOpen
  \bibfield  {author} {\bibinfo {author} {\bibfnamefont {N.}~\bibnamefont
  {Lordi}}, \bibinfo {author} {\bibfnamefont {E.~J.}\ \bibnamefont {Tsao}},
  \bibinfo {author} {\bibfnamefont {A.~J.}\ \bibnamefont {Lind}}, \bibinfo
  {author} {\bibfnamefont {S.~A.}\ \bibnamefont {Diddams}},\ and\ \bibinfo
  {author} {\bibfnamefont {J.}~\bibnamefont {Combes}},\ }\bibfield  {title}
  {\bibinfo {title} {Quantum theory of temporally mismatched homodyne
  measurements with applications to optical-frequency-comb metrology},\ }\href
  {https://doi.org/10.1103/physreva.109.033722} {\bibfield  {journal} {\bibinfo
   {journal} {Physical Review A}\ }\textbf {\bibinfo {volume} {109}},\ \bibinfo
  {pages} {033722} (\bibinfo {year} {2024})}\BibitemShut {NoStop}%
\bibitem [{\citenamefont {Benea-Chelmus}\ \emph {et~al.}(2025)\citenamefont
  {Benea-Chelmus}, \citenamefont {Faist}, \citenamefont {Leitenstorfer},
  \citenamefont {Moskalenko}, \citenamefont {Pupeza}, \citenamefont
  {Seletskiy},\ and\ \citenamefont {Vodopyanov}}]{BeneaChelmus2025}%
  \BibitemOpen
  \bibfield  {author} {\bibinfo {author} {\bibfnamefont {I.-C.}\ \bibnamefont
  {Benea-Chelmus}}, \bibinfo {author} {\bibfnamefont {J.}~\bibnamefont
  {Faist}}, \bibinfo {author} {\bibfnamefont {A.}~\bibnamefont
  {Leitenstorfer}}, \bibinfo {author} {\bibfnamefont {A.~S.}\ \bibnamefont
  {Moskalenko}}, \bibinfo {author} {\bibfnamefont {I.}~\bibnamefont {Pupeza}},
  \bibinfo {author} {\bibfnamefont {D.~V.}\ \bibnamefont {Seletskiy}},\ and\
  \bibinfo {author} {\bibfnamefont {K.~L.}\ \bibnamefont {Vodopyanov}},\
  }\bibfield  {title} {\bibinfo {title} {Electro-optic sampling of classical
  and quantum light},\ }\href {https://doi.org/10.1364/optica.544333}
  {\bibfield  {journal} {\bibinfo  {journal} {Optica}\ }\textbf {\bibinfo
  {volume} {12}},\ \bibinfo {pages} {546} (\bibinfo {year} {2025})}\BibitemShut
  {NoStop}%
\bibitem [{\citenamefont {Riek}\ \emph {et~al.}(2015)\citenamefont {Riek},
  \citenamefont {Seletskiy}, \citenamefont {Moskalenko}, \citenamefont
  {Schmidt}, \citenamefont {Krauspe}, \citenamefont {Eckart}, \citenamefont
  {Eggert}, \citenamefont {Burkard},\ and\ \citenamefont
  {Leitenstorfer}}]{Riek2015}%
  \BibitemOpen
  \bibfield  {author} {\bibinfo {author} {\bibfnamefont {C.}~\bibnamefont
  {Riek}}, \bibinfo {author} {\bibfnamefont {D.~V.}\ \bibnamefont {Seletskiy}},
  \bibinfo {author} {\bibfnamefont {A.~S.}\ \bibnamefont {Moskalenko}},
  \bibinfo {author} {\bibfnamefont {J.~F.}\ \bibnamefont {Schmidt}}, \bibinfo
  {author} {\bibfnamefont {P.}~\bibnamefont {Krauspe}}, \bibinfo {author}
  {\bibfnamefont {S.}~\bibnamefont {Eckart}}, \bibinfo {author} {\bibfnamefont
  {S.}~\bibnamefont {Eggert}}, \bibinfo {author} {\bibfnamefont
  {G.}~\bibnamefont {Burkard}},\ and\ \bibinfo {author} {\bibfnamefont
  {A.}~\bibnamefont {Leitenstorfer}},\ }\bibfield  {title} {\bibinfo {title}
  {Direct sampling of electric-field vacuum fluctuations},\ }\href
  {https://doi.org/10.1126/science.aac9788} {\bibfield  {journal} {\bibinfo
  {journal} {Science}\ }\textbf {\bibinfo {volume} {350}},\ \bibinfo {pages}
  {420} (\bibinfo {year} {2015})}\BibitemShut {NoStop}%
\bibitem [{\citenamefont {Riek}\ \emph {et~al.}(2017)\citenamefont {Riek},
  \citenamefont {Sulzer}, \citenamefont {Seeger}, \citenamefont {Moskalenko},
  \citenamefont {Burkard}, \citenamefont {Seletskiy},\ and\ \citenamefont
  {Leitenstorfer}}]{Riek2017}%
  \BibitemOpen
  \bibfield  {author} {\bibinfo {author} {\bibfnamefont {C.}~\bibnamefont
  {Riek}}, \bibinfo {author} {\bibfnamefont {P.}~\bibnamefont {Sulzer}},
  \bibinfo {author} {\bibfnamefont {M.}~\bibnamefont {Seeger}}, \bibinfo
  {author} {\bibfnamefont {A.~S.}\ \bibnamefont {Moskalenko}}, \bibinfo
  {author} {\bibfnamefont {G.}~\bibnamefont {Burkard}}, \bibinfo {author}
  {\bibfnamefont {D.~V.}\ \bibnamefont {Seletskiy}},\ and\ \bibinfo {author}
  {\bibfnamefont {A.}~\bibnamefont {Leitenstorfer}},\ }\bibfield  {title}
  {\bibinfo {title} {Subcycle quantum electrodynamics},\ }\href
  {https://doi.org/10.1038/nature21024} {\bibfield  {journal} {\bibinfo
  {journal} {Nature}\ }\textbf {\bibinfo {volume} {541}},\ \bibinfo {pages}
  {376} (\bibinfo {year} {2017})}\BibitemShut {NoStop}%
\bibitem [{\citenamefont {Benea-Chelmus}\ \emph {et~al.}(2019)\citenamefont
  {Benea-Chelmus}, \citenamefont {Settembrini}, \citenamefont {Scalari},\ and\
  \citenamefont {Faist}}]{BeneaChelmus2019}%
  \BibitemOpen
  \bibfield  {author} {\bibinfo {author} {\bibfnamefont {I.-C.}\ \bibnamefont
  {Benea-Chelmus}}, \bibinfo {author} {\bibfnamefont {F.~F.}\ \bibnamefont
  {Settembrini}}, \bibinfo {author} {\bibfnamefont {G.}~\bibnamefont
  {Scalari}},\ and\ \bibinfo {author} {\bibfnamefont {J.}~\bibnamefont
  {Faist}},\ }\bibfield  {title} {\bibinfo {title} {Electric field correlation
  measurements on the electromagnetic vacuum state},\ }\href
  {https://doi.org/10.1038/s41586-019-1083-9} {\bibfield  {journal} {\bibinfo
  {journal} {Nature}\ }\textbf {\bibinfo {volume} {568}},\ \bibinfo {pages}
  {202} (\bibinfo {year} {2019})}\BibitemShut {NoStop}%
\bibitem [{\citenamefont {Moskalenko}\ \emph {et~al.}(2015)\citenamefont
  {Moskalenko}, \citenamefont {Riek}, \citenamefont {Seletskiy}, \citenamefont
  {Burkard},\ and\ \citenamefont {Leitenstorfer}}]{Moskalenko2015}%
  \BibitemOpen
  \bibfield  {author} {\bibinfo {author} {\bibfnamefont {A.~S.}\ \bibnamefont
  {Moskalenko}}, \bibinfo {author} {\bibfnamefont {C.}~\bibnamefont {Riek}},
  \bibinfo {author} {\bibfnamefont {D.~V.}\ \bibnamefont {Seletskiy}}, \bibinfo
  {author} {\bibfnamefont {G.}~\bibnamefont {Burkard}},\ and\ \bibinfo {author}
  {\bibfnamefont {A.}~\bibnamefont {Leitenstorfer}},\ }\bibfield  {title}
  {\bibinfo {title} {Paraxial theory of direct electro-optic sampling of the
  quantum vacuum},\ }\href {https://doi.org/10.1103/physrevlett.115.263601}
  {\bibfield  {journal} {\bibinfo  {journal} {Phys. Rev. Lett.}\ }\textbf
  {\bibinfo {volume} {115}},\ \bibinfo {pages} {263601} (\bibinfo {year}
  {2015})}\BibitemShut {NoStop}%
\bibitem [{\citenamefont {Kizmann}\ \emph {et~al.}(2019)\citenamefont
  {Kizmann}, \citenamefont {Guedes}, \citenamefont {Seletskiy}, \citenamefont
  {Moskalenko}, \citenamefont {Leitenstorfer},\ and\ \citenamefont
  {Burkard}}]{Kizmann2019}%
  \BibitemOpen
  \bibfield  {author} {\bibinfo {author} {\bibfnamefont {M.}~\bibnamefont
  {Kizmann}}, \bibinfo {author} {\bibfnamefont {T.~L.~M.}\ \bibnamefont
  {Guedes}}, \bibinfo {author} {\bibfnamefont {D.~V.}\ \bibnamefont
  {Seletskiy}}, \bibinfo {author} {\bibfnamefont {A.~S.}\ \bibnamefont
  {Moskalenko}}, \bibinfo {author} {\bibfnamefont {A.}~\bibnamefont
  {Leitenstorfer}},\ and\ \bibinfo {author} {\bibfnamefont {G.}~\bibnamefont
  {Burkard}},\ }\bibfield  {title} {\bibinfo {title} {Subcycle squeezing of
  light from a time flow perspective},\ }\href
  {https://doi.org/10.1038/s41567-019-0560-2} {\bibfield  {journal} {\bibinfo
  {journal} {Nat. Phys.}\ }\textbf {\bibinfo {volume} {15}},\ \bibinfo {pages}
  {960} (\bibinfo {year} {2019})}\BibitemShut {NoStop}%
\bibitem [{\citenamefont {Guedes}\ \emph {et~al.}(2019)\citenamefont {Guedes},
  \citenamefont {Kizmann}, \citenamefont {Seletskiy}, \citenamefont
  {Leitenstorfer}, \citenamefont {Burkard},\ and\ \citenamefont
  {Moskalenko}}]{Guedes2019}%
  \BibitemOpen
  \bibfield  {author} {\bibinfo {author} {\bibfnamefont {T.~L.~M.}\
  \bibnamefont {Guedes}}, \bibinfo {author} {\bibfnamefont {M.}~\bibnamefont
  {Kizmann}}, \bibinfo {author} {\bibfnamefont {D.~V.}\ \bibnamefont
  {Seletskiy}}, \bibinfo {author} {\bibfnamefont {A.}~\bibnamefont
  {Leitenstorfer}}, \bibinfo {author} {\bibfnamefont {G.}~\bibnamefont
  {Burkard}},\ and\ \bibinfo {author} {\bibfnamefont {A.~S.}\ \bibnamefont
  {Moskalenko}},\ }\bibfield  {title} {\bibinfo {title} {Spectra of
  ultrabroadband squeezed pulses and the finite-time {U}nruh-{D}avies effect},\
  }\href {https://doi.org/10.1103/physrevlett.122.053604} {\bibfield  {journal}
  {\bibinfo  {journal} {Phys. Rev. Lett.}\ }\textbf {\bibinfo {volume} {122}},\
  \bibinfo {pages} {053604} (\bibinfo {year} {2019})}\BibitemShut {NoStop}%
\bibitem [{\citenamefont {Kizmann}\ \emph {et~al.}(2022)\citenamefont
  {Kizmann}, \citenamefont {Moskalenko}, \citenamefont {Leitenstorfer},
  \citenamefont {Burkard},\ and\ \citenamefont {Mukamel}}]{Kizmann2022}%
  \BibitemOpen
  \bibfield  {author} {\bibinfo {author} {\bibfnamefont {M.}~\bibnamefont
  {Kizmann}}, \bibinfo {author} {\bibfnamefont {A.~S.}\ \bibnamefont
  {Moskalenko}}, \bibinfo {author} {\bibfnamefont {A.}~\bibnamefont
  {Leitenstorfer}}, \bibinfo {author} {\bibfnamefont {G.}~\bibnamefont
  {Burkard}},\ and\ \bibinfo {author} {\bibfnamefont {S.}~\bibnamefont
  {Mukamel}},\ }\bibfield  {title} {\bibinfo {title} {Quantum susceptibilities
  in time-domain sampling of electric field fluctuations},\ }\href
  {https://doi.org/10.1002/lpor.202100423} {\bibfield  {journal} {\bibinfo
  {journal} {Laser Photonics Rev.}\ }\textbf {\bibinfo {volume} {16}},\
  \bibinfo {pages} {2100423} (\bibinfo {year} {2022})}\BibitemShut {NoStop}%
\bibitem [{\citenamefont {Onoe}\ \emph {et~al.}(2022)\citenamefont {Onoe},
  \citenamefont {Guedes}, \citenamefont {Moskalenko}, \citenamefont
  {Leitenstorfer}, \citenamefont {Burkard},\ and\ \citenamefont
  {Ralph}}]{Onoe2022}%
  \BibitemOpen
  \bibfield  {author} {\bibinfo {author} {\bibfnamefont {S.}~\bibnamefont
  {Onoe}}, \bibinfo {author} {\bibfnamefont {T.~L.~M.}\ \bibnamefont {Guedes}},
  \bibinfo {author} {\bibfnamefont {A.~S.}\ \bibnamefont {Moskalenko}},
  \bibinfo {author} {\bibfnamefont {A.}~\bibnamefont {Leitenstorfer}}, \bibinfo
  {author} {\bibfnamefont {G.}~\bibnamefont {Burkard}},\ and\ \bibinfo {author}
  {\bibfnamefont {T.~C.}\ \bibnamefont {Ralph}},\ }\bibfield  {title} {\bibinfo
  {title} {Realizing a rapidly switched {U}nruh-{DeWitt} detector through
  electro-optic sampling of the electromagnetic vacuum},\ }\href
  {https://doi.org/10.1103/physrevd.105.056023} {\bibfield  {journal} {\bibinfo
   {journal} {Phys. Rev. D}\ }\textbf {\bibinfo {volume} {105}},\ \bibinfo
  {pages} {056023} (\bibinfo {year} {2022})}\BibitemShut {NoStop}%
\bibitem [{\citenamefont {Guedes}\ \emph {et~al.}(2023)\citenamefont {Guedes},
  \citenamefont {Vakulchyk}, \citenamefont {Seletskiy}, \citenamefont
  {Leitenstorfer}, \citenamefont {Moskalenko},\ and\ \citenamefont
  {Burkard}}]{Guedes2023}%
  \BibitemOpen
  \bibfield  {author} {\bibinfo {author} {\bibfnamefont {T.~L.~M.}\
  \bibnamefont {Guedes}}, \bibinfo {author} {\bibfnamefont {I.}~\bibnamefont
  {Vakulchyk}}, \bibinfo {author} {\bibfnamefont {D.~V.}\ \bibnamefont
  {Seletskiy}}, \bibinfo {author} {\bibfnamefont {A.}~\bibnamefont
  {Leitenstorfer}}, \bibinfo {author} {\bibfnamefont {A.~S.}\ \bibnamefont
  {Moskalenko}},\ and\ \bibinfo {author} {\bibfnamefont {G.}~\bibnamefont
  {Burkard}},\ }\bibfield  {title} {\bibinfo {title} {Back action in quantum
  electro-optic sampling of electromagnetic vacuum fluctuations},\ }\href
  {https://doi.org/10.1103/physrevresearch.5.013151} {\bibfield  {journal}
  {\bibinfo  {journal} {Phys. Rev. Research}\ }\textbf {\bibinfo {volume}
  {5}},\ \bibinfo {pages} {013151} (\bibinfo {year} {2023})}\BibitemShut
  {NoStop}%
\bibitem [{\citenamefont {Namba}(1961)}]{Namba1961}%
  \BibitemOpen
  \bibfield  {author} {\bibinfo {author} {\bibfnamefont {S.}~\bibnamefont
  {Namba}},\ }\bibfield  {title} {\bibinfo {title} {Electro-optical effect of
  zincblende},\ }\href {http://opg.optica.org/abstract.cfm?URI=josa-51-1-76}
  {\bibfield  {journal} {\bibinfo  {journal} {J. Opt. Soc. Am.}\ }\textbf
  {\bibinfo {volume} {51}},\ \bibinfo {pages} {76} (\bibinfo {year}
  {1961})}\BibitemShut {NoStop}%
\bibitem [{\citenamefont {Gallot}\ and\ \citenamefont
  {Grischkowsky}(1999)}]{Gallot1999}%
  \BibitemOpen
  \bibfield  {author} {\bibinfo {author} {\bibfnamefont {G.}~\bibnamefont
  {Gallot}}\ and\ \bibinfo {author} {\bibfnamefont {D.}~\bibnamefont
  {Grischkowsky}},\ }\bibfield  {title} {\bibinfo {title} {Electro-optic
  detection of terahertz radiation},\ }\href
  {https://doi.org/10.1364/josab.16.001204} {\bibfield  {journal} {\bibinfo
  {journal} {J. Opt. Soc. Am. B.}\ }\textbf {\bibinfo {volume} {16}},\ \bibinfo
  {pages} {1204} (\bibinfo {year} {1999})}\BibitemShut {NoStop}%
\bibitem [{\citenamefont {Leitenstorfer}\ \emph {et~al.}(1999)\citenamefont
  {Leitenstorfer}, \citenamefont {Hunsche}, \citenamefont {Shah}, \citenamefont
  {Nuss},\ and\ \citenamefont {Knox}}]{Leitenstorfer1999}%
  \BibitemOpen
  \bibfield  {author} {\bibinfo {author} {\bibfnamefont {A.}~\bibnamefont
  {Leitenstorfer}}, \bibinfo {author} {\bibfnamefont {S.}~\bibnamefont
  {Hunsche}}, \bibinfo {author} {\bibfnamefont {J.}~\bibnamefont {Shah}},
  \bibinfo {author} {\bibfnamefont {M.~C.}\ \bibnamefont {Nuss}},\ and\
  \bibinfo {author} {\bibfnamefont {W.~H.}\ \bibnamefont {Knox}},\ }\bibfield
  {title} {\bibinfo {title} {Detectors and sources for ultrabroadband
  electro-optic sampling: Experiment and theory},\ }\href
  {https://doi.org/10.1063/1.123601} {\bibfield  {journal} {\bibinfo  {journal}
  {Appl. Phys. Lett.}\ }\textbf {\bibinfo {volume} {74}},\ \bibinfo {pages}
  {1516} (\bibinfo {year} {1999})}\BibitemShut {NoStop}%
\bibitem [{\citenamefont {Sulzer}\ \emph {et~al.}(2020)\citenamefont {Sulzer},
  \citenamefont {Oguchi}, \citenamefont {Huster}, \citenamefont {Kizmann},
  \citenamefont {Guedes}, \citenamefont {Liehl}, \citenamefont {Beckh},
  \citenamefont {Moskalenko}, \citenamefont {Burkard}, \citenamefont
  {Seletskiy},\ and\ \citenamefont {Leitenstorfer}}]{Sulzer2020}%
  \BibitemOpen
  \bibfield  {author} {\bibinfo {author} {\bibfnamefont {P.}~\bibnamefont
  {Sulzer}}, \bibinfo {author} {\bibfnamefont {K.}~\bibnamefont {Oguchi}},
  \bibinfo {author} {\bibfnamefont {J.}~\bibnamefont {Huster}}, \bibinfo
  {author} {\bibfnamefont {M.}~\bibnamefont {Kizmann}}, \bibinfo {author}
  {\bibfnamefont {T.~L.~M.}\ \bibnamefont {Guedes}}, \bibinfo {author}
  {\bibfnamefont {A.}~\bibnamefont {Liehl}}, \bibinfo {author} {\bibfnamefont
  {C.}~\bibnamefont {Beckh}}, \bibinfo {author} {\bibfnamefont {A.~S.}\
  \bibnamefont {Moskalenko}}, \bibinfo {author} {\bibfnamefont
  {G.}~\bibnamefont {Burkard}}, \bibinfo {author} {\bibfnamefont {D.~V.}\
  \bibnamefont {Seletskiy}},\ and\ \bibinfo {author} {\bibfnamefont
  {A.}~\bibnamefont {Leitenstorfer}},\ }\bibfield  {title} {\bibinfo {title}
  {Determination of the electric field and its {H}ilbert transform in
  femtosecond electro-optic sampling},\ }\href
  {https://doi.org/10.1103/physreva.101.033821} {\bibfield  {journal} {\bibinfo
   {journal} {Phys. Rev. A}\ }\textbf {\bibinfo {volume} {101}},\ \bibinfo
  {pages} {033821} (\bibinfo {year} {2020})}\BibitemShut {NoStop}%
\bibitem [{\citenamefont {Kempf}\ \emph {et~al.}(2024)\citenamefont {Kempf},
  \citenamefont {Muraviev}, \citenamefont {Breuning}, \citenamefont
  {Schunemann}, \citenamefont {Tenne}, \citenamefont {Leitenstorfer},\ and\
  \citenamefont {Vodopyanov}}]{Kempf2024}%
  \BibitemOpen
  \bibfield  {author} {\bibinfo {author} {\bibfnamefont {H.}~\bibnamefont
  {Kempf}}, \bibinfo {author} {\bibfnamefont {A.}~\bibnamefont {Muraviev}},
  \bibinfo {author} {\bibfnamefont {F.}~\bibnamefont {Breuning}}, \bibinfo
  {author} {\bibfnamefont {P.~G.}\ \bibnamefont {Schunemann}}, \bibinfo
  {author} {\bibfnamefont {R.}~\bibnamefont {Tenne}}, \bibinfo {author}
  {\bibfnamefont {A.}~\bibnamefont {Leitenstorfer}},\ and\ \bibinfo {author}
  {\bibfnamefont {K.}~\bibnamefont {Vodopyanov}},\ }\bibfield  {title}
  {\bibinfo {title} {Direct sampling of femtosecond electric-field waveforms
  from an optical parametric oscillator},\ }\bibfield  {journal} {\bibinfo
  {journal} {APL Photonics}\ }\textbf {\bibinfo {volume} {9}},\ \href
  {https://doi.org/10.1063/5.0189059} {10.1063/5.0189059} (\bibinfo {year}
  {2024})\BibitemShut {NoStop}%
\bibitem [{\citenamefont {Lindel}\ \emph {et~al.}(2020)\citenamefont {Lindel},
  \citenamefont {Bennett},\ and\ \citenamefont {Buhmann}}]{Lindel2020}%
  \BibitemOpen
  \bibfield  {author} {\bibinfo {author} {\bibfnamefont {F.}~\bibnamefont
  {Lindel}}, \bibinfo {author} {\bibfnamefont {R.}~\bibnamefont {Bennett}},\
  and\ \bibinfo {author} {\bibfnamefont {S.~Y.}\ \bibnamefont {Buhmann}},\
  }\bibfield  {title} {\bibinfo {title} {Theory of polaritonic quantum-vacuum
  detection},\ }\href {https://doi.org/10.1103/physreva.102.041701} {\bibfield
  {journal} {\bibinfo  {journal} {Physical Review A}\ }\textbf {\bibinfo
  {volume} {102}},\ \bibinfo {pages} {041701} (\bibinfo {year}
  {2020})}\BibitemShut {NoStop}%
\bibitem [{\citenamefont {Lindel}\ \emph {et~al.}(2021)\citenamefont {Lindel},
  \citenamefont {Bennett},\ and\ \citenamefont {Buhmann}}]{Lindel2021}%
  \BibitemOpen
  \bibfield  {author} {\bibinfo {author} {\bibfnamefont {F.}~\bibnamefont
  {Lindel}}, \bibinfo {author} {\bibfnamefont {R.}~\bibnamefont {Bennett}},\
  and\ \bibinfo {author} {\bibfnamefont {S.~Y.}\ \bibnamefont {Buhmann}},\
  }\bibfield  {title} {\bibinfo {title} {Macroscopic quantum electrodynamics
  approach to nonlinear optics and application to polaritonic quantum-vacuum
  detection},\ }\href {https://doi.org/10.1103/physreva.103.033705} {\bibfield
  {journal} {\bibinfo  {journal} {Phys. Rev. A}\ }\textbf {\bibinfo {volume}
  {103}},\ \bibinfo {pages} {033705} (\bibinfo {year} {2021})}\BibitemShut
  {NoStop}%
\bibitem [{\citenamefont {Virally}\ \emph {et~al.}(2021)\citenamefont
  {Virally}, \citenamefont {Cusson},\ and\ \citenamefont
  {Seletskiy}}]{Virally2021}%
  \BibitemOpen
  \bibfield  {author} {\bibinfo {author} {\bibfnamefont {S.}~\bibnamefont
  {Virally}}, \bibinfo {author} {\bibfnamefont {P.}~\bibnamefont {Cusson}},\
  and\ \bibinfo {author} {\bibfnamefont {D.~V.}\ \bibnamefont {Seletskiy}},\
  }\bibfield  {title} {\bibinfo {title} {Enhanced electro-optic sampling with
  quantum probes},\ }\href {https://doi.org/10.1103/physrevlett.127.270504}
  {\bibfield  {journal} {\bibinfo  {journal} {Phys. Rev. Lett.}\ }\textbf
  {\bibinfo {volume} {127}},\ \bibinfo {pages} {270504} (\bibinfo {year}
  {2021})}\BibitemShut {NoStop}%
\bibitem [{\citenamefont {Beckh}\ \emph {et~al.}(2021)\citenamefont {Beckh},
  \citenamefont {Sulzer}, \citenamefont {Fritzsche}, \citenamefont {Riek},\
  and\ \citenamefont {Leitenstorfer}}]{Beckh2021}%
  \BibitemOpen
  \bibfield  {author} {\bibinfo {author} {\bibfnamefont {C.}~\bibnamefont
  {Beckh}}, \bibinfo {author} {\bibfnamefont {P.}~\bibnamefont {Sulzer}},
  \bibinfo {author} {\bibfnamefont {N.}~\bibnamefont {Fritzsche}}, \bibinfo
  {author} {\bibfnamefont {C.}~\bibnamefont {Riek}},\ and\ \bibinfo {author}
  {\bibfnamefont {A.}~\bibnamefont {Leitenstorfer}},\ }\bibfield  {title}
  {\bibinfo {title} {Analysis of subcycle electro-optic sampling without
  background},\ }\href {https://doi.org/10.1007/s10762-021-00789-4} {\bibfield
  {journal} {\bibinfo  {journal} {J Infrared Millim Terahertz Waves}\ }\textbf
  {\bibinfo {volume} {42}},\ \bibinfo {pages} {701} (\bibinfo {year}
  {2021})}\BibitemShut {NoStop}%
\bibitem [{\citenamefont {Settembrini}\ \emph {et~al.}(2022)\citenamefont
  {Settembrini}, \citenamefont {Lindel}, \citenamefont {Herter}, \citenamefont
  {Buhmann},\ and\ \citenamefont {Faist}}]{Settembrini2022}%
  \BibitemOpen
  \bibfield  {author} {\bibinfo {author} {\bibfnamefont {F.~F.}\ \bibnamefont
  {Settembrini}}, \bibinfo {author} {\bibfnamefont {F.}~\bibnamefont {Lindel}},
  \bibinfo {author} {\bibfnamefont {A.~M.}\ \bibnamefont {Herter}}, \bibinfo
  {author} {\bibfnamefont {S.~Y.}\ \bibnamefont {Buhmann}},\ and\ \bibinfo
  {author} {\bibfnamefont {J.}~\bibnamefont {Faist}},\ }\bibfield  {title}
  {\bibinfo {title} {Detection of quantum-vacuum field correlations outside the
  light cone},\ }\bibfield  {journal} {\bibinfo  {journal} {Nat. Commun.}\
  }\textbf {\bibinfo {volume} {13}},\ \href
  {https://doi.org/10.1038/s41467-022-31081-1} {10.1038/s41467-022-31081-1}
  (\bibinfo {year} {2022})\BibitemShut {NoStop}%
\bibitem [{\citenamefont {Settembrini}\ \emph {et~al.}(2023)\citenamefont
  {Settembrini}, \citenamefont {Herter},\ and\ \citenamefont
  {Faist}}]{Settembrini2023}%
  \BibitemOpen
  \bibfield  {author} {\bibinfo {author} {\bibfnamefont {F.~F.}\ \bibnamefont
  {Settembrini}}, \bibinfo {author} {\bibfnamefont {A.}~\bibnamefont
  {Herter}},\ and\ \bibinfo {author} {\bibfnamefont {J.}~\bibnamefont
  {Faist}},\ }\href {https://doi.org/10.48550/ARXIV.2310.00364} {\bibinfo
  {title} {Third order nonlinear correlation of the electromagnetic vacuum at
  near-infrared frequencies}} (\bibinfo {year} {2023})\BibitemShut {NoStop}%
\bibitem [{\citenamefont {Lindel}\ \emph {et~al.}(2023)\citenamefont {Lindel},
  \citenamefont {Herter}, \citenamefont {Faist},\ and\ \citenamefont
  {Buhmann}}]{Lindel2023}%
  \BibitemOpen
  \bibfield  {author} {\bibinfo {author} {\bibfnamefont {F.}~\bibnamefont
  {Lindel}}, \bibinfo {author} {\bibfnamefont {A.~M.}\ \bibnamefont {Herter}},
  \bibinfo {author} {\bibfnamefont {J.}~\bibnamefont {Faist}},\ and\ \bibinfo
  {author} {\bibfnamefont {S.~Y.}\ \bibnamefont {Buhmann}},\ }\bibfield
  {title} {\bibinfo {title} {Probing vacuum field fluctuations and source
  radiation separately in space and time},\ }\href
  {https://doi.org/10.1103/physrevresearch.5.043207} {\bibfield  {journal}
  {\bibinfo  {journal} {Physical Review Research}\ }\textbf {\bibinfo {volume}
  {5}},\ \bibinfo {pages} {043207} (\bibinfo {year} {2023})}\BibitemShut
  {NoStop}%
\bibitem [{\citenamefont {Lindel}\ \emph {et~al.}(2024)\citenamefont {Lindel},
  \citenamefont {Herter}, \citenamefont {Gebhart}, \citenamefont {Faist},\ and\
  \citenamefont {Buhmann}}]{Lindel2024}%
  \BibitemOpen
  \bibfield  {author} {\bibinfo {author} {\bibfnamefont {F.}~\bibnamefont
  {Lindel}}, \bibinfo {author} {\bibfnamefont {A.}~\bibnamefont {Herter}},
  \bibinfo {author} {\bibfnamefont {V.}~\bibnamefont {Gebhart}}, \bibinfo
  {author} {\bibfnamefont {J.}~\bibnamefont {Faist}},\ and\ \bibinfo {author}
  {\bibfnamefont {S.~Y.}\ \bibnamefont {Buhmann}},\ }\bibfield  {title}
  {\bibinfo {title} {Entanglement harvesting from electromagnetic quantum
  fields},\ }\href {https://doi.org/10.1103/physreva.110.022414} {\bibfield
  {journal} {\bibinfo  {journal} {Physical Review A}\ }\textbf {\bibinfo
  {volume} {110}},\ \bibinfo {pages} {022414} (\bibinfo {year}
  {2024})}\BibitemShut {NoStop}%
\bibitem [{\citenamefont {Schubert}\ \emph {et~al.}(2014)\citenamefont
  {Schubert}, \citenamefont {Hohenleutner}, \citenamefont {Langer},
  \citenamefont {Urbanek}, \citenamefont {Lange}, \citenamefont {Huttner},
  \citenamefont {Golde}, \citenamefont {Meier}, \citenamefont {Kira},
  \citenamefont {Koch},\ and\ \citenamefont {Huber}}]{Schubert2014}%
  \BibitemOpen
  \bibfield  {author} {\bibinfo {author} {\bibfnamefont {O.}~\bibnamefont
  {Schubert}}, \bibinfo {author} {\bibfnamefont {M.}~\bibnamefont
  {Hohenleutner}}, \bibinfo {author} {\bibfnamefont {F.}~\bibnamefont
  {Langer}}, \bibinfo {author} {\bibfnamefont {B.}~\bibnamefont {Urbanek}},
  \bibinfo {author} {\bibfnamefont {C.}~\bibnamefont {Lange}}, \bibinfo
  {author} {\bibfnamefont {U.}~\bibnamefont {Huttner}}, \bibinfo {author}
  {\bibfnamefont {D.}~\bibnamefont {Golde}}, \bibinfo {author} {\bibfnamefont
  {T.}~\bibnamefont {Meier}}, \bibinfo {author} {\bibfnamefont
  {M.}~\bibnamefont {Kira}}, \bibinfo {author} {\bibfnamefont {S.~W.}\
  \bibnamefont {Koch}},\ and\ \bibinfo {author} {\bibfnamefont
  {R.}~\bibnamefont {Huber}},\ }\bibfield  {title} {\bibinfo {title} {Sub-cycle
  control of terahertz high-harmonic generation by dynamical bloch
  oscillations},\ }\href {https://doi.org/10.1038/nphoton.2013.349} {\bibfield
  {journal} {\bibinfo  {journal} {Nature Photonics}\ }\textbf {\bibinfo
  {volume} {8}},\ \bibinfo {pages} {119} (\bibinfo {year} {2014})}\BibitemShut
  {NoStop}%
\bibitem [{\citenamefont {Langer}\ \emph {et~al.}(2016)\citenamefont {Langer},
  \citenamefont {Hohenleutner}, \citenamefont {Schmid}, \citenamefont
  {Poellmann}, \citenamefont {Nagler}, \citenamefont {Korn}, \citenamefont
  {Schüller}, \citenamefont {Sherwin}, \citenamefont {Huttner}, \citenamefont
  {Steiner}, \citenamefont {Koch}, \citenamefont {Kira},\ and\ \citenamefont
  {Huber}}]{Langer2016}%
  \BibitemOpen
  \bibfield  {author} {\bibinfo {author} {\bibfnamefont {F.}~\bibnamefont
  {Langer}}, \bibinfo {author} {\bibfnamefont {M.}~\bibnamefont
  {Hohenleutner}}, \bibinfo {author} {\bibfnamefont {C.~P.}\ \bibnamefont
  {Schmid}}, \bibinfo {author} {\bibfnamefont {C.}~\bibnamefont {Poellmann}},
  \bibinfo {author} {\bibfnamefont {P.}~\bibnamefont {Nagler}}, \bibinfo
  {author} {\bibfnamefont {T.}~\bibnamefont {Korn}}, \bibinfo {author}
  {\bibfnamefont {C.}~\bibnamefont {Schüller}}, \bibinfo {author}
  {\bibfnamefont {M.~S.}\ \bibnamefont {Sherwin}}, \bibinfo {author}
  {\bibfnamefont {U.}~\bibnamefont {Huttner}}, \bibinfo {author} {\bibfnamefont
  {J.~T.}\ \bibnamefont {Steiner}}, \bibinfo {author} {\bibfnamefont {S.~W.}\
  \bibnamefont {Koch}}, \bibinfo {author} {\bibfnamefont {M.}~\bibnamefont
  {Kira}},\ and\ \bibinfo {author} {\bibfnamefont {R.}~\bibnamefont {Huber}},\
  }\bibfield  {title} {\bibinfo {title} {Lightwave-driven quasiparticle
  collisions on a subcycle timescale},\ }\href
  {https://doi.org/10.1038/nature17958} {\bibfield  {journal} {\bibinfo
  {journal} {Nature}\ }\textbf {\bibinfo {volume} {533}},\ \bibinfo {pages}
  {225} (\bibinfo {year} {2016})}\BibitemShut {NoStop}%
\bibitem [{\citenamefont {Kopylov}\ \emph {et~al.}(2024)\citenamefont
  {Kopylov}, \citenamefont {Meier},\ and\ \citenamefont
  {Sharapova}}]{Kopylov2024}%
  \BibitemOpen
  \bibfield  {author} {\bibinfo {author} {\bibfnamefont {D.~A.}\ \bibnamefont
  {Kopylov}}, \bibinfo {author} {\bibfnamefont {T.}~\bibnamefont {Meier}},\
  and\ \bibinfo {author} {\bibfnamefont {P.~R.}\ \bibnamefont {Sharapova}},\
  }\href {https://doi.org/10.48550/ARXIV.2403.05259} {\bibinfo {title} {Theory
  of multimode squeezed light generation in lossy media}} (\bibinfo {year}
  {2024})\BibitemShut {NoStop}%
\bibitem [{\citenamefont {Tziperman}\ \emph {et~al.}(2024)\citenamefont
  {Tziperman}, \citenamefont {Christiansen}, \citenamefont {Kaminer},\ and\
  \citenamefont {Mølmer}}]{Tziperman2024}%
  \BibitemOpen
  \bibfield  {author} {\bibinfo {author} {\bibfnamefont {O.}~\bibnamefont
  {Tziperman}}, \bibinfo {author} {\bibfnamefont {V.~R.}\ \bibnamefont
  {Christiansen}}, \bibinfo {author} {\bibfnamefont {I.}~\bibnamefont
  {Kaminer}},\ and\ \bibinfo {author} {\bibfnamefont {K.}~\bibnamefont
  {Mølmer}},\ }\bibfield  {title} {\bibinfo {title} {Parametric amplification
  of a quantum pulse},\ }\href {https://doi.org/10.1103/physreva.110.053712}
  {\bibfield  {journal} {\bibinfo  {journal} {Physical Review A}\ }\textbf
  {\bibinfo {volume} {110}},\ \bibinfo {pages} {053712} (\bibinfo {year}
  {2024})}\BibitemShut {NoStop}%
\bibitem [{\citenamefont {Yurke}\ and\ \citenamefont
  {Denker}(1984)}]{Yurke1984}%
  \BibitemOpen
  \bibfield  {author} {\bibinfo {author} {\bibfnamefont {B.}~\bibnamefont
  {Yurke}}\ and\ \bibinfo {author} {\bibfnamefont {J.~S.}\ \bibnamefont
  {Denker}},\ }\bibfield  {title} {\bibinfo {title} {Quantum network theory},\
  }\href {https://doi.org/10.1103/physreva.29.1419} {\bibfield  {journal}
  {\bibinfo  {journal} {Physical Review A}\ }\textbf {\bibinfo {volume} {29}},\
  \bibinfo {pages} {1419} (\bibinfo {year} {1984})}\BibitemShut {NoStop}%
\bibitem [{\citenamefont {Weiss}\ \emph {et~al.}(2023)\citenamefont {Weiss},
  \citenamefont {Herbst}, \citenamefont {Schlegel}, \citenamefont {Dannegger},
  \citenamefont {Evers}, \citenamefont {Donges}, \citenamefont {Nakajima},
  \citenamefont {Leitenstorfer}, \citenamefont {Goennenwein}, \citenamefont
  {Nowak},\ and\ \citenamefont {Kurihara}}]{Weiss2023}%
  \BibitemOpen
  \bibfield  {author} {\bibinfo {author} {\bibfnamefont {M.~A.}\ \bibnamefont
  {Weiss}}, \bibinfo {author} {\bibfnamefont {A.}~\bibnamefont {Herbst}},
  \bibinfo {author} {\bibfnamefont {J.}~\bibnamefont {Schlegel}}, \bibinfo
  {author} {\bibfnamefont {T.}~\bibnamefont {Dannegger}}, \bibinfo {author}
  {\bibfnamefont {M.}~\bibnamefont {Evers}}, \bibinfo {author} {\bibfnamefont
  {A.}~\bibnamefont {Donges}}, \bibinfo {author} {\bibfnamefont
  {M.}~\bibnamefont {Nakajima}}, \bibinfo {author} {\bibfnamefont
  {A.}~\bibnamefont {Leitenstorfer}}, \bibinfo {author} {\bibfnamefont
  {S.~T.~B.}\ \bibnamefont {Goennenwein}}, \bibinfo {author} {\bibfnamefont
  {U.}~\bibnamefont {Nowak}},\ and\ \bibinfo {author} {\bibfnamefont
  {T.}~\bibnamefont {Kurihara}},\ }\bibfield  {title} {\bibinfo {title}
  {Discovery of ultrafast spontaneous spin switching in an antiferromagnet by
  femtosecond noise correlation spectroscopy},\ }\bibfield  {journal} {\bibinfo
   {journal} {Nature Communications}\ }\textbf {\bibinfo {volume} {14}},\ \href
  {https://doi.org/10.1038/s41467-023-43318-8} {10.1038/s41467-023-43318-8}
  (\bibinfo {year} {2023})\BibitemShut {NoStop}%
\bibitem [{\citenamefont {Moore}(1920)}]{Moore1920}%
  \BibitemOpen
  \bibfield  {author} {\bibinfo {author} {\bibfnamefont {E.~H.}\ \bibnamefont
  {Moore}},\ }\bibfield  {title} {\bibinfo {title} {On the reciprocal of the
  general algebraic matrix},\ }\href
  {https://doi.org/10.1090/s0002-9904-1920-03322-7} {\bibfield  {journal}
  {\bibinfo  {journal} {Bulletin of the American Mathematical Society}\
  }\textbf {\bibinfo {volume} {26}},\ \bibinfo {pages} {394} (\bibinfo {year}
  {1920})}\BibitemShut {NoStop}%
\bibitem [{\citenamefont {Penrose}(1955)}]{Penrose1955}%
  \BibitemOpen
  \bibfield  {author} {\bibinfo {author} {\bibfnamefont {R.}~\bibnamefont
  {Penrose}},\ }\bibfield  {title} {\bibinfo {title} {A generalized inverse for
  matrices},\ }\href {https://doi.org/10.1017/s0305004100030401} {\bibfield
  {journal} {\bibinfo  {journal} {Mathematical Proceedings of the Cambridge
  Philosophical Society}\ }\textbf {\bibinfo {volume} {51}},\ \bibinfo {pages}
  {406} (\bibinfo {year} {1955})}\BibitemShut {NoStop}%
\bibitem [{\citenamefont {Yang}\ \emph {et~al.}(2025)\citenamefont {Yang},
  \citenamefont {Sharma},\ and\ \citenamefont {Moskalenko}}]{Yang2025}%
  \BibitemOpen
  \bibfield  {author} {\bibinfo {author} {\bibfnamefont {G.}~\bibnamefont
  {Yang}}, \bibinfo {author} {\bibfnamefont {S.}~\bibnamefont {Sharma}},\ and\
  \bibinfo {author} {\bibfnamefont {A.~S.}\ \bibnamefont {Moskalenko}},\ }\href
  {https://doi.org/10.48550/ARXIV.2506.01730} {\bibinfo {title} {Electro-optic
  sampling of the electric-field operator for ultrabroadband pulses of gaussian
  quantum light}} (\bibinfo {year} {2025}),\ \Eprint
  {https://arxiv.org/abs/2506.01730} {arXiv:2506.01730 [quant-ph]} \BibitemShut
  {NoStop}%
\bibitem [{\citenamefont {Boyd}(2019)}]{Boyd2019}%
  \BibitemOpen
  \bibfield  {author} {\bibinfo {author} {\bibfnamefont {R.~W.}\ \bibnamefont
  {Boyd}},\ }\href@noop {} {\emph {\bibinfo {title} {Nonlinear optics}}}\
  (\bibinfo  {publisher} {Elsevier Science \& Technology},\ \bibinfo {year}
  {2019})\BibitemShut {NoStop}%
\bibitem [{\citenamefont {Marple}(1964)}]{Marple1964}%
  \BibitemOpen
  \bibfield  {author} {\bibinfo {author} {\bibfnamefont {D.~T.~F.}\
  \bibnamefont {Marple}},\ }\bibfield  {title} {\bibinfo {title} {Refractive
  index of {ZnSe}, {ZnTe}, and {CdTe}},\ }\href
  {https://doi.org/10.1063/1.1713411} {\bibfield  {journal} {\bibinfo
  {journal} {J. Appl. Phys.}\ }\textbf {\bibinfo {volume} {35}},\ \bibinfo
  {pages} {539} (\bibinfo {year} {1964})}\BibitemShut {NoStop}%
\bibitem [{\citenamefont {Adesso}\ \emph {et~al.}(2004)\citenamefont {Adesso},
  \citenamefont {Serafini},\ and\ \citenamefont {Illuminati}}]{Adesso2004}%
  \BibitemOpen
  \bibfield  {author} {\bibinfo {author} {\bibfnamefont {G.}~\bibnamefont
  {Adesso}}, \bibinfo {author} {\bibfnamefont {A.}~\bibnamefont {Serafini}},\
  and\ \bibinfo {author} {\bibfnamefont {F.}~\bibnamefont {Illuminati}},\
  }\bibfield  {title} {\bibinfo {title} {Extremal entanglement and mixedness in
  continuous variable systems},\ }\href
  {https://doi.org/10.1103/physreva.70.022318} {\bibfield  {journal} {\bibinfo
  {journal} {Physical Review A}\ }\textbf {\bibinfo {volume} {70}},\ \bibinfo
  {pages} {022318} (\bibinfo {year} {2004})}\BibitemShut {NoStop}%
\bibitem [{\citenamefont {Adesso}\ and\ \citenamefont
  {Datta}(2010)}]{Adesso2010}%
  \BibitemOpen
  \bibfield  {author} {\bibinfo {author} {\bibfnamefont {G.}~\bibnamefont
  {Adesso}}\ and\ \bibinfo {author} {\bibfnamefont {A.}~\bibnamefont {Datta}},\
  }\bibfield  {title} {\bibinfo {title} {Quantum versus classical correlations
  in gaussian states},\ }\href {https://doi.org/10.1103/physrevlett.105.030501}
  {\bibfield  {journal} {\bibinfo  {journal} {Physical Review Letters}\
  }\textbf {\bibinfo {volume} {105}},\ \bibinfo {pages} {030501} (\bibinfo
  {year} {2010})}\BibitemShut {NoStop}%
\bibitem [{\citenamefont {Sperling}\ \emph {et~al.}(2019)\citenamefont
  {Sperling}, \citenamefont {Perez-Leija}, \citenamefont {Busch},\ and\
  \citenamefont {Silberhorn}}]{Sperling2019}%
  \BibitemOpen
  \bibfield  {author} {\bibinfo {author} {\bibfnamefont {J.}~\bibnamefont
  {Sperling}}, \bibinfo {author} {\bibfnamefont {A.}~\bibnamefont
  {Perez-Leija}}, \bibinfo {author} {\bibfnamefont {K.}~\bibnamefont {Busch}},\
  and\ \bibinfo {author} {\bibfnamefont {C.}~\bibnamefont {Silberhorn}},\
  }\bibfield  {title} {\bibinfo {title} {Mode-independent quantum entanglement
  for light},\ }\href {https://doi.org/10.1103/physreva.100.062129} {\bibfield
  {journal} {\bibinfo  {journal} {Physical Review A}\ }\textbf {\bibinfo
  {volume} {100}},\ \bibinfo {pages} {062129} (\bibinfo {year}
  {2019})}\BibitemShut {NoStop}%
\bibitem [{\citenamefont {Law}\ \emph {et~al.}(2000)\citenamefont {Law},
  \citenamefont {Walmsley},\ and\ \citenamefont {Eberly}}]{Law2000}%
  \BibitemOpen
  \bibfield  {author} {\bibinfo {author} {\bibfnamefont {C.~K.}\ \bibnamefont
  {Law}}, \bibinfo {author} {\bibfnamefont {I.~A.}\ \bibnamefont {Walmsley}},\
  and\ \bibinfo {author} {\bibfnamefont {J.~H.}\ \bibnamefont {Eberly}},\
  }\bibfield  {title} {\bibinfo {title} {Continuous frequency entanglement:
  Effective finite hilbert space and entropy control},\ }\href
  {https://doi.org/10.1103/physrevlett.84.5304} {\bibfield  {journal} {\bibinfo
   {journal} {Physical Review Letters}\ }\textbf {\bibinfo {volume} {84}},\
  \bibinfo {pages} {5304} (\bibinfo {year} {2000})}\BibitemShut {NoStop}%
\bibitem [{\citenamefont {Bernstein}\ and\ \citenamefont
  {So}(1993)}]{Bernstein1993}%
  \BibitemOpen
  \bibfield  {author} {\bibinfo {author} {\bibfnamefont {D.}~\bibnamefont
  {Bernstein}}\ and\ \bibinfo {author} {\bibfnamefont {W.}~\bibnamefont {So}},\
  }\bibfield  {title} {\bibinfo {title} {Some explicit formulas for the matrix
  exponential},\ }\href {https://doi.org/10.1109/9.233156} {\bibfield
  {journal} {\bibinfo  {journal} {IEEE Transactions on Automatic Control}\
  }\textbf {\bibinfo {volume} {38}},\ \bibinfo {pages} {1228} (\bibinfo {year}
  {1993})}\BibitemShut {NoStop}%
\bibitem [{\citenamefont {Horoshko}\ \emph {et~al.}(2024)\citenamefont
  {Horoshko}, \citenamefont {Kolobov}, \citenamefont {Parigi},\ and\
  \citenamefont {Treps}}]{Horoshko2024}%
  \BibitemOpen
  \bibfield  {author} {\bibinfo {author} {\bibfnamefont {D.~B.}\ \bibnamefont
  {Horoshko}}, \bibinfo {author} {\bibfnamefont {M.~I.}\ \bibnamefont
  {Kolobov}}, \bibinfo {author} {\bibfnamefont {V.}~\bibnamefont {Parigi}},\
  and\ \bibinfo {author} {\bibfnamefont {N.}~\bibnamefont {Treps}},\ }\bibfield
   {title} {\bibinfo {title} {Few-mode squeezing in type-i parametric
  downconversion by complete group velocity matching},\ }\href
  {https://doi.org/10.1364/ol.528280} {\bibfield  {journal} {\bibinfo
  {journal} {Optics Letters}\ }\textbf {\bibinfo {volume} {49}},\ \bibinfo
  {pages} {4078} (\bibinfo {year} {2024})}\BibitemShut {NoStop}%
\bibitem [{\citenamefont {Roman-Rodriguez}\ \emph {et~al.}(2021)\citenamefont
  {Roman-Rodriguez}, \citenamefont {Brecht}, \citenamefont {K}, \citenamefont
  {Silberhorn}, \citenamefont {Treps}, \citenamefont {Diamanti},\ and\
  \citenamefont {Parigi}}]{RomanRodriguez2021}%
  \BibitemOpen
  \bibfield  {author} {\bibinfo {author} {\bibfnamefont {V.}~\bibnamefont
  {Roman-Rodriguez}}, \bibinfo {author} {\bibfnamefont {B.}~\bibnamefont
  {Brecht}}, \bibinfo {author} {\bibfnamefont {S.}~\bibnamefont {K}}, \bibinfo
  {author} {\bibfnamefont {C.}~\bibnamefont {Silberhorn}}, \bibinfo {author}
  {\bibfnamefont {N.}~\bibnamefont {Treps}}, \bibinfo {author} {\bibfnamefont
  {E.}~\bibnamefont {Diamanti}},\ and\ \bibinfo {author} {\bibfnamefont
  {V.}~\bibnamefont {Parigi}},\ }\bibfield  {title} {\bibinfo {title}
  {Continuous variable multimode quantum states via symmetric group velocity
  matching},\ }\href {https://doi.org/10.1088/1367-2630/abef96} {\bibfield
  {journal} {\bibinfo  {journal} {New Journal of Physics}\ }\textbf {\bibinfo
  {volume} {23}},\ \bibinfo {pages} {043012} (\bibinfo {year}
  {2021})}\BibitemShut {NoStop}%
\end{thebibliography}%

\end{document}